\documentclass[reprint,nofootinbib,onecolumn,amsmath,amssymb,aps]{revtex4-2}
\usepackage{graphicx}  
\usepackage{dcolumn}   
\usepackage{bm}        
\usepackage{amssymb}   
\usepackage{amsmath}
\usepackage{comment}
\usepackage{graphicx}
\usepackage{color}
\usepackage{slashed}
\usepackage{hyperref}
\usepackage[dvipsnames]{xcolor}
\usepackage{wasysym}
\usepackage[braket, qm]{qcircuit}
\usepackage{braket}
\usepackage{soul}
\usepackage{multirow}

\newcommand{\Eq}[1]{Eq.(\ref{#1})}
\newcommand{\Eqs}[2]{Eqs.(\ref{#1}-\ref{#2})}
\newcommand{\Fig}[1]{Fig. (\ref{#1})}

\newcommand{\xb}{\mathbf{x}}
\newcommand{\pb}{{\mathbf{p}}}

\newcommand{\nb}{{\mathbf{n}}}

\newcommand{\ua}{\mathord{\uparrow}}
\newcommand{\da}{\mathord{\downarrow}}

\newcommand{\nn}{\nonumber\\}

\begin{document}
\author{Jo\~{a}o Barata}
\email{joaolourenco.henriques@usc.es}
\affiliation{Instituto Galego de F\'{i}sica de Altas Enerx\'{i}as (IGFAE), Universidade de Santiago de Compostela, E-15782 Galicia, Spain}
\author{Niklas Mueller}
\email{niklasmu@umd.edu}
\affiliation{Department of Physics, University of Maryland, College Park, MD 20742, USA}
\author{Andrey Tarasov}
\email{tarasov.3@osu.edu}
\affiliation{Department of Physics, The Ohio State University, Columbus, OH 43210, USA}
\affiliation{Joint BNL-SBU Center for Frontiers in Nuclear Science (CFNS) at Stony Brook University, Stony Brook, NY 11794, USA}
\author{Raju Venugopalan}
\email{raju@bnl.gov}
\affiliation{Physics Department, Brookhaven National Laboratory, Bldg. 510A, Upton, NY 11973, USA}

\title{Single-particle digitization strategy for quantum computation \\of  a $\phi^4$ scalar field theory}
\begin{abstract}

Motivated by the parton picture of high energy quantum chromodynamics, we develop a single-particle digitization strategy for the efficient quantum simulation of relativistic scattering processes in a $d+1$ dimensional scalar $\phi^4$ field theory. We work out quantum algorithms for initial state preparation, time evolution and final state measurements. We outline  a non-perturbative renormalization strategy in this single-particle framework. 
\end{abstract}
\maketitle
%
%
\section{Introduction}
Significant effort has been invested towards studying problems in quantum chemistry~\cite{whitfield2011simulation,kassal2011simulating,o2016scalable,hempel2018quantum}, condensed matter physics~\cite{lewenstein2007ultracold,bloch2012quantum,bao2015universal}, cosmology~\cite{fischer2004quantum,jain2007analog,chatrchyan2020analog}, and in high energy and nuclear physics~\cite{carlson2018quantum,mccaskey2019quantum,dumitrescu2018cloud,rico2018so,cloet2019opportunities,matchev2020quantum},
with digital quantum computers and analog quantum simulators~\cite{kielpinski2002architecture,monroe2002quantum,blais2004cavity,cirac2012goals,hauke2012can,preskill2018quantum}. A major motivation is to deepen our understanding of conventionally intractable features of the ground state properties of strongly correlated many-body systems such as the spectrum of bound states. Another is to advance the state-of-the art in scattering problems, which provide dynamical information on such complex systems. 

In this work, our focus will be on the problem of developing quantum algorithms for high energy scattering  and multi-particle production in relativistic quantum field theory.  Underlying our work is the promising yet distant goal of extracting  dynamical information on the properties of hadrons and nuclei in quantum chromodynamics (QCD).  

Examples of scattering problems in QCD where quantum information science can accelerate our present computational capabilities are low energy scattering in nuclear many-body systems ~\cite{roggero2018linear,roggero2020quantum}, the thermalization process in ultrarelativistic ion-ion collisions~\cite{Berges:2020fwq}, studies of the structure of nuclear matter probed in Deeply Inelastic Scattering (DIS) of electrons off protons and nuclei~\cite{Breidenbach:1969kd,Bjorken:1968dy,Bjorken:1969ja,Gross:1973id,Blumlein:2012bf,mueller2020deeply,lamm2020parton,kreshchuk2020quantum} and the fragmentation of quarks and gluons into jets of hadrons~\cite{Bassetto:1984ik,Dokshitzer:1987nm}. 
For instance, both jet fragmentation functions and DIS structure functions require one to 
compute autocorrelation functions of currents in Minkowski spacetime; this poses a challenge to classical Monte Carlo methods that are constructed to compute Euclidean spacetime correlators~\cite{Metz:2016swz,Winter:2017bfs,Ji:2013dva,Alexandrou:2015rja,Chen:2016utp,Radyushkin:2016hsy,Lin:2017snn,Detmold:2019ghl}.

 Quantum devices have the potential to overcome the limitations of classical computers in addressing many of the above problems. However presently their  limitation is that scattering problems involve a vast range of spatial (momentum) and temporal (energy) scales and require that a large number of (local) quantum field operators be quantum simulated. This is challenging with present day NISQ era technology restricted to few tens of non-error-corrected qubits~\cite{preskill2018quantum}.
 
As discussed in seminal papers by Jordan, Lee and Preskill~\cite{jordan2011quantum,jordan2012quantum},
quantum simulating scattering problems in relativistic quantum field theories requires a lattice discretization and, in the case of a bosonic theory, the truncation of the local Hilbert space of field operators. One can view such a digitization as defining a low energy effective theory, in the sense of a generalized renormalization group (RG)~\cite{klco2020fixed}. We will argue here that, from this viewpoint, a digitization scheme does not necessarily need to be based on a decomposition of local field operators but, more generally, should be based on the most economical implementation of the relevant directions of the RG for a specific problem.

Pursuant to this goal, we will explore a digitization strategy for the bosonic field theory of a real scalar field in $d+1$ spacetime dimensions based on a generalization of relativistic Bose-symmetrized ``single particle states" previously discussed by us in~\cite{mueller2020deeply}.  For a wide class of problems, this digitization requires resources that are only logarithmic in volume $\mathcal{V}$ (but linear in particle number), as opposed to field operator based approaches that depend linearly on the volume.
We will discuss a strategy for initial state preparation, time evolution and measurement for scattering processes in high energy physics. Our time evolution algorithm has a gate complexity similar to that of~\cite{jordan2011quantum,jordan2012quantum}; however because the basis states are eigenstates of the free Hamiltonian and of the particle number operator,  initial state preparation and measurement are particularly simple.

Our approach is unusual in the sense that relativistic many body systems are usually described by field operators within quantum field theory. In relativistic theories, particle number is not conserved and relativistic single particle states in position space are  not Fourier conjugates of single particle states in momentum space. While it seems that these properties render quantum simulation of relativistic quantum field theory fundamentally different to those in quantum chemistry~\cite{whitfield2011simulation,kassal2011simulating,o2016scalable,hempel2018quantum} or in nonrelativistic
low energy nuclear physics~\cite{carlson2018quantum,mccaskey2019quantum,dumitrescu2018cloud,roggero2019dynamic}, we will demonstrate here that this is not the case and that we are able to utilize 
algorithms that are conceptually similar.  

A powerful motivation underlying our approach is the single particle picture~\cite{Strassler:1992zr} behind the well known Feynman diagram techniques to compute scattering cross-sections in high energy physics at weak coupling. Because the computational complexity of Feynman diagram computations grows factorially with the required precision, their computation presents another opportunity for a quantum advantage~\cite{jordan2011quantum,jordan2012quantum}. Since as noted, scattering problems 
can be formulated in terms of  Minkowski space field correlators, a first principles path integral computation with classical Monte Carlo techniques is challenging. Albeit considering a simpler theory, our ultimate aim is to apply this approach to quantum simulate scattering problems in quantum chromodynamics; a first step towards this goal would be a hybrid strategy combining a quantum treatment of some of the scattering degrees of freedom with a classical treatment of the rest~\cite{mueller2020deeply}. A relevant analogy in this regard is the simulation of quantum impurities in strongly correlated condensed matter systems~\cite{PhysRevX.6.031045}, or the simulation of open quantum systems in heavy-ion collisions \cite{deJong:2020tvx}. In light of the many challenges of NISQ era computing, the digitization strategy we will present may therefore offer a useful compromise between being able to make progress in a limited class of problems in high energy physics with restricted resources and conceptual simplicity versus simulating any possible problem in quantum field theory in full generality.  

This manuscript is organized as follows: In Section \ref{sec:DIS_Smatrix}, we will discuss the conceptual basis of our approach to quantum computing scattering cross-sections in high energy physics. Our digitization strategy is discussed in Section \ref{sec:strategies_for_quantum_simulation}. In Section \ref{sec:algorithm}, we will present the single particle digitization algorithm for quantum computing scattering cross-sections:  state preparation is discussed in Section \ref{sec:InitialStatePrep}, the implementation of the time evolution operator as a quantum circuit in Section \ref{sec:TimeEvolution}, the strategy to extract cross-sections through measurements in Section \ref{sec:Measurement} and renormalization aspects of the problem in Section \ref{sec:Renormalization}.
In Section \ref{sec:outlook}, we summarize our results and discuss extensions of this approach to include theories with fermion and gauge fields with internal symmetries.

We elaborate on several of ths discussions in the main text in multiple appendices. In Appendix \ref{app:detailsSingleParticle}, we provide details of the single particle digitization strategy. In Appendix \ref{app:StochasticStatePrep}, we discuss
the state preparation algorithm in greater detail. Appendices \ref{app:kineticterm}, \ref{app:squeezing} and 
\ref{app:interactionterm} contain details of 
the algorithm for the time evolution operator. 
Finally in Appendix \ref{app:renormalization}, we provide further details of the renormalization procedure.

\section{High Energy Scattering}\label{sec:DIS_Smatrix}
Understanding the structure of matter at the sub-nucleon scales of nuclear and particle physics requires a wide range of scattering experiments. 
The theoretical foundations of these scattering problems is well developed within the framework of relativistic quantum field theory. The simplest formulation of a scattering process is through the S-matrix,
\begin{align}
S_{\beta\alpha} \equiv \langle \Psi^{\rm out}_\beta | \Psi^{\rm in}_\alpha \rangle\,,
\end{align}
defined as the overlap of asymptotic in- ($|\Psi^{\rm in}_\alpha\rangle$) and out- ($|\Psi_\beta^{\rm out}\rangle$) states, that are time-independent  eigenstates of the Hamiltonian $H=H_0 + V$. 

In the Heisenberg picture, all non-trivial information on these states is encoded in the  Lippmann-Schwinger equation~\cite{lippmann1950variational,newton2013scattering},
 \begin{align}\label{eq:LippmannSchwingerEquation}
|\Psi_\alpha^{\rm in/out}\rangle=| \phi_\alpha\rangle +  G_0  V |\Psi_\alpha^{\rm in/out}\rangle\nonumber 
=(V- VG_0 V)^{-1}V| \phi_\alpha\rangle\,,
 \end{align}
where $|\phi_\alpha\rangle$ are single particle eigenstates of the free Hamiltonian $H_0$ satisfying 
$H_0 |\phi_\alpha\rangle = E_\alpha | \phi_\alpha\rangle $, 
 $G_0 \equiv (E_\alpha-H_0 \pm i\epsilon)^{-1}$ and $ V- VG_0 V$ is the Schwinger operator. The $S$-matrix can also be expressed as 
 \begin{align}
S_{\beta\alpha}=\delta_{\alpha\beta} - 2\pi i\, \delta(E_\alpha - E_\beta) \, T_{\beta\alpha}\,,
 \end{align}
 where  energy  conservation is explicit, and the T-matrix is defined as 
 \begin{align}\label{eq:TMatrixDefinition}
T_{\beta\alpha} &= \langle \phi_\beta | V | \Psi^{\rm out}_\alpha\rangle = \langle \Psi^{\rm in}_\beta | V | \phi_\alpha\rangle
=  \langle \Psi^{\rm in}_\beta | (V- VG_0 V) | \Psi^{\rm out}_\alpha\rangle\,.
 \end{align} 
The cross-section for a scattering process $\alpha \rightarrow \beta$ is given by the modulus squared of  $T_{\beta\alpha}$ (multiplied by kinematic factors),
 \begin{align}\label{eq:TmatrixSquared}
 |T_{\beta\alpha}|^2 = \langle  \Psi^{\rm in}_\alpha|\,  ( V- VG_0 V)  P_\beta^{\rm out} ( V- VG_0 V)^\dagger \,  |\Psi^{\rm in}_\alpha\rangle\,,
 \end{align}
 with  $P_\beta^{\rm out} = | \Psi^{\rm out}_\beta\rangle \langle \Psi^{\rm out}_\beta|$. 
The $T$-matrix elements in Eq.~(\ref{eq:TMatrixDefinition}) can be computed by solving the Lippmann-Schwinger equation. This can be achieved using analytic perturbative techniques such as the Born expansion~\cite{newton2013scattering} or non-perturbatively using Schwinger's variational principle~\cite{lippmann1950variational}, the Schwinger-Lanczos~\cite{meyer1991schwinger} or R-matrix approaches~\cite{wigner1947higher}. 

Quantum variants of these methods are currently under development; for an implementation of the Quantum-Lanczos algorithm in a scattering problem, see ~\cite{Yeter-Aydeniz:2020jte}.
We will proceed here with the formulation of the quantum scattering problem in the time-dependent Schr\"{o}dinger picture~\cite{Jordan:2011ci,Jordan:2011ne}.

Before we proceed in that direction, we note that our single-particle digitization strategy for the S-matrix can be mapped on to a virial expansion, which is a ``cluster" expansion in powers of the density that captures the many-body properties of a system at low particle densities. It is particularly successful in reproducing their ground state properties\footnote{The extension of the virial expansion to non-equilibrium autocorrelation functions is highly non-trivial; an excellent review of this topic can be found in \cite{Dorfman,Dorfman2}.}, which are expressed as a density expansion in the n-th order virial coefficients\footnote{This expression, originally formulated as a quantum many-body extension~\cite{Huang:1957im,Huang:1957zz},  to the famous Beth-Uhlenbeck formula was later  generalized to discuss relativistic many-body $n\leftrightarrow m$ inelastic processes~\cite{Dashen:1969ep,Dashen:1974yy}.}
\begin{equation}
   b_n \propto \left[S^\dagger \frac{\partial S}{\partial E}\right]_n \,,
\end{equation}
where $\left[S^\dagger \frac{\partial S}{\partial E}\right]_n$  of $n\rightarrow n$ scattering particles. Thus because our single particle strategy is optimal for capturing the many-body dynamics of a relativistic theory at low occupancies, a computation of $n\rightarrow n$ scattering matrix elements will allow us to determine ground state properties in our framework with the same range of validity as the virial expansion. 
Indeed, one can in principle go further and test the validity of this expansion relative to a direct computation of ground state properties of relativistic many-body systems in our 
framework.




 \subsection{Schr\"{o}dinger picture of S-matrix scattering}

In the Schr\"{o}dinger picture, the scattering process is described in terms of time-dependent wavepackets 
 \begin{align}
 \label{eq:wavepacket}
 | \Psi_g^{\rm in/out}(t) \rangle \equiv \int d\alpha \,  g(\alpha) e^{-i E_\alpha t}\,  | \Psi^{\rm in/out}_\alpha \rangle\,,
 \end{align}
where $g(\alpha)$ is a function that describes the localization of the wavepacket. 
In this approach, the Lippmann-Schwinger equation can be expressed as
 \begin{align}\label{eq:timedependentLSeq}
 | \Psi_g^{\rm in/out}(t)\rangle = &| \phi_g(t)\rangle + \int_{0}^{\infty}dT \, e^{\pm i(H_0 \mp  i\epsilon)T} \,  
 V  | \Psi_g^{\rm in/out}(t\mp T)\rangle\,,
 \end{align}
 where $|\phi_g(t)\rangle$ is defined identically as in Eq.~(\ref{eq:wavepacket}). The in-wavepacket satisfies the boundary condition $| \Psi_g^{\rm in}(- \infty)\rangle= | \phi_g(-\infty)\rangle $ at negative infinity and the out-wavepacket  satisfies a similar condition at positive infinity.
In the Schr\"{o}dinger picture, one may interpret $V(T)\equiv V e^{-  \epsilon |T|} $ as adiabatically turning on the interaction to obtain $| \Psi^{\rm in}(t)\rangle$ from  evolution of the initial condition $|\phi_g(-\infty)\rangle$ using  \Eq{eq:timedependentLSeq} and likewise, in reverse, for $|\Psi^{\rm out}_g \rangle$. This approach, employing single particle wavepackets,  will form the basis of our algorithm in Section \ref{sec:algorithm}.

 \subsection{Spacetime picture of scattering experiments at high energies}\label{sec:SpaceTimePictureScattering}
 
At high energies, the gap of single particle states to continuum particle-antiparticle pairs becomes small, and a description of scattering in terms of the second quantized language of quantum field operators appears natural. Quantum simulation of this problem is desirable because of the well-known challenges of classical computation.

However interestingly at high energies, for a wide class of scattering problems, single particle digitization strategies applied at lower energies may be viable and indeed desirable. The latter can be understood straightforwardly in the context of the scattering of two protons at the ultrarelativistic energies of the Large Hadron Collider (LHC). The wavepackets of the two colliding protons can be constructed formally, along the lines of Eq.~(\ref{eq:wavepacket}); however such wavepackets, as observed by Bjorken and Feynman, for many final states of interest in scattering at high energies are accurately described in terms of the scattering of pointlike ``parton" (quark, antiquark and gluon) constituents within the protons that are eigenstates of the free QCD Hamiltonian~\cite{Bjorken:1969ja,feynman2018photon}. In this parton picture of high energy scattering, as we will now discuss, the switch-on/off time $\tau_0$ and the interaction time $\tau_I$ can be related to physical time scales.

These time scales are best understood in the context\footnote{We refer readers unfamiliar with DIS to our paper \cite{mueller2020deeply} for some of the key references and for a discussion of aspects of this scattering problem from a quantum computing perspective.} of the deeply inelastic scattering (DIS) of electrons (and other leptons) off protons and nuclei. In DIS, the incoming electron emits a virtual photon that strikes a quark or antiquark within the hadron, thereby providing information on the quark and (indirectly) gluon distributions within. The relevant DIS kinematic variables are the momentum resolution $Q$ of the probe (with $Q^2 \gg \Lambda_{\rm QCD}^2$, the QCD confinement scale) and the Bjorken variable $x_{\rm Bj}\approx Q^2/s$, where $\sqrt{s}$ is the DIS center-of-mass energy. 

In the QCD parton model,  $x_{\rm Bj}\sim x$ is the momentum fraction of the hadron carried by the struck quark or antiquark. The DIS cross-section at large $x_{\rm Bj}$ corresponds to the projection of the hadron wavefunction into a Fock state that is a direct product state of single-particle parton states that make up the hadron's quantum numbers. In contrast, the small  $x_{\rm Bj}$ (high energy) cross-section corresponds to the scattering of the virtual photon off a Fock state containing a large number of partons, most of which carry a small fraction ($x\ll 1$) of the hadron's momentum. 

The physically motivated time required to probe fluctuations of the proton into differing parton configurations is the Ioffe time~\cite{Ioffe:1969kf} $\tau_0\sim \tau_{\rm Ioffe} = 1/(2 M_p\,x_{\rm Bj})$, with $M_p$ the proton mass; in the DIS example, this gives the coherence time of the fluctuation of the virtual photon into a parton state in the rest frame of the proton or nuclear target\footnote{In general, the coherence time is a distribution, with the stated value being the upper bound. For fluctuations of the virtual photon into a highly excited QCD Fock state, coherence time estimates are considerably shorter~\cite{Kovchegov:2001dh}.}. Likewise in DIS, the interaction time of the probe is the typically much shorter time scale $\tau_I \sim 1/Q$.  A minimal bound on this time scale is $\tau_W$, the Wigner time delay defined as $\partial S /\partial E$, where $E$ denotes energy, in the virial expansion, we discussed previously, of a scattering process of $n\rightarrow m$ particles
\cite{Dorfman,Dorfman2,Huang:1957im,Huang:1957zz,Wigner:1955zz,Dashen:1969ep,Dashen:1974yy}~\footnote{The generalization of these ideas relating asymptotic scattering phase shifts to differences in energy levels of static  quantities in a finite box was pioneered by Luscher~\cite{Luscher:1986pf,Luscher:1990ux}. It is an active area of research in lattice gauge theory~\cite{Hansen:2019nir,Briceno:2017max}, recently discussed in the context of quantum computing~\cite{Briceno:2020rar}.}

The parton picture is manifest when field theories are quantized~\cite{Brodsky:1997de} on a lightlike surface $x^+=0$, with the lightcone Hamiltonian $P^- = P_0^-+V$, defined as the generator of translations in $x^+$. The Galilean subgroup of the lightfront Poincar\'{e} group is isomorphic to the symmetry group of two dimensional quantum mechanics~\cite{Susskind:1967rg}, allowing one to formulate scattering problems in quantum field theory in the language of nonrelativistic quantum mechanics. In particular, due to time dilation at high energies, the lightfront potential is suppressed (by powers of the energy) relative to the kinetic term; Fock states, which are single particle direct product states of partons, therefore provide a good eigenbasis for high energy scattering~\cite{Bjorken:1970ah}. 

Even though the single particle picture of high energy scattering finds an elegant representation in lightfront quantization, it is not restricted to it. It is a generic feature of Feynman diagrams in perturbation theory~\cite{Weinberg:1966fm} and more recently of so-called ``conformal truncation"  methods~\cite{Fitzpatrick:2018xlz,Anand:2020gnn,Liu:2020eoa} introduced in the context of conformal field theory~\cite{James_2018}. This property of high energy scattering motivates exploring a single particle digitization strategy, which we will discuss at length in the rest of this paper in conventional equal time quantization\footnote{For recent work on quantum computing in lightcone quantization, see \cite{Kreshchuk:2020dla,Kreshchuk:2020kcz}.}. Further, as detailed in Section~\ref{sec:Measurement}, this approach is particularly valuable in performing measurements on a quantum computer. 

Our single particle digitization strategy will encounter significant challenges when applied to gauge theories.  Concretely, when applied to the digitization of theories coupled to gauge fields, the presented time-evolution strategy must be modified, as we discuss further in section~\ref{sec:outlook}. Nevertheless one may be able to make progress employing this strategy in physical problems where hybrid quantum/classical techniques are applicable; in QCD, these include these include  Effective Field Theories (EFTs) for jet physics~\cite{Bauer:2000yr},  high parton densities (small $x$)~\cite{Gelis:2010nm} and at finite temperature~\cite{Braaten:1989mz}, and a lattice EFT for computing parton distributions~\cite{Ji:2013dva}.

\section{Single-Particle Strategy }
\label{sec:strategies_for_quantum_simulation}

In this Section, and in the next, we will develop a single particle digitization strategy for a relativistic (real) scalar field theory with local quartic interactions in $d+1$ spacetime dimensions. The  Hamiltonian for this theory is given by 
\begin{align}\label{eq:HamiltonianPhi4Cont}
{\bar H}=\int {\rm d}^d\textbf{x} \Big[ \frac{\pi_\mathbf{x}^2}{2}+ \frac{1}{2}(\nabla \phi_\mathbf{x})^2 + \frac{\overline{m}^2}{2} \phi_\mathbf{x}^2 + \frac{\overline{\lambda}}{4!} \phi^4_\mathbf{x} \Big]\,,
\end{align}
where $\overline{m}$ and  $\overline{\lambda}$  are the (bare) mass and quartic coupling, and $\nabla $ is the spatial gradient operator in $d$ dimensions.
The Heisenberg field operators are
\begin{align}\label{eq:PhiOperatorCont}
\phi_\xb =\int  \frac{{\rm d}^d \pb}{(2\pi)^d} \frac{1}{\sqrt{2 \, \overline{\omega}_\pb}} \Big[ a_\pb+ a_{-\pb}^\dagger \Big]e^{i\pb\cdot \xb}\,,
\end{align}
which satisfies, with its canonical conjugate operator $\pi_\xb$, the equal-time commutation relations $[\phi_\xb,\pi_\mathbf{y}]=i \delta^{(d)}(\xb-\mathbf{y})$. The annihilation (creation) operators $a_\pb$ ( $a^\dagger_\pb$) are momentum-space Fock operators, corresponding to a set of harmonic oscillators with frequency $\overline{\omega}_\pb=\sqrt{\pb^2+\overline{m}^2}$  and commutation relations $[a_\pb, a^\dagger_{\mathbf{k}}]=(2\pi)^d \delta^{(d)}(\pb-\mathbf{k})$,  $[a_\pb, a_{\mathbf{k}}]=[a^\dagger_\pb, a^\dagger_{\mathbf{k}}]=0$.  
Single particle states are defined as 
\begin{align}\label{eq:stateDimLessLorentz}
{| \pb}\rangle^{\rm phys} \equiv \sqrt{2 \,\overline{\omega}_\pb} a^\dagger_{\pb} | {\rm{vac}}\rangle\,,
\end{align}
which satisfy the relativistic  normalization condition $\langle \pb | \mathbf{k} \rangle^{\rm phys} = 2 \, {\overline{\omega}}_\pb \, \delta^{(3)}(\pb-\mathbf{k}) $, where $|\rm vac\rangle$ denotes the Fock vacuum.

We discretize the theory on a spatial lattice of size $N_s^d$ 
and express the Hamiltonian (in dimensionless units) as 
\begin{align}\label{eq:latticeHamiltonian}
H \equiv {a_s\bar H}=  \sum_{\textbf{n}} \Big[ \frac{1}{2}\pi^2_{\textbf{n}} + \frac{1}{2} ({\nabla}\phi_{{\textbf{n}}})^2  + \frac{{m}^2}{2} \phi^2_{\textbf{n}} + \frac{{\lambda}}{4!}\phi^4_{\textbf{n}} \Big]\,,
\end{align}
where ${m} = \overline{m} \, a_s$, ${\lambda} = \overline{\lambda} \, a^{4-d}_s$ are dimensionless bare mass and coupling parameters, $a_s$ the lattice spacing and ${\textbf{n}}=( n_1,\dots , n_d)$, $n_i\in[0,N_s-1]$ labels a point $\xb = \mathbf{n}a_s$ on the lattice.  We will likewise define a momentum space lattice vector  $\mathbf{q}=(q_1,\dots,q_d)$, $q_i\in[-\frac{N_s}{2},\frac{N_s}{2}-1]$. The lattice field operators are
\begin{align}
\label{eq:lattice-Fock}
\phi_{\mathbf{n}}  = \frac{1}{\sqrt{\mathcal{V}}} \sum_{\mathbf{q}} \frac{1}{\sqrt{2 {\omega}_{\mathbf{q}}} } 
\Big[ a_{\mathbf{q} } +  a^\dagger_{-\mathbf{q} } \Big] e^{i {2\pi}{} {\mathbf{n}\cdot \mathbf{q}}/N_s{}}\,,
\nn
\pi_{\mathbf{n}}  = \frac{-i}{\sqrt{\mathcal{V}}} \sum_{\mathbf{q}}   \sqrt{\frac{\omega_\mathbf{q}}{2}}
\Big[ a_{\mathbf{q} } -  a^\dagger_{-\mathbf{q}} \Big] e^{i {2\pi}{} {\mathbf{n}\cdot \mathbf{q}}/N_s{}}\,,
\end{align}
where $\mathcal{V}=N_s^d$ and  ${\omega}_\mathbf{q} = \overline{\omega}_\mathbf{q} \, a_s^{-1}$ is the dimensionless energy. 
Note that we use the same notation for the dimensionless lattice Fock operators $a_{\mathbf{q}}$ and the dimensionful continuum operators in \Eq{eq:PhiOperatorCont}. 

We will implement below the time evolution operator of the free Hamiltonian (setting $\lambda=0$ in \Eq{eq:latticeHamiltonian}) in the momentum representation. This allows us to use the continuum dispersion relation $\omega_\mathbf{q} = \sqrt{\mathbf{p}^2+m^2}$ ($\mathbf{p}\equiv\mathbf{p}(\mathbf{q})$), as opposed to the lattice dispersion relation 
that one has when working in position space; this potentially reduces discretization errors significantly.

The key idea in our digitization scheme is to decompose the many-particle Hilbert space into single particle sectors $\mathcal{H} = \bigotimes_{l=0}^\infty \mathcal{H}^{l}$, where a number of qubits are used to represent either momentum or position eigenstates in a binary decomposition. Since we are dealing with a relativistic theory where particle number is not conserved, an additional qubit is used to indicate whether or not a particle ``exists”. With this in mind, the single particle Hilbert space is spanned by
\begin{align}\label{eq:defsingleparticleH}
\mathcal{H}^{l} =\text{span} \{ |\Omega \rangle^{(l)},  \{ | \mathbf{q}\rangle^{(l)} \}  \}\,,
\end{align}
where $\Omega$ denotes ``empty states”, and $|\mathbf{q}\rangle$ ``occupied states”.
Further, a  ``register" of $N\equiv \log_2{\mathcal{V}}+1 $ spins (qubits)  
 represents a relativistic single-particle state with momentum $\mathbf{q}=(\textbf{q}_1,\dots,\textbf{q}_d)$ in $d$ dimensions,
\begin{align}\label{eq:stateBinaryDecomp}
| \mathbf{q}\rangle^{(l)}  \equiv  \big| \textbf{q}_1,\dots, \textbf{q}_d\rangle  \big| \ua\big\rangle \,,
\end{align} 
where one qubit $| \mathfrak{n}\rangle=|\ua\rangle$ denotes that the single-particle state is occupied.   Each momentum component of the occupied single-particle state
\begin{align}
| q_i\rangle \equiv | s_i \rangle | | q_i| \rangle \, ,
\end{align}
is represented by $(N-1)/d$ qubits, where $s_i = \text{sign} (q_i)$ is the sign (one qubit) and $|q_i|$ the absolute value (abs).
Likewise, 
we define an unoccupied single-particle state as a state where abs, sign and occupation number qubits are all in the $|\downarrow \rangle$ state, 
\begin{align}\label{eq:emptystate}
|\Omega \rangle^{(l)} \equiv | \da^{\otimes  d\cdot N^{\rm abs}} , \da^{\otimes d} , \da \rangle\,,
\end{align}
and the Fock vacuum is defined as  $| {\rm vac} \rangle = \bigotimes_l  |\Omega \rangle^{(l)}$.
We will represent these momentum states using a binary encoding with the qubits representing the digits.
In this case, $N^{\rm abs}=\frac{N-1}{d}-1 = \frac{\log_2{(\mathcal{V}/2^d)}}{d}$ qubits\footnote{To avoid a sign ambiguity, we choose the lattice such that $q_i=0$ is excluded. Then $s_i=\ua$($\da$) is a positive (negative) sign. We use a physical convention $|\ua/\da\rangle$ of up/down spins to label states, instead of the more common $|0/1\rangle$ notation.}. States with zero occupation number but finite $\mathbf{q}$ are unphysical and are excluded. Concrete examples of this single-particle digitization scheme are given in Appendix \ref{app:detailsSingleParticle}. The normalization $\langle \mathbf{q}| \mathbf{q}' \rangle = \delta_{\mathbf{q},\mathbf{q}'}$ of these basis states differs from the  relativistic normalization in \Eq{eq:stateDimLessLorentz}, with  $| \mathbf{q}\rangle =  | \pb \rangle^{\rm phys}/{\sqrt{2{\omega}_\mathbf{q}}}$.
A generic state $| \psi\rangle^{(l)}\in\mathcal{H}^l$ can be written as
\begin{align}\label{eq:l_state}
| \psi\rangle^{(l)} = \mathfrak{a}_0 | \Omega\rangle^{(l)}+\sum_{\mathbf{q}} \mathfrak{a}_{\mathbf{q}} | \mathbf{q} \rangle^{(l)}\,,
\end{align}
with $|\mathfrak{a}_0|^2+\sum_{\mathbf{q}} |\mathfrak{a}_{\mathbf{q}}|^2=1 $. The free part of the Hamiltonian ($H_0$) is block diagonal with the blocks labeled by the number of particles. Particle number eigenstates are on-shell single-particle states and those that are not correspond to virtual particles. 

The Fock operators in \Eq{eq:lattice-Fock} for the $M$ many-particle states on the combined Hilbert space $\mathcal{H} = \bigotimes_{l=0}^\infty \mathcal{H}^{l}$, are
\begin{align}\label{eq:DefinitionCreationOperators}
a_{\mathbf{q}} \equiv \lim_{M\rightarrow \infty} \frac{1}{\sqrt{M}} \sum_{l=0}^{M-1} a^{(l)}_{\mathbf{q}}\,,
\end{align}
with $a^{(l)}_{\mathbf{q}},a^{(l)\dagger}_{\mathbf{q}}$ denoting chains of spin raising and lowering operators for each $\mathbf{q}$, and  $(a^{(l)\dagger}_{\mathbf{q}})^2=(a^{(l)}_{\mathbf{q}})^2=0$. In practice, one 
 truncates the number of single-particle registers at a finite $M$. If $M$ is large compared to the typical occupancy of a state  $\mathfrak{n} \equiv \sum_i \mathfrak{n}^{(i)}$, the bosonic commutation algebra is realized, $[a_{\mathbf{q}},a^\dagger_{\mathbf{q}'} ] = \delta_{{\mathbf{q}},{\mathbf{q}}'} + O(\frac{\mathfrak{n}}{M}) $. Additional details of the construction are presented in Appendix \ref{app:detailsSingleParticle}. 

In the single-particle digitization of the Hilbert space of the scalar field theory, its dimension grows logarithmically with the volume $\mathcal{V}$ and linearly with $M$. This is ideal for high energy scattering problems, where the particle number density is small, such as the Bjorken limit~\cite{Breidenbach:1969kd,Bjorken:1968dy,Bjorken:1969ja,Gross:1973id} of the DIS problem we discussed earlier.  The digitization is not economical for a very dense system because, as we will discuss, Bose symmetrization creates a large overhead of unphysical states. However, as we also noted, our digitization strategy could potentially be extended to such dense systems employing a hybrid quantum/classical approach.

\section{Quantum algorithm}\label{sec:algorithm}
\begin{figure}[h]
\begin{center}
\includegraphics[width=0.75\textwidth]{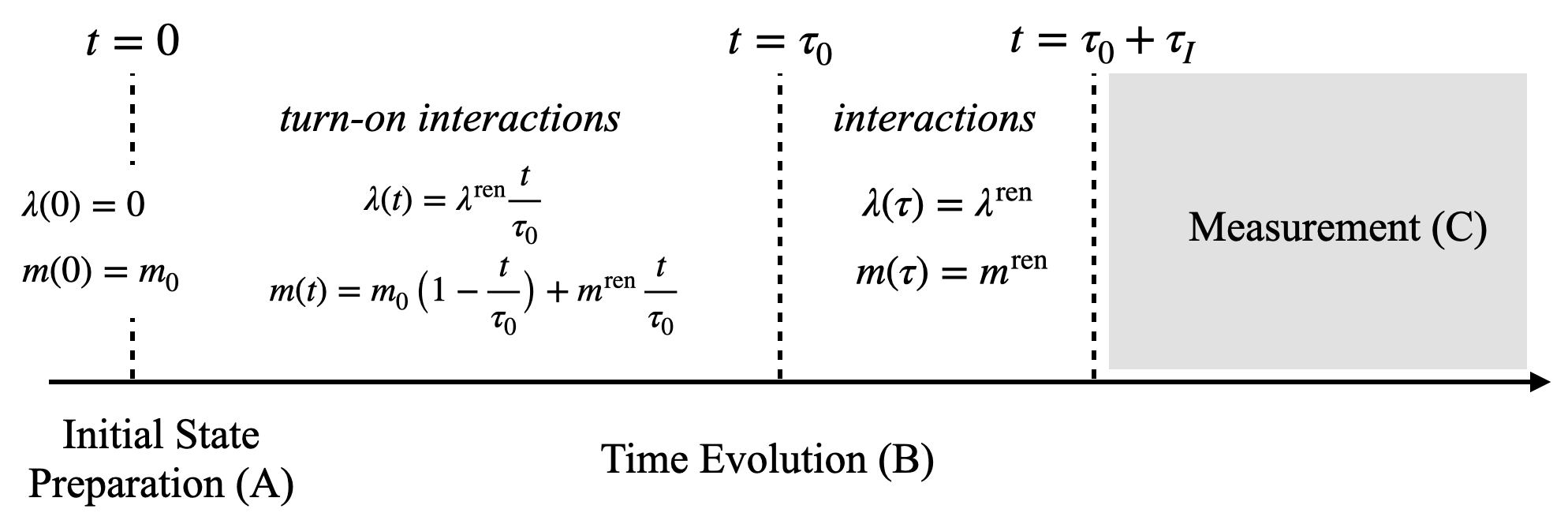}
\end{center}
\caption{Overview of the general algorithm to quantum compute high energy scattering cross-sections, including
the values of the bare couplings $\lambda$ and $m$ for simulation time $t$.
Initial state preparation is discussed in Section \ref{sec:InitialStatePrep}, time evolution in Section \ref{sec:TimeEvolution}, and measurement of particle cross-sections in Section \ref{sec:Measurement}.
The choice of (renormalized) couplings $\lambda(t)$, $m(t)$ is discussed in Section \ref{sec:Renormalization}.} 
\label{fig:algorithm}
\end{figure}

In this Section, the single-particle digitization strategy will be formulated as a concrete algorithm to quantum compute scattering cross-sections. In line with the spacetime picture discussed in Section \ref{sec:DIS_Smatrix}, and paralleling the approach of Jordan, Lee and Preskill \cite{Jordan:2011ci,Jordan:2011ne}, the components of our algorithm are
\begin{enumerate}
\item[A.] {Initial state preparation}, discussed in Section \ref{sec:InitialStatePrep}.
\item [B.] Simulating the time evolution, discussed in Section~\ref{sec:TimeEvolution}.
\item[C.]  Measurement of observables and their relation to scattering cross-sections, discussed in Section~\ref{sec:Measurement}.
\item[D.] Renormalization, discussed in Section \ref{sec:Renormalization}.
\end{enumerate}
These different elements are compactly summarized in \Fig{fig:algorithm}.

We will first discuss the preparation of the initial state of non-interacting particles in spatially separated wavepackets.
Their preparation is particularly simple  using the digitization presented in Section \ref{sec:strategies_for_quantum_simulation} compared to the field based approach of \cite{Jordan:2011ci,Jordan:2011ne,klco2019digitization}, because single-particle states and the vacuum are computational basis states.
Our algorithm consists of preparing a quantum mechanical superposition of these basis states to form wavepackets, placing them in separated regions of phase space and finally Bose-symmetrization of the resulting few/many-body wavefunction. 

To implement the time evolution operator, we will employ a Suzuki-Trotter scheme\footnote{For practical applications, it is important to note that more efficient algorithms for time evolution exist. One such example is the linear combination of unitaries~\cite{childs2017quantum}, later generalized to the method of quantum singular transformations~\cite{gilyen2019quantum}. The latter class of quantum algorithms not only significantly speeds up time evolution, but is also an efficient replacement of the classic phase estimation algorithm discussed in the context of measurements in Section \ref{sec:Measurement}.}~\cite{Trotter:1959,Suzuki:1976}.
We will treat the time evolution of the free and interacting parts of the Hamiltonian in \Eq{eq:latticeHamiltonian} separately. We first evolve the wavepacket with the free Hamiltonian $H_0$, which is diagonal in the momentum representation.  This is  followed by a squeezing operation (analogous to that performed in quantum optics~\cite{gerry2000quantum,yeter2018quantum}), a quantum Fourier transformation~\cite{NielsenChuang}
from momentum space to position space, and lastly, an implementation of the interaction term in position space, where it is local.

This algorithm differs from the field based approach of~\cite{Jordan:2011ci,Jordan:2011ne,klco2019digitization} where the time evolution operator is split into a part diagonal in the $\phi_{\mathbf{x}}$-basis and one diagonal in the conjugate $\pi_{\mathbf{x}}$-basis. While the overall Trotter complexity scales as $O(\mathcal{V})$  in both cases, an important difference is that we avoid the lattice discretization of the Laplacian in \Eq{eq:latticeHamiltonian} by working directly in momentum space.

Time evolution involves a switch-on of interactions from the non-interacting theory in the infinite past, $m(0)\equiv m_0, \, \lambda(0)=0$ (in practice at some finite time $t=0$) towards acquiring the renormalized, physical couplings $m(\tau_0)\equiv m^{\rm ren}, \, \lambda(\tau_0)=\lambda^{\rm ren}$ at $t=\tau_0$ right before the particles collide. The non-perturbative renormalization in the single-particle framework, which differs from that of~\cite{Jordan:2011ci,Jordan:2011ne}, is  discussed in Section \ref{sec:Renormalization}. 

An important practical issue for the quantum algorithm is the spreading of the wavepackets during the switch-on time of interactions, which may potentially cause the wavepackets
to interact before the coupling is turned to its final value. For this one may use the ``forward-backward'' evolution scheme outlined in~\cite{Jordan:2011ci,Jordan:2011ne}. Note however that because at large energies the dispersion is approximately linear $\omega\sim |\mathbf{p}|$, the spreading of the wavepackets is anticipated to be small~\cite{Su}.

Another relevant point is that the adiabatic preparation of single-particle states~\cite{Jordan:2011ci,Jordan:2011ne}  will require a very large number of Trotter steps at high energies and likewise, for the turn-off of interactions. This can be understood by  considering the energy gap between single-particle states with momentum $\mathbf{p}$ and energy $E=({\mathbf{p}^2+m^2})^{\frac{1}{2}}$,
and the lowest of the two-particle states with total momentum  $\tilde{\mathbf{p}}\equiv\mathbf{p}_1+\mathbf{p}_2$ (with relative momentum $\tilde{\mathbf{q}}\equiv \mathbf{p}_1-\mathbf{p}_2=0$) and energy $E=({\tilde{\textbf{p}}^2 + (2m)^2})^{\frac{1}{2}}$ at weak coupling. Because this gap vanishes as $\mathbf{p}\rightarrow\infty$ and  $\tilde{\mathbf{p}}\rightarrow\infty$, adiabatic state preparation is all but impractical at high energies. 

Different state preparation algorithms have been suggested~\cite{lamm2018simulation,kokail2019self,bapat2019bang,harmalkar2020quantum,gustafson2020toward,choi2020rodeo} which are potentially faster than adiabatic state preparation. We note however that, departing from the strictly idealized S-matrix picture, in scattering processes  such as Deeply Inelastic Scattering in QCD discussed in Section \ref{sec:SpaceTimePictureScattering}, the Ioffe time and like physical scales are the relevant time scales for state preparation and may allow for quicker non-adiabatic state preparation.  Because the algorithm discussed below is general,  and one may also make use of alternative state preparation algorithms ~\cite{lamm2018simulation,kokail2019self,bapat2019bang,harmalkar2020quantum,gustafson2020toward,choi2020rodeo},  we will not say anything further beyond noting this interesting possibility.

We will discuss finally in this Section the determination of scattering cross-sections, utilizing a natural connection of our digitization strategy to particle physics concepts. In contrast to field based digitizations \cite{Jordan:2011ci,Jordan:2011ne},  particle number measurements do not require any additional gate operations. Measurement of energy density or momentum, for example, via a phase estimation algorithm, have a simple gate complexity. Some of the ``classical analysis" in high energy experiments, of binning data or imposing kinematic cuts, can be incorporated directly in the quantum algorithm. We argue that, using novel techniques such as oblivious amplitude amplification \cite{brassard2002quantum,NielsenChuang,Berry_2014}, a quantum computer could possibly ``beat" an actual particle physics experiment by producing rare events with a higher probability.

\subsection{Initial State Preparation}\label{sec:InitialStatePrep}

We will now discuss the state preparation of 
a Bose-symmetric state of single particle wavepackets at $t=0$ and zero coupling
that are well separated in position space. As a first step, we
create wavepackets separately in $\mathfrak{n}$ of the $M$ particle registers
(where $\mathfrak{n}$ is the number of initial scatterers, typically $\mathfrak{n}=2$). Each wavepacket $i=0,\dots,\mathfrak{n}-1$ is centered at  $(\bar{\mathbf{x}}_i,\bar{\mathbf{p}}_i)$ and is Gaussian distributed with a width $(\Delta \mathbf{x}, \Delta \mathbf{p})$ around this center, where $|\Delta \mathbf{x}|\ll |\bar{\mathbf{x}}_i-\bar{\mathbf{x}}_j|$ for all $i\neq j$ (here assumed to be identical for all particles). Typically one chooses $|\Delta \mathbf{x}| \sim 1/m$, and $|\mathbf{p}|\gg |\Delta \mathbf{p}|\sim m$, where $m$ is the mass (in dimensionless units), so that particles are well localized on macroscopic scales.

Wavepackets comprised of  single-particle  states $| \mathbf{q}\rangle$, located at the origin $( \bar{\mathbf{x}}_i,  \bar{\mathbf{p}}_i)=(0,0)$, are written\footnote{We will henceforth drop the label $(l)$ denoting a particular single-particle state, as in \Eq{eq:l_state}.}  in a momentum space representation as
\begin{align}
| \Psi \rangle  
= \frac{1}{\sqrt{\mathcal{V}}} \sum_{\mathbf{q}} \, \Psi_\mathbf{q} | \mathbf{q}\rangle\,,
\end{align}
where $ \Psi_\textbf{q} $ is a real, positive and strongly localized distribution such as a Gaussian distribution. Each such wavepacket can be translated to $( \bar{\mathbf{x}}_i,  \bar{\mathbf{p}}_i)\neq(0,0)$ such that as previously,  $|\bar{\mathbf{x}}_i -  \bar{\mathbf{x}}_j|\gg |\Delta\mathbf{x}|$, and $\mathbf{p}_i$ corresponds to projectile kinematics, using  circuits we will discuss shortly.

To create a wavepacket in the momentum space representation,  centered at  $( \bar{\mathbf{x}}_i,  \bar{\mathbf{p}}_i)=(0,0)$,
and with width $\Delta \mathbf{p}$ ($|\Delta \mathbf{x}|\sim |\Delta \mathbf{p}|^{-1}$), from the vacuum state $|\Omega \rangle$ (\Eq{eq:emptystate}) we use a simple variant of the algorithm in~\cite{grover2002creating,kaye2004quantum} which we illustrate below
for $d=1$ spatial dimensions.  First, accounting for the $\mathbf{q} \rightarrow -\mathbf{q}$ symmetry of $\Psi_{\mathbf{q}}$,  we first flip the occupation number qubit and then apply the Hadamard gate ($H$) to the sign qubit, 
\begin{align}
 | \da^{\otimes   N^{\rm abs}} , \da , \da \rangle \xrightarrow{\sigma^x,H} \frac{1}{\sqrt{2}} \Big[
  | \da^{\otimes  N^{\rm abs}} , \ua , \ua \rangle +  | \da^{\otimes  N^{\rm abs}} , \da , \ua \rangle\Big]\,.
\end{align}
Subsequently, we rotate all remaining $N^{\rm abs}\sim \log_2(N_s)$ qubits by an angle $\theta_k = \pi/4 - \epsilon_k$,
\begin{align}\label{eq:isp_ouralg}
| \da \rangle^{(k)} \rightarrow \cos(\theta_k) | \da \rangle^{(k)}  + \sin(\theta_k)| \ua \rangle^{(k)}\,.
\end{align}
where  $k\in[0,N^{\rm abs}-1] $ and $\epsilon_k\in [0,\pi/4)$. Thus for each $\ket{\da}^{(k)}$ the state gets a $\cos(\theta_k)$ coefficient, while each $\ket{\ua}^{(k)}$ receives a $\sin(\theta_k)$ contribution. One can relate each $\epsilon_k$ to a specific distribution\footnote{This distribution should be one that is probabilistic, namely, efficiently integrable with importance sampling techniques~\cite{grover2002creating}.}.

Subsequently, we displace the centers of every single-particle wavepacket in position and momentum space, such that they are widely separated $|\bar{\mathbf{x}}_i -  \bar{\mathbf{x}}_j|\gg |\Delta\mathbf{x}|$, with $\bar{\mathbf{p}}_i$ denoting the initial momentum of each projectile. To achieve this, we need to use the translation operator $T_{\mathbf{n}}$ ($T_{\mathbf{q} }$) in position space (momentum space), defined as
\begin{align}\label{eq:definfiniteTranslation}
&T_{\mathbf{n}}|\mathbf{q}\rangle = e^{ - i\, 2\pi {\mathbf{n}\cdot \mathbf{q}}/{N_s}  }|\mathbf{q}\rangle\,, 
\qquad T_{\mathbf{n}} | \Omega \rangle = |\Omega \rangle\,,
\end{align}
where $\mathbf{x}=\mathbf{n}a_s$ and $\mathbf{n}=(n_1,\dots n_d)$.
It can be decomposed in terms of one-dimensional translation operators
$T_{\textbf{n}}\equiv \bigotimes_{k=1}^d T^{(k)}_{{n}_k}$. To illustrate the circuit
implementation of $T^{(k)}_{n_k}$, we will consider the operator for a translation by one lattice site in the positive direction $T^{(k)}_1$. A finite translation can
then be achieved by successive applications of $T^{(k)}_{n_k} = (T^{(k)}_1)^{n_k}$ ( $T^{(k)}_{n_k} = (T^{(k)\dagger}_1)^{|n_k|}$)  if $n_k>0$ ($n_k<0$), or directly via a simple modification of the algorithm for $T_1^{(k)}$, with identical gate complexity.

\begin{figure}[t]
\begin{center}
\includegraphics[width=0.52\textwidth]{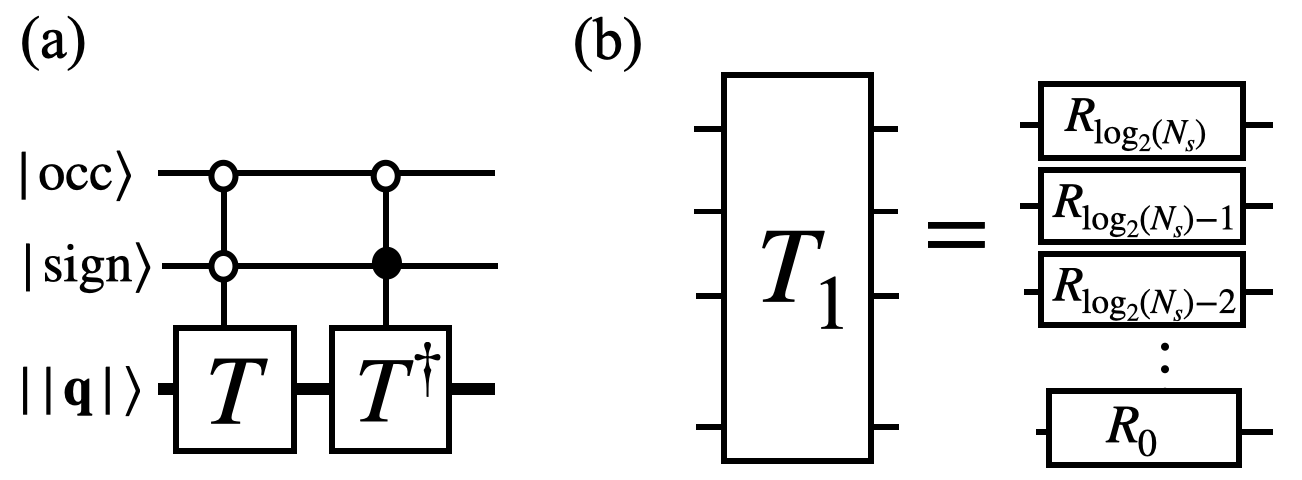}
\end{center}
\caption{(a) Translation operator for $d=1$, where we abbreviate $T\equiv T^{(1)}_{{n}_1}=(T^{(1)}_{1})^{|n_1|}$ for $n_1>0$ and $(T^{(1)\dagger}_{1})^{|n_1|}$ for $n_1<0$. A white (black) circle indicates control by the $| \ua \rangle$ ($|\da\rangle$) state. (b) Decomposition of the circuit-see text for details.}
\label{fig:translationcircuit}
\end{figure}

The  circuit for an infinitesimal translation\footnote{We will consider now the $d=1$ case and drop the label $k$.} $T_{{1}}$  is given in \Fig{fig:translationcircuit}
acting on a state in the momentum space basis, using the gate $R_t \equiv {\rm diag}(1,\exp\{-2\pi i/2^t\})$. Because the operators $T$ and $T^\dagger$ act on the register containing $|\mathbf{q}|$, their action is controlled by the sign
qubit to account for the sign in the exponent of \Eq{eq:definfiniteTranslation}.
%
%
%
%
The momentum translation operator $T_\mathbf{q}$ can be implemented using exactly the same circuit, preceded by a change of basis $| \mathbf{q}\rangle \rightarrow | \mathbf{n}\rangle$ (via a quantum Fourier transform, as will be discussed in Section \ref{sec:qFT}). The generalization to arbitrary $d$ is straightforward and has $O(M\log(\mathcal{V}))$ circuit complexity\footnote{Assuming large volumes, we will not discuss the action of the translation operator on the spatial boundaries.}. 

The result of this procedure are multi-particle initial states comprised of widely separated, non-overlapping, wavepackets $| \Psi_{i}\rangle$ and ``empty" vacuum registers $| \Omega \rangle$,
\begin{align}\label{eq:unsymmwavep}
|\phi\rangle \equiv | \Psi_{0}, \Psi_{1},\dots \Psi_{\mathfrak{n}-1},\Omega, \dots, \Omega,\dots \rangle\,.
\end{align}
The corresponding  Bose-symmetrized state is given by
\begin{align}\label{eq:definitionBoseSym}
|\phi_B\rangle \equiv  \frac{1}{\sqrt{\mathcal{N}}} \sum_P \hat{P} |\phi\rangle \,,
\end{align}
where $\hat{P}$ is the Bose permutation operator and $\mathcal{N}=M!/(M-\mathfrak{n})!$. 

To prepare $|\phi_B\rangle$ (\Eq{eq:definitionBoseSym}) from $|\phi\rangle$ (\Eq{eq:unsymmwavep}), we will use a variant of an algorithm which, for the case $\mathfrak{n}=1$, $M=2$, we can illustrate as, 
\begin{align}
 | \Psi ,\Omega \rangle | 0 \rangle \xrightarrow{H}  | \Psi,\Omega \rangle \frac{1}{\sqrt{2}} \big[  | 0 \rangle + |1  \rangle \big]
\xrightarrow{\rm CSWAP}   \frac{1}{\sqrt{2}} \big[   | \Psi,\Omega\rangle | 0 \rangle + | \Omega,\Psi\rangle |1  \rangle \big]
\xrightarrow{\rm CNOT}   \frac{1}{\sqrt{2}} \big[   | \Psi,\Omega\rangle  + | \Omega,\Psi\rangle  \big]| 0 \rangle
=|\phi_B\rangle \,.
\end{align}
The basic idea is to introduce $s\equiv\log_2{(M!/(M-\mathfrak{n})!)}\sim O(M^{\mathfrak{n}})$ ancilla qubits ($s=1$ in this example), that are prepared in a symmetric Bell superposition state. Each term in this superposition controls a specific SWAP operation between pairs of particle registers. The  CNOT operation uses the occupation number qubits of the registers to un-compute the ancilla. Circuits for  arbitrary $\mathfrak{n}$ and $M$
do not differ fundamentally from this example but are slightly more complicated and are discussed in Appendix \ref{app:StochasticStatePrep}. 

In particular, if  $M$ and $\mathfrak{n}$ cannot be chosen such that $s$ is an integer, one must choose $s=\lceil \log_2{(M!/(M-\mathfrak{n})!)} \rceil~\sim O(M^\mathfrak{n})$, where the symbol $\lceil\,y \rceil$ denotes the smallest integer larger than $y$. As discussed in Appendix \ref{app:StochasticStatePrep}, the symmetrization yields some unwanted permutations in this case which
are eliminated through measurements and the symmetrization procedure becomes probabilistic as opposed to exact if $\log_2{(M!/(M-\mathfrak{n})!)}$ is an integer.  The chance of returning the desired state is
$p_{\rm success}=\mathcal{N}/2^s\ge 1/2$.  As shown in Appendix \ref{app:StochasticStatePrep}, one can always pick $M$  for fixed $\mathfrak{n}$ such that the probability of success is maximized.
Note that the un-computation of the $s$ ancilla qubits for $\mathfrak{n}\geq 2$ requires using
information stored in the momentum/position registers as control qubits. 
Fermionic states can be prepared along similar lines\footnote{The authors plan to return to this case in future work, with the expectation that this leads to novel fermion-qubit mappings~\cite{tranter2018comparison}.}.
 
Particles generated during the time evolution of the initial state are accommodated by a large number of empty registers $\mathfrak{n}_\Omega \equiv M-\mathbf{n} \approx M \gg \mathfrak{n}$ initially. 
A rough estimate for $M$ is the number of particles in the final state, ranging widely with energy from a few to few tens to few hundreds, an upper bound for which is the ratio of the collision energy to the  particle mass  $\sqrt{s}/\, \overline{m}$.
This estimate does not include virtual states the system could be fluctuating into over shorter time scales. In weak coupling, there is a one-to-one correspondence between the Fock space explored in our digitization and that 
described by Feynman diagrams, allowing us to estimate that M should scale as the number of all internal and external lines. In the strong coupling limit, no such estimate is available and thus explicit numerical analysis, including a non-perturbative renormalization procedure, will be required.

 This algorithm for initial state preparation can 
 be contrasted with the corresponding one in the field based digitization \cite{Jordan:2011ci,Jordan:2011ne}. In the latter case, one first prepares the non-interacting vacuum state in a Gaussian basis state using the algorithm of \cite{kitaev2008wavefunction} while in our case the vacuum is a computational basis state. Secondly, one employs a Suzuki-Trotter scheme to realize  the application of position space Fock operators onto the vacuum state approximated by a linear combination of the field operators $\phi_{\mathbf{x}}$ and $\pi_{\mathbf{x}}$ in a region of space. 
 In this case, Bose-symmetrization is built into the realization of  operators $\phi_{\mathbf{x}}$ and $\pi_{\mathbf{x}}$ and does not need to be enforced explicitly. In our case, a superselection rule specifies the physical sector removing unsymmetrized states.

\subsection{Time Evolution}\label{sec:TimeEvolution}
We will follow a Trotter-Suzuki scheme with $N_\delta=(t-t_0)/\delta$ steps to implement the time evolution operator, 
\begin{align}
U(t,t_0)&\equiv e^{-i H (t-t_0)} = \big(e^{-i H \delta }\big)^{N_{\delta}} + O(\delta^2)
= (e^{-i H_I  \delta} e^{-i H_0 \delta})^{N_{\delta}}+ O(\delta^2)
\equiv (U_I U_0)^{N_{\delta}}+ O(\delta^2)\,,
\end{align}
separating the evolution operator into  free $U_0\equiv\exp{\{-i H_0 \delta\}}$ and interacting $U_I\equiv\exp\{-iH_I \delta\}$ parts, where $H_0$ is given by the quadratic terms and $H_I$ by the $\phi^4$ interaction term in \Eq{eq:latticeHamiltonian}.

\begin{figure}[t]
\begin{center}
\includegraphics[width=0.7\textwidth]{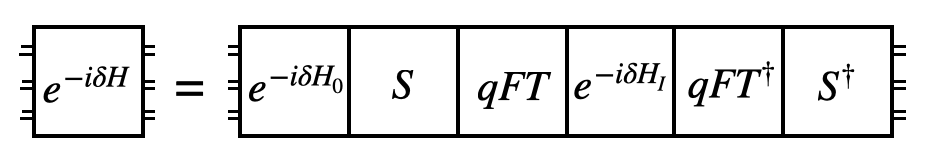}
\end{center}
\caption{Overview of the time evolution scheme for one Trotter-Suzuki step $\delta$. Here $S$ (discussed in Section \ref{sec:squeezing}) and $qFT$ (discussed in Section \ref{sec:qFT}) denote the squeezing operation and the quantum Fourier transformation, respectively.}
\label{fig:timeevolutionscheme}
\end{figure}

   We implement $U_0$ in the momentum space basis of Bose-symmetrized states \Eqs{eq:defsingleparticleH}{eq:l_state}, where it is diagonal. Using a combined squeezing operation and Fourier transformation, the interaction part $U_I$ is then implemented in position space where it is local. Our strategy is summarized in \Fig{fig:timeevolutionscheme}, and the different elements are worked out below.

\subsubsection{Free part $U_0$}\label{sec:freepart}
The infinitesimal (normal-ordered) time evolution operator $U_0$ is given by
\begin{align}\label{eq:freevo}
U_0&\equiv \exp\big\{-i \delta \sum_{\mathbf{q}} \omega_{\mathbf{q}}\, a_{\mathbf{q}}^\dagger a_{\mathbf{q}} \big\}
=\exp\big\{- \frac{i\delta}{M} \sum_{\mathbf{q}}\omega_{\mathbf{q}}\big[  \sum_{i=0}^{M-1} a^{(i)\dagger}_{\mathbf{q}} a^{(i)}_{\mathbf{q}}+\sum_{i\neq j=0}^{M-1} a^{(i)\dagger}_{\mathbf{q}} a^{(j)}_{\mathbf{q}}   \big]  \big\}\,,
\end{align}
where $U_0$ is diagonal when acting on a state $| \psi \rangle$ in the representation discussed above. It can be written as multiplication by a phase factor,
\begin{align}\label{eq:FreeTimeEvoDetails}
U_0  \ket{\psi} = e^{ -\frac{i\delta}{M} \sum_{\mathbf{q}} \omega_{\mathbf{q}}  \mathfrak{n}_{\mathbf{q}}(1+\mathfrak{n}_\Omega)  }| \psi \rangle
= S_\varphi^{1+\mathfrak{n}_\Omega }| \psi \rangle\,,
\end{align}
where $S_\varphi \equiv \exp\{-i\frac{\delta}{M}\varphi\}$,  $\varphi\equiv\sum_{\bar{n}} \omega_{\mathbf{q}} \mathfrak{n}_{\mathbf{q}}$
is the total energy of all occupied states, and  $\mathfrak{n}_{\mathbf{q}}$ ($\mathfrak{n}_\Omega$) the number of registers with momentum $\mathbf{q}$ (empty registers), while $\omega_{\mathbf{q}}$ is the continuum dispersion relation. The factor $\mathfrak{n}_{\mathbf{q}}(1+\mathfrak{n}_\Omega)$
reflects the two terms in the exponent of \Eq{eq:freevo}.

 \begin{figure}[t]
\begin{center}
\includegraphics[width=0.4\textwidth]{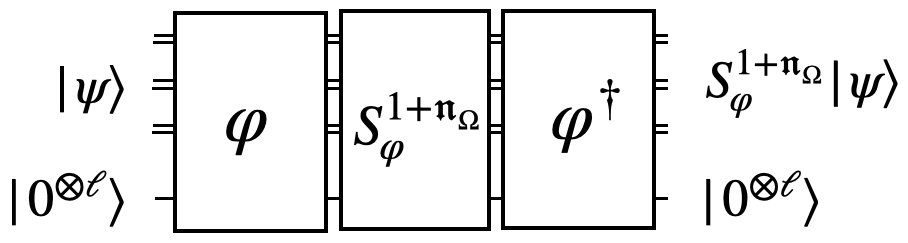}
\end{center}
\caption{Quantum circuit for $U_0$, using $O(M\text{poly}\log(\mathcal{V}))$ operations and $2\ell$ ancilla qubits. Double lines indicate particle registers (including $|\mathbf{q}|$, sign and occupation number qubits).}
\label{fig:kineticcircuit1}
\end{figure}

\begin{figure}[t]
\begin{center}
\includegraphics[width=0.67\textwidth]{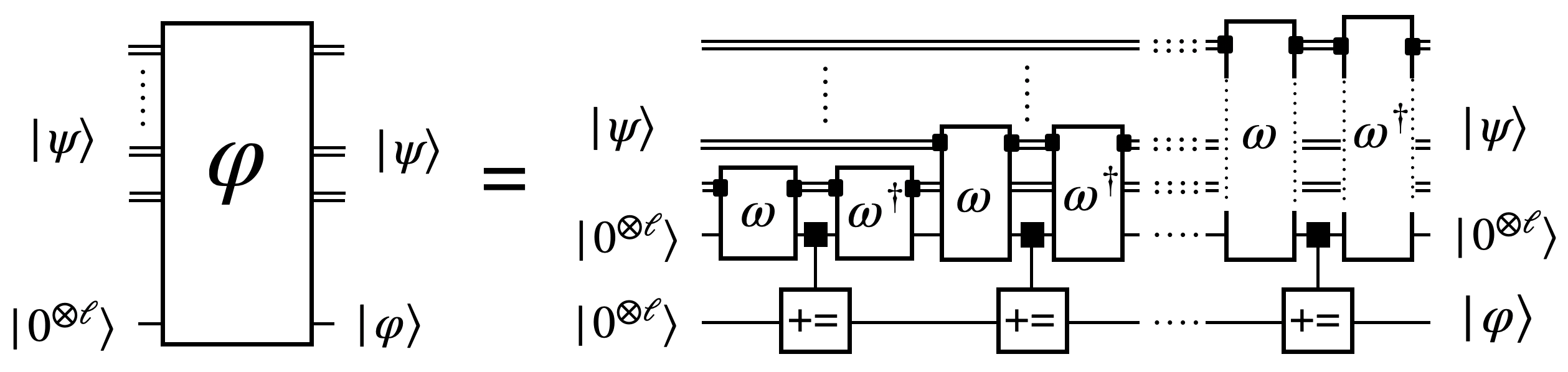}
\end{center}
\caption{Quantum circuit to compute $\varphi$, based on the algorithm in~\cite{Zalka:1996st}. \fbox{$\omega$} is an oracle to compute $\omega(\mathbf{q})$ for input $| \mathbf{q}\rangle$, and \fbox{\tiny+=} is a quantum-addition circuit~\cite{vedral1996quantum,draper2000addition}. The $\blacksquare$ symbol appearing in the gate \fbox{\tiny+=} denotes that the associated register is an input into the gate. The relevant particle register input for the \fbox{$\omega$} gates is denoted by (small) black boxes accordingly. }
\label{fig:kineticcircuit2}
\end{figure}

The algorithm for computing \Eq{eq:FreeTimeEvoDetails} is summarized in \Fig{fig:kineticcircuit1}. It involves first computing the phase $\varphi$. This is done by the sub-circuit depicted in \Fig{fig:kineticcircuit2}, with two auxiliary registers of $\ell$ qubits. Here $\ell$ is determined by the precision of the algorithm to compute $\varphi$, \fbox{+=} is the quantum-addition operation \cite{vedral1996quantum,draper2000addition} and we treat the circuit \fbox{$\omega$} to compute $\omega_{\mathbf{q}}$ as a quantum ``oracle". The number of ancilla registers $2\ell$ is determined by the precision with which we wish to compute $\omega_{\mathbf{q}}$ from $\mathbf{q}$. It should be taken to be similar to the number of qubits $\ell\sim O(\log(\mathcal{V})/d)$ that are necessary to realizing $\mathbf{q}$ in one dimension. The number of gate operators included
in \fbox{$\omega$} is $\text{poly}\log(\mathcal{V})$. 
Efficient algorithms to compute simple arithmetic functions can be found in the literature~\cite{Cao_2013,munozcoreas2018tcount,bhaskar2015quantum,Hner2018OptimizingQC}.

Once $\varphi$ is computed, one follows with $O(M)$  diagonal phase rotations $S_\varphi^{1+\mathfrak{n}_\Omega}$, using the occupation number qubits of each register as control qubits. (The detailed circuit is  shown in Appendix \ref{app:kineticterm}.).  Finally, we un-compute  $| \varphi \rangle$, so that in total we use $O(M)$ \fbox{+=} and \fbox{$\omega$} gates. As a consequence, the algorithm for $U_0$ has an overall complexity of $O(M\, \text{poly}\log(\mathcal{V}))$ gate operations per Trotter step.

%
%
\subsubsection{Squeezing Transformation}\label{sec:squeezing}
In order to implement the interaction piece of the time evolution operator $U_I$, we first perform a transformation 
from the single-particle representation in momentum space to position space.
In a relativistic theory, single particle states in position and momentum space are not simply Fourier conjugates. Therefore to obtain one from the other requires a combined squeezing operation~\cite{yeter2018quantum} followed by a (quantum-) Fourier transformation. To illustrate this, note
that position space Fock operators are given by
\begin{align}\label{eq:Fock_operators_x_space}
&a_{\mathbf{n}} \equiv \frac{1}{\sqrt{2}} \big( \phi_{\mathbf{n}} + i\pi_{\mathbf{n}} \big)\, ,\qquad
a^\dagger_{\mathbf{n}} \equiv \frac{1}{\sqrt{2}} \big( \phi_{\mathbf{n}} - i \pi_{\mathbf{n}} \big) \, ,
\end{align}
with the commutation relations $[a_{\mathbf{n}}, a^\dagger_{\mathbf{n}^\prime}] = \delta_{\mathbf{n},\mathbf{n}^\prime}$, and the single-particle decomposition $a_{\mathbf{n}} \equiv \sum_i a_{\mathbf{n}} ^{(i)}/\sqrt{M}$,  $a^\dagger_{\mathbf{n}} \equiv  \sum_i a_{\mathbf{n}} ^{(i)\dagger}/\sqrt{M}$.
We can define the Fourier conjugates $A_{\mathbf{q}}$  of $a_{\mathbf{n}}$ as 
\begin{align} \label{eq:quantFT}
a_{\mathbf{n}}\equiv \frac{1}{\sqrt{\mathcal{V}}} \sum_{\mathbf{q}} \, A_\mathbf{q}\,  e^{i {2\pi}{} {\mathbf{n}\cdot \mathbf{q}}/N_s{}}\, ,
\end{align}
and likewise for their Hermitian conjugate counterparts. These are related~\cite{yeter2018quantum} to the momentum space Fock operators $a_{\mathbf{q}}$, $a_{\mathbf{q}}^\dagger$ by
\begin{align}\label{eq:squeezingTrafo}
A_{\mathbf{q}} \equiv \frac{1}{2}\big[ \omega_{\mathbf{q}}^{-\frac{1}{2}}+\omega_{\mathbf{q}}^{\frac{1}{2}} \big] a_{\mathbf{q}}
+ \frac{1}{2}\big[ \omega_{\mathbf{q}}^{-\frac{1}{2}}-\omega_{\mathbf{q}}^{\frac{1}{2}} \big] a^\dagger_{-\mathbf{q}}\,,
\end{align}
and likewise for $A_{\mathbf{q}} ^\dagger$. Such squeezing operations
are well-known in quantum optics~\cite{nieto1997holstein,gerry2000quantum,marshall2015quantum}, where they are natural in the preparation of squeezed states.  We will work out here their implementation on a digital quantum computer.  To do so,  note that \Eq{eq:squeezingTrafo} is realized by  
\begin{align}\label{eq:squeezdef1}
&A_\mathbf{q}=S a_{\mathbf{q}}S^\dagger,\qquad A^\dagger_\mathbf{q}=S a^\dagger_{\mathbf{q}}S^\dagger,
\end{align}
where  $S\equiv \prod_{\mathbf{q}}S_\mathbf{q}$ and 
\begin{align}\label{eq:squeezingdef}
S_\mathbf{q}\equiv \exp\big\{-z_\mathbf{q} [ a_{\mathbf{q}}^\dagger a_{-\mathbf{q}}^\dagger - a_{-\mathbf{q}}a_\mathbf{q}  ]  \big\}\,,
\end{align}
is a unitary operator with $z_\mathbf{q}\equiv \frac{1}{2}\log(\omega_\mathbf{q})$. See also  Appendix \ref{app:squeezing} where we derive \Eq{eq:squeezingTrafo} from \Eq{eq:squeezdef1} and \Eq{eq:squeezingdef}.

The circuit implementation of $S_\mathbf{q}$ is compactly summarized  in \Fig{fig:squeezing1}. We can use a Trotter scheme to implement $S$, splitting the operation  into $\mathcal{V}$ modes $\mathbf{q}$ and $M(M-1)/2$ steps over all possible register pairs $i\neq j$, $i,j=0,\dots M-1$, with a Trotter error of $O([{ \mathfrak{n}_\mathbf{q}z_\mathbf{q}}/{M}]^2)$,
where $\mathfrak{n}_\mathbf{q}$ is the occupation number of the mode $\mathbf{q}$ of the state the operator acts on. We can then write
\begin{align}\label{eq:full_S}
S= \prod_{\mathbf{q},\langle i\neq j \rangle} S_{\mathbf{q},ij} \,,
\end{align}
and 
\begin{align}
S_{\mathbf{q},ij} \equiv \exp\big\{-\frac{z_\mathbf{q}}{M} [ a^{(i)\dagger}_{\mathbf{q}} a_{-\mathbf{q}}^{(j)\dagger} - a^{(j)}_{-\mathbf{q}}a^{(i)}_\mathbf{q}  ]  \big\}\,.
\end{align}

 \begin{figure}[t]
\begin{center}
\includegraphics[width=0.57\textwidth]{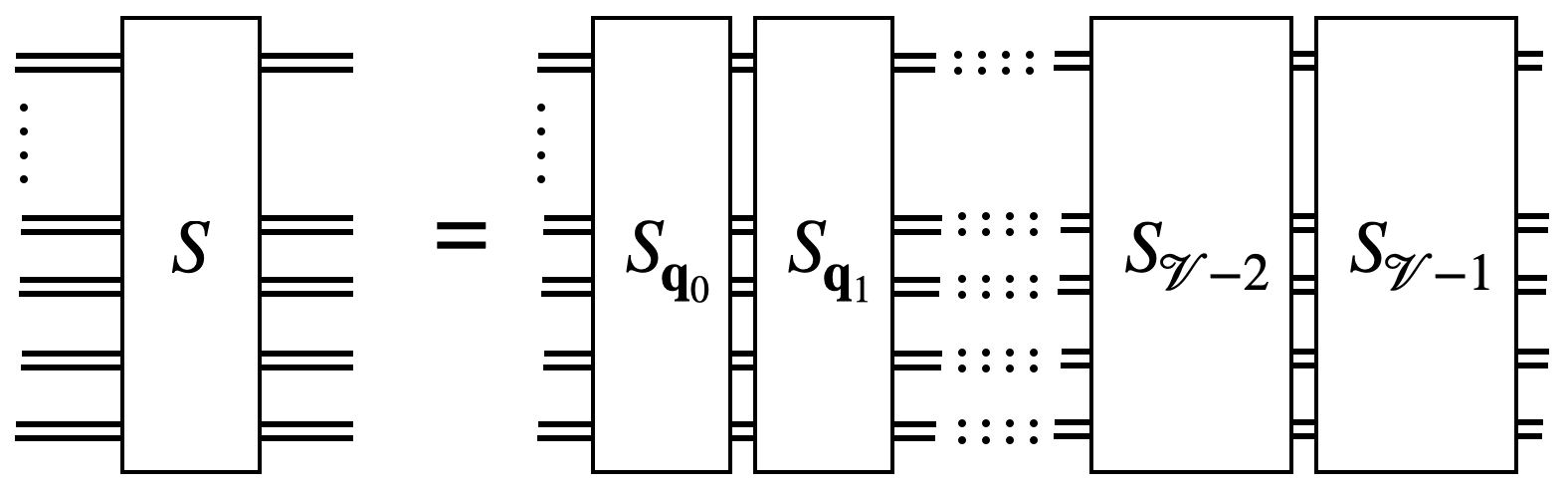}
\end{center}
\caption{Squeezing operator decomposition  $S=\prod_{\textbf{q}=\textbf{q}_0}^{\textbf{q}=\textbf{q}_{\mathcal{V}-1}} S_\textbf{q}$. Notice that since creation and annihilation operators of different momentum modes commute, there is no Trotter error associated to this decomposition. See \Eq{eq:squeezingdef}.}
\label{fig:squeezing0}
\end{figure}

 \begin{figure}[t]
\begin{center}
\includegraphics[width=0.5\textwidth]{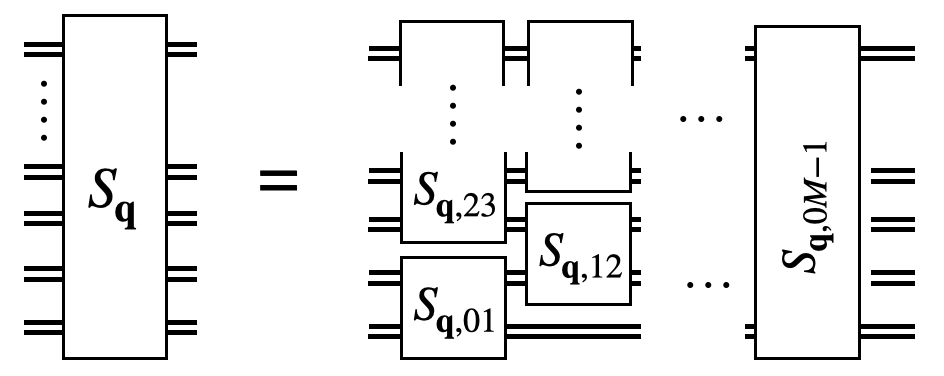}
\end{center}
\caption{Trotter decomposition of the squeezing operator $S$ into $M(M-1)/2$ operations {$S_{\mathbf{q},i j}$} ($i\neq j$). Note that because $S_{\mathbf{q},ij}=S_{\mathbf{q},ji}$, we can simplify  $S_{\mathbf{q},ij}(z_\mathbf{q})S_{\mathbf{q},ji}(z_\mathbf{q}) = S_{\mathbf{q},ij}(2z_\mathbf{q})$.}
\label{fig:squeezing1}
\end{figure}

To implement $S_{\mathbf{q},ij}$, we decompose the single particle Fock operators into spin raising and lowering operators (see Appendix \ref{app:detailsSingleParticle}), 
\begin{align}\label{eq:squezingCirc}
S_{\mathbf{q},ij} \equiv \exp\big\{- i \frac{z_\mathbf{q}}{M} \sigma^y_{\mathbf{q},ij}\big\} \, ,
\end{align}
where $\sigma^y_{\mathbf{q},ij}\equiv (-i)[a^{(i)\dagger}_{\mathbf{q}}a^{(j)\dagger}_{-\mathbf{q}} - a^{(j)}_{-\mathbf{q}}a^{(i)}_{\mathbf{q}} ]$. In the matrix representation of the $\mathfrak{N}$ occupation and momentum qubits spanning $\{|\mathbf{q}\rangle\otimes|-\mathbf{q}\rangle$ , $| \Omega \rangle\otimes| \Omega \rangle\} $, this can be written as
\begin{align}\label{eq:Sy}
\sigma^y_{\mathbf{q},ij} = \begin{pmatrix}
0 &\dots& 0 &-i \\
0 & \ddots & &  0\\
\vdots  &  &  \ddots& \vdots   \\
i & 0 & \dots & 0 
\end{pmatrix}  \equiv \sigma_\mathfrak{N}^y\,.
\end{align}

\begin{figure}[t]
\begin{center}
\includegraphics[width=0.5\textwidth]{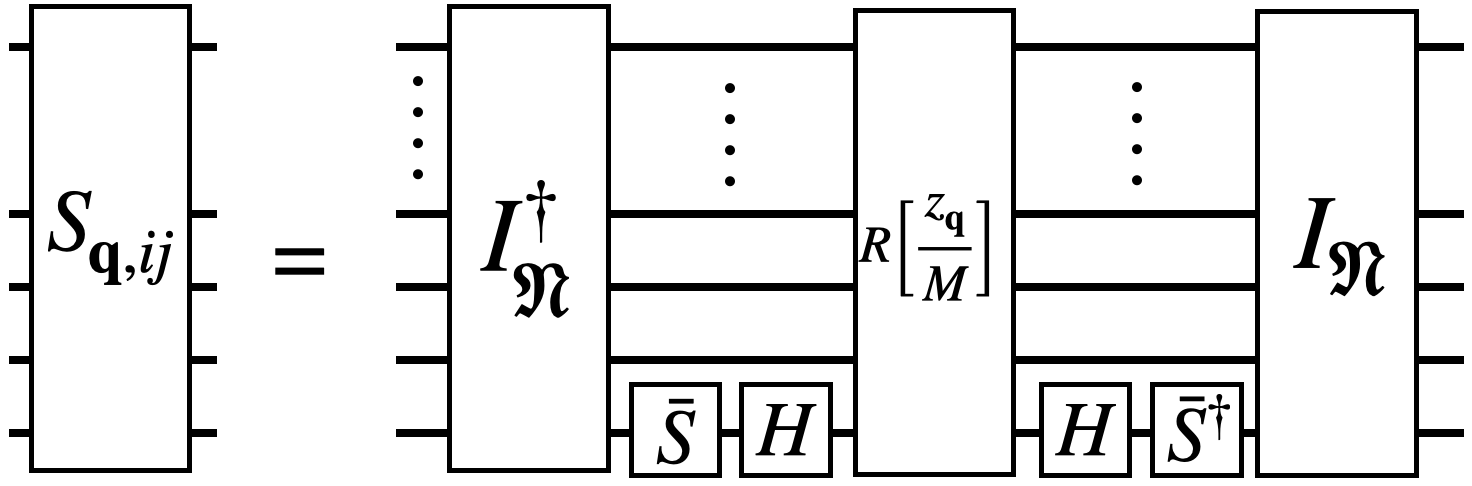}
\end{center}
\caption{Circuit implementation of $S_{\mathbf{q},ij}$ (\Eq{eq:squezingCirc}), using the bit-increment operator $I_{\mathfrak{N}}$ and the diagonal single qubit rotation $\exp\{i \frac{z_\mathbf{q}}{M} 
\sigma^z \}$. It acts on the $\mathfrak{N}$ qubits that make up $(-i)[a^{(i)\dagger}_{\mathbf{q}}a^{(j)\dagger}_{-\mathbf{q}} - a^{(j)}_{-\mathbf{q}}a^{(i)}_{\mathbf{q}} ]$.}
\label{fig:squeezing2}
\end{figure}

Following a similar strategy as in~\cite{shaw2020quantum},
we block-diagonalize $\sigma_\mathfrak{N}^y$, using the (periodic) binary increment operator $I_\mathfrak{N}$ ($I_1=\sigma^x$) 
\begin{align}\label{eq:SyInc}
I_{\mathfrak{N}}^{\dagger}\sigma_\mathfrak{N}^y I_\mathfrak{N}= 
\begin{pmatrix}
0 &\dots & \dots &0 \\
\vdots& 0 & &  0\\ 
\vdots  &  &  \ddots& i   \\
0 & 0 & -i & 0 
\end{pmatrix}\equiv  \tilde{\sigma}_\mathfrak{N}^y\,.
\end{align}
The binary increment operator is a simple circuit and can be found in the literature (for example,  in Fig.~(2) of \cite{shaw2020quantum}), and is given explicitly in Appendix \ref{app:squeezing}.
The recursion relation 
\begin{align}
\tilde{\sigma}_\mathfrak{N}^y = \frac{1}{2}(1-\sigma^z) \otimes \tilde{\sigma}^y_{\mathfrak{N}-1}\,,
\end{align}
where $\tilde{\sigma}^y_1 = - \sigma^y$, allows us to write
\begin{align}\label{eq:ful_rec}
\tilde{\sigma}_{\mathfrak{N}}^y = \Big[ \bigotimes_{i=2}^{{\mathfrak{N}}} \frac{1}{2}( 1- \sigma^z )  \Big]\otimes \tilde{\sigma}^y_1\,.
\end{align}
Because $(1-\sigma^z)$ is diagonal, the problem reduces to diagonalizing  $ \tilde{\sigma}^y_1=-\sigma^y=-\bar{S}H\sigma^z H\bar{S}^\dagger$, using the Hadamard $H$ and phase gate $\bar{S}$ acting on one qubit. Consequently, we can write
\begin{align}
S_{\mathbf{q},ij} = &I_{\mathfrak{N}} \, (1\otimes \dots 1\otimes H \bar{S}^\dagger) \, R\left[\frac{z_\mathbf{q}}{M} \right] 
 (1\otimes \dots 1\otimes  \bar{S} H) \, I_{\mathfrak{N}}^\dagger\,,
\end{align}
where  $R[\frac{z_\mathbf{q}}{M} ] \equiv \exp\{i\frac{z_\mathbf{q}}{M} \left[\otimes_{i=2}^{\mathfrak{N}} \frac{1}{2}[1-\sigma^z]_i \right]\otimes  
\sigma^z \}$ is a simple controlled (diagonal) $\sigma^z$-rotation. The algorithm is compactly summarized in \Fig{fig:squeezing2}. 

The circuit implementation of the squeezing transformation contains $O(M^2 \mathcal{V}\, \text{poly}\log(\mathcal{V}))$ elementary gate operations per Trotter time-step, where
$\text{poly}\log(\mathcal{V})$ stands for the the complexity of the bit increment $I_\mathfrak{N}$ and controlled $z$-rotation $R(z_\mathbf{q}/M)$. The $M^2$ factor is due to iterations over pairs of particle registers, while $\mathcal{V}$ reflects
the operation being performed for all modes $\mathbf{q}$.

%
%
\subsubsection{Quantum Fourier Transform}\label{sec:qFT}
Because the quantum Fourier transformation in \Eq{eq:quantFT} is a standard transformation and can be found in many textbooks~\cite{NielsenChuang}, our 
discussion here will be brief. Within our digitization framework, it is performed separately for each register and dimension, conditional on whether the corresponding register is occupied.  Towards this end, we first bring states (\Eqs{eq:defsingleparticleH}{eq:l_state}) into  a form where we can apply known algorithms for the symmetric quantum Fourier transform.
This is done by first flipping the sign qubits which we then use to control $\sigma^x$-operations of all remaining qubits making up $\mathbf{q}_i$, $i=1,\dots,d$.   Interpreting the sign qubits as the  major qubits of the decomposition of each $\mathbf{q}_i$, this allows us to apply the algorithm of~\cite{klco2019digitization}, with $O(M\text{poly}\log{\mathcal{(V)}})$ elementary gate operations.

%
%
\subsubsection{Interaction part $U_I$}\label{sec:interactionpart}
We now turn to the final quantum circuit for the time evolution operator, that of the interaction term $U_I$.
The $\phi^4$ interaction term is local in position space and can be decomposed
into $\mathcal{V}$ Trotter steps per time step $\delta$,
 \begin{align}
 U_I =\exp{\{ -i \delta  \sum_{\mathbf{n}}  \frac{\lambda}{4!}  \phi^4_\mathbf{n}  \}} = \prod_\mathbf{n} \exp\{ -i \frac{\delta \lambda}{4!} \phi_\mathbf{n}^4 \} \equiv \prod_\mathbf{n} U_{I,\mathbf{n}}\,.
 \end{align} 
To implement the circuit for this operator, we write the field operator as $\phi_\mathbf{n} \equiv \sum_{i=0}^{M-1} \phi_\mathbf{n}^{(i)} /\sqrt{M}$, where
 \begin{align}\label{eq:matrixEqPos}
 \phi_\mathbf{n}^{(i)} \equiv\frac{ a^{(i)}_\mathbf{n} + a^{(i)\dagger}_\mathbf{n}}{\sqrt{2}}= \frac{1}{\sqrt{2}}
\begin{pmatrix}
0 &\dots& 0 &1 \\
0 & \ddots & &  0\\
\vdots  &  &  \ddots& \vdots   \\
1 & 0 & \dots & 0 
\end{pmatrix} \equiv \frac{1}{\sqrt{2}} \sigma_\mathfrak{N}^x\,,
\end{align}
with $\sigma^x_\mathfrak{N}$ being the $\mathfrak{N}$-qubit operator decomposition of $\phi^{(i)}_\mathbf{n}$, comprised of the qubits that span $\{ | \mathbf{n}\rangle , | \Omega \rangle \}$, as outlined in Appendix \ref{app:detailsSingleParticle}.
Following a similar strategy as before for the implementation of the squeezing operation in Section~\ref{sec:squeezing}, we write
\begin{align}
U_{I,\mathbf{n}}\equiv V_\mathbf{n} U^{\rm diag}_{I,\mathbf{n}} V_\mathbf{n}^\dagger
\end{align}
where $U^{\rm diag}_{I,\mathbf{n}}  $ is a diagonal rotation matrix given by 
 \begin{align}\label{eq:phi4_single_particle_basis_final}
U^{\rm diag}_{I,\nb}&\equiv 
e^{-i \Delta \sum_{\langle i,j,k,l\rangle} \phi_{\mathbf{n}}^{(i) \, \rm diag}\phi_{\mathbf{n}}^{(j) \,  \rm diag } \phi_{\mathbf{n}}^{(k) \, \rm diag } \phi_{\mathbf{n}}^{(l) \, \rm diag }}\,,
\end{align}
with $\Delta \equiv \delta \lambda / (96M^2)$
and $V_\mathbf{n} \equiv \prod_{i=0}^{M-1}V_\mathbf{n}^{(i)}$ where
\begin{align}\label{eq:V}
V^{(i)}_\textbf{n} = I_\mathfrak{N} (1\otimes \dots 1\otimes H)\,.
\end{align}
 Here $I_\mathfrak{N}$ is the bit-increment operator and $H$ the Hadamard gate, while  $\phi^{(i) \, \rm diag}_\nb \equiv
V_\mathbf{n}^{(i)\dagger} \phi^{(i)}_\mathbf{n}V_\mathbf{n}^{(i)}$ satisfies $\phi^{(i)\, \rm diag}_\nb=  \bigotimes_{j} \frac{1}{2}( 1- \sigma^z )_j  \otimes \sigma^z$, in analogy to the previous Section.\par
\begin{figure}[t]
\begin{center}
\includegraphics[width=0.4\textwidth]{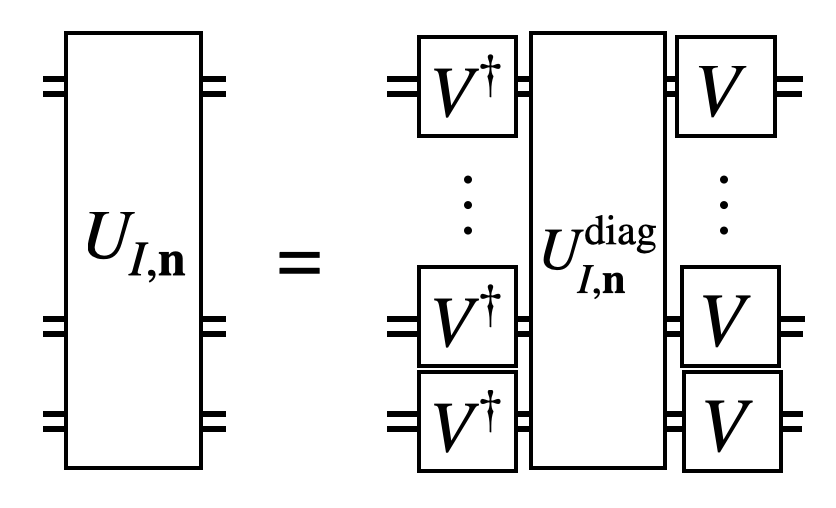}
\end{center}
\caption{Circuit implementation of $U_{I,\mathbf{n}}$. Double lines indicate particle registers. The operator $V\equiv V^{(i)}_\textbf{n}$($i=0,\dots,M-1$) is given in \Eq{eq:V}.
}
\label{fig:interaction1}
\end{figure}

 
The algorithm to implement $U_{I,\mathbf{n}}$ is compactly summarized in \Fig{fig:interaction1}, where $U^{\rm diag}_{I,\mathbf{n}}$ can be realized using standard techniques for quantum simulation \cite{NielsenChuang}. The exact form of $U^{\rm diag}_{I,\mathbf{n}}$ can be obtained by performing the summation over $\langle i,j  ,k,l \rangle$ in \Eq{eq:phi4_single_particle_basis_final}. There are five distinct cases in this sum: either the four particle's indices match, three indices match, two indices match, two pairs of indices match independently or they all differ; to exemplify how this summation is carried out, we explicitly compute $U^{\rm diag}_{I,\mathbf{n}}$ for $M=4$ and $\textbf{n}=-1/2$ in Appendix \ref{app:interactionterm}.

The algorithm for $U_I$ involves $O(M^4\mathcal{V}\, \text{poly}\log(\mathcal{V}))$ elementary gate operations. The $M^4$ dependence originates from the need to account for all the possible ways to form four-tuples with $M$ particles, and reflects the brute force approach  detailed in Appendix \ref{app:interactionterm}. This bound can be lowered (presumably down to $O(M)$), provided one finds an efficient algorithm to deal with the combinatorics in computing the respective phases by summing over $\langle i,j,k,l\rangle$; unfortunately we have not been able to construct such a simplified algorithm thus far. The linear dependence on volume is due to the fact that one has to loop over all positions while performing, for each one, $O(\text{poly}\log(\mathcal{V}))$ gate operations.

The Trotter complexity of the single-particle algorithm presented  scales linearly with volume, similar to that of the field operator based strategy in ~\cite{Jordan:2011ci,Jordan:2011ne,Klco:2018zqz}.
A meaningful comparison between the approaches will depend on the problem under consideration. Determining the precision required to quantum simulate a simple scattering process, including taking the continuum limit, both in terms of the qubit representation of states and in the accuracy of the time evolution operator, will require a detailed numerical study using larger systems.

Moreover, error mitigation techniques~\cite{NielsenChuang,brun2019quantum} need to be applied should one attempt a quantum simulation with presently available devices. This is particularly important for the scheme presented because the size of the unphysical Hilbert space of  non-Bose-symmetric states grows with $M$. This is similar to the problem of quantum simulating gauge theories where simulation errors may drive the system away from the physical
Hilbert space defined by Gauss law. It has been suggested that one can detect
such violations of symmetries without compromising the information encoded in the system and thereby correct for them~\cite{stryker2019oracles,tran2020faster,lamm2020suppressing,halimeh2020gauge,tran2020destructive}.

\subsection{Measurement}\label{sec:Measurement}

In the spacetime picture of S-matrix scattering developed thus far, we first discussed the preparation of wavepackets in the interacting theory  by adiabatically turning on the interaction over a time scale $\tau_0$. After this time scale, the wavepackets overlap and interact over a time scale $\tau_I$, determined  given by their spatial overlap. We will discuss here the algorithm for the measurement process subsequent to the scattering.

After the scattering, the wavefunction of the system can be written in the most general form\footnote{The position space representation has an identical form and is used interchangeably in the forthcoming discussion. In fact, when we use the word ``localized" here it  can equally well mean ``in position space", albeit the formulas we give are in the momentum space representation.}
\begin{align}\label{eq:StateAfterEvolution}
| \Psi(t) \rangle = \sum_{\ell} \alpha_\ell (t) | \Psi_\ell \rangle 
\equiv\sum_{\rm {basis \, states}} \frac{\alpha_{( \mathbf{q}, \mathbf{q}',\dots)}(t)}{\sqrt{\mathcal{N}_{( \mathbf{q}, \mathbf{q}',\dots)}(t)}}  \, \big[| \mathbf{q}, \mathbf{q}' ,\dots, \Omega \rangle + \rm{symm.} \big]\,,
\end{align}
with unknown coefficients $\alpha_{( \mathbf{q}, \mathbf{q}',\dots)}(t)$. Here `symm' denotes Bose-symmetric permutations and  $\mathcal{N}_{ (\mathbf{q}, \mathbf{q}',\dots)}\equiv M!/[\mathfrak{n}_\Omega! \prod_{\mathbf{q}} \mathfrak{n}_{\mathbf{q}}!]$ is a generalization of the Bose-symmetric factor $\mathcal{N}$ introduced in Section \ref{sec:InitialStatePrep} for the $M$ single-particle registers, now also accounting for the possibility of degenerate momenta among particle registers.

Upon measurement of all qubits\footnote{This is to be contrasted with the procedure in \cite{jordan2011quantum,Jordan:2011ne} where particle number measurement requires additional gate operations.}, the wavefunction in \Eq{eq:StateAfterEvolution} will collapse to a state with well defined particle number for every mode $\mathbf{q}$ (a Fock state) with probability $|\alpha_{( \mathbf{q}, \mathbf{q}',\dots)}|^2$.
Despite this, it is important to note that \Eq{eq:StateAfterEvolution}  does not imply any kind of localization/clustering of the particles measured in a detector, if measured at $t=\tau_0 + \tau_I$. One may further evolve the system over a time $\tau_f$ during which one turns off the interaction slowly to avoid interactions between separated wavepackets\footnote{Obviously, one can go beyond this picture by extracting information on the scattering process through measurements at \textit{any} time, as we shall discuss below.}  until 
one ends up with localized particles over macroscopic scales that are theory specific. These are then straightforward to measure due to \Eq{eq:StateAfterEvolution}.

Measurements of identical Bose particles, with different orderings amongst the particle registers, are physically equivalent. Up to kinematic factors, this measurement defines the differential cross-section
\begin{align}
 \frac{\text{d}^{d\mathfrak{n} }\sigma }{\text{d}^d \mathbf{p}_0 \dots \text{d}^d \mathbf{p}_\mathfrak{n}}\,,
\end{align}
of $\mathfrak{n}=\sum_{\mathbf{q}} \mathfrak{n}_\mathbf{q}$ particles for a given outcome. From this perspective, running the quantum computer multiple times is very similar to accumulating events in an actual particle physics experiment -- followed by a classical analysis of events. However on the quantum computer, every outcome allowed by energy-momentum conservation, as well other conserved quantities corresponding to symmetries of the system, is contained in the state \Eq{eq:StateAfterEvolution}. For example, one can  simply  measure only occupancy qubits, but not their corresponding momentum counterparts, to obtain an integrated cross-section, 
\begin{align}
\sigma_\mathfrak{n} \equiv \int\, \text{d}^d \mathbf{p}_0 \dots \text{d}^d \mathbf{p}_\mathfrak{n} \frac{\text{d}^{d\mathfrak{n} }\sigma }{\text{d}^d \mathbf{p}_0 \dots \text{d}^d \mathbf{p}_\mathfrak{n}} \,,
\end{align}
directly. Similarly, in more complicated theories, one can introduce single-particle registers with qubits corresponding to electric charge, spin or color and directly project on to desired values of these for a specific measurement.

One can also instruct the quantum algorithm to impose kinematical cuts such as measuring localized particle number in some region $\pb\in [\pb^{\rm min},\pb^{\rm max}]$. To achieve this, one requires $2d$ auxiliary registers (of size $\log_2(\mathcal{V})$) set to kinematic bounds $\pb^{\rm min/max}$ in $d$ dimensions. One further requires a unitary comparator circuit~\cite{cuccaro2004new,oliveira2007quantum,xia2018efficient} (using  $\log_2(\mathcal{V})$ ancilla qubits and $O(\log(\mathcal{V}))$ gate operations) which computes whether $\pb_i \leq \pb_i^{\rm max}$ and $\pb_i \ge \pb_i^{\rm min}$ ($i=1,\dots,d$) and stores the information in $2d$ ancilla qubits with outcome $| 11 \rangle^{\otimes d}$ if the momentum is within the kinematical range. This provides a way to efficiently split the Hilbert space into two non-overlapping regions while tagging each component of the final state $\ket{\Psi_l}$ accordingly. As a consequence, techniques like (Oblivious) Amplitude Amplification~\cite{brassard2002quantum,NielsenChuang,Berry_2014,Guerreschi_2019,Berry_2014,paetznick2014repeatuntilsuccess} might be employed to boost the probability of measuring the rare final state that satisfies the kinematical cuts imposed. Alternately, generalizing to other theories, one can use this method to identify 
states with unusual particle number content.

We note however that the regime at $\lambda=0$, and that of its physical value, may not be adiabatically connected because the spectrum of the latter 
may consist of bound states. In this case, one omits the evolution over a time $\tau_f$ where one turns off the interaction and instead should keep the interaction time $\tau_I$ long enough to include the physical time it takes to form such a bound state.
One example where the spectrum of the free and interacting theory are not adiabatically connected is that of QCD. While color charged quark and gluon states arguably form a good basis to represent the proton wavefunction at high energies and short time intervals, at large distances and time intervals they are not contained in the physical spectrum 
because of the confinement/deconfinement phase transition, as is manifest in the dynamical process of hadronization/fragmentation between these regimes~\cite{mueller1981multiplicity,dokshitzer1991basics,webber1999fragmentation,andersson2005lund}. Such difficulties are also present in a field-based digitization, and explicit numerical analysis is required to investigate how well the proposed basis can approximate such states in the continuum limit. 

 As we discussed previously, the minimal time scale for the formation of a bound state is the Wigner time delay -- for a discussion of resonance formation in the S-matrix picture, see \cite{Dashen:1974yy}. Once this is done, and bound states are sufficiently separated, one can make local measurements of quantum 
numbers such as particle number or momentum (electric charge and spin can also be measured in more complicated theories, for example), the operator for the latter defined as 
\begin{align}\label{eq:momentumeigenvalue}
{P}^i_{ \tilde{\mathcal{V}}_{\mathbf{p}} }\equiv \int_{\tilde{\mathcal{V}}_{\mathbf{p}}} d^d \textbf{p}\,  \textbf{p}^i \, a^\dagger_{\mathbf{p}} a_{\mathbf{p}}\,,
\end{align}
where $i=1,\dots,d$ and $\tilde{\mathcal{V}}_{\mathbf{p}}$ stands for a region in momentum space.  Its expectation value can be obtained using variants of the phase estimation algorithm (PEA)~\cite{cleve1998quantum,abrams1999quantum,knill2007optimal,roggero2020short}. 
The idea is to act on the state with $U\equiv \exp{(-i {P}^i_{ \tilde{\mathcal{V}}_{\mathbf{p}} } )}$  to determine the operator expectation value $\langle {P}^i_{ \tilde{\mathcal{V}}_{\mathbf{p}} } \rangle$. 
The PEA determines, with high probability, this expectation value to within precision $\varepsilon$. 
It requires extra $O(\log(\varepsilon^{-1}))\sim O(\log(\mathcal{V})/d)$ ancilla qubits\footnote{We require that the precision of the PEA should be the same as that for the momentum space discretization.} and $O(\log(\varepsilon^{-1}))$ applications of the controlled-$U$ operations. In our digitization scheme, it is straightforward to obtain $\langle {P}^i_{ \tilde{\mathcal{V}}_{\mathbf{p}} } \rangle$ because 
the circuits of Section \ref{sec:freepart} can be applied with small modifications. Concretely, one replaces $\omega_{\mathbf{p}}$ by $\mathbf{p}^i$ 
in this algorithm and also uses a comparator circuit to check if $\mathbf{p}^{i}$ is in $\tilde{\mathcal{V}}_{\mathbf{p}}$ controlling the execution of the circuit.

Likewise, the energy operator, restricted to $\tilde{\mathcal{V}}_{\mathbf{p}}$, is
\begin{align}\label{eq:MeasureHamiltonian}
H_{ \tilde{\mathcal{V}}_{\mathbf{p}} }\equiv \int_{\tilde{\mathcal{V}}_{\mathbf{p}}} d^d \textbf{p}\, H_\mathbf{p}  =\int_{\tilde{\mathcal{V}}_{\mathbf{p}}}  d^d \textbf{p}\, H_{0,\mathbf{p}}+  
\int_{\tilde{\mathcal{V}}_{\mathbf{p}}}  d^d \textbf{p}\, H_{I,\mathbf{p}}\,,
\end{align}
where $H_{0,\mathbf{p}}$ and $H_{I,\mathbf{p}}$ are the Fourier transforms of the Hamiltonian densities $H_{0,\mathbf{x}}$ and $H_{I,\mathbf{x}}$, with $H_0=\int d^d \mathbf{x}\,  H_{0,\mathbf{x}} $ and $H_I=\int d^d\mathbf{x} \, H_{I,\mathbf{x}}$.
One can measure the contribution to the expectation value $\langle H_{ \tilde{\mathcal{V}}_{\mathbf{p}} } \rangle$ from the first term just as in  \Eq{eq:momentumeigenvalue}. To obtain the second term, we write
\begin{align}\label{eq:HamiltInt}
\int_{\tilde{\mathcal{V}}_{\mathbf{p}}}  d^d \mathbf{p}\, H_{I,\mathbf{p}} = \int  d^d \textbf{p}\, H_{I,\mathbf{p}}\,  \theta_{\tilde{\mathcal{V}}_{\mathbf{p}}}({\mathbf{p}})\, , 
\end{align}
where $\theta_{\tilde{\mathcal{V}}_{\mathbf{p}}}({\mathbf{p}})$ is a (smooth) envelope
function restricting the integrand to $\tilde{\mathcal{V}}_{\mathbf{p}}$. To illustrate 
the procedure, we now assume for simplicity that $\theta_{\tilde{\mathcal{V}}_{\mathbf{p}}}({\mathbf{p})}$ is a sharp envelope function, e.g. a $d$-dimensional box function with equal length, i.e. $\theta_{\tilde{\mathcal{V}}_{\mathbf{p}}}({\mathbf{p})}=1$ if $\mathbf{p}\in \tilde{\mathcal{V}}_{\mathbf{p}}$ and zero otherwise, where $\tilde{\mathcal{V}}_{\mathbf{p}} \equiv (L_{\mathbf{p}})^d $ is centered at some $\bar{\mathbf{p}}$. We can make use of the Fourier convolution theorem to compute this term. First, using the momentum space translation operator introduced in Section \ref{sec:InitialStatePrep}, we translate the state such that $\tilde{\mathcal{V}}_{\mathbf{p}}$ is centered around zero. After performing the squeezing and Fourier transformations discussed in sections \ref{sec:squeezing} and \ref{sec:qFT}, \Eq{eq:HamiltInt} can be written as
\begin{align}
 \int  d^d \textbf{x}\, H_{I,\mathbf{x}}\,  g(-\mathbf{x})\,,
\end{align}
where the Fourier transform of the box function (centered around zero) is real,  $g(\mathbf{x})\equiv (2\pi)^{d/2} \prod_{i=1}^d  \frac{\sin(\textbf{x}_i L_{\mathbf{p}} / 2  )}{\mathbf{x}_i}$. The PEA~\cite{cleve1998quantum,abrams1999quantum,knill2007optimal,roggero2020short} can be applied again, replicating the algorithm of Section \ref{sec:interactionpart}, albeit with the replacement $\lambda \rightarrow \lambda  g(-\mathbf{x})$. For this specific envelope function, the measurement has a gate complexity of $O( \mathcal{V} M^4 \, \text{poly}\log(\mathcal{V}) )$. A sharp envelope function is not ideal because it requires evaluating also the side-bands of the $\sin(x)/x$ function. In practice, one should use a smooth cutoff function, whose Fourier transform is known analytically or numerically, which falls off exponentially.
In this case, the estimate will only depend on the much smaller sub-volume $\mathcal{V}_{\mathbf{x}}\subset \mathcal{V}$ over which the Fourier transform of the envelope function is supported, instead of the full volume $\mathcal{V}$. Similar algorithms are applied to compute energy and momentum densities restricted in position space.

In general, being able to control the wavefunction of a many-body system
at any time $t$ one can in principle follow the entire spacetime evolution
of a particular collision system, instead of measuring just its asymptotic outcome, and thereby  obtain snapshots of the collision process. This is important for systems such as ultrarelativistic heavy ion collisions where the primary interest lies in the thermalization and hydrodynamization of the produced matter~\cite{Berges:2020fwq} as opposed to the asymptotic final states. Likewise, following Feynman's idea of quantum simulating a particle physics experiment in its entirety,
having full control over the time evolution allows one to measure arbitrary (non-equal time) correlation functions directly. (See also \cite{Mueller:2019gjj} where this point is discussed.) This will allow for a more   
direct comparison with current theoretical efforts such as computing parton distribution functions \cite{Ji:2013dva} or hadronic and Compton scattering amplitudes~\cite{briceno2020role}
from correlation functions.

Moreover, quantum computation allows one to address the question of entanglement in nuclear physics \cite{Robin:2020aeh} and in high energy physics. With regard to the latter, the single particle basis described here may be useful to quantify entanglement between partons as probed in DIS experiments~\cite{kharzeev2017deep,hagiwara2018classical,kovner2019entanglement,tu2020einstein}, its role in thermalization of the quark-gluon plasma, in hadronization~\cite{Berges:2017zws,Berges:2017hne,Bauer:2019qxa}, or in the composition of the proton's spin~\cite{beane2020chiral,Tarasov:2020cwl}.

\subsection{Renormalization}\label{sec:Renormalization}

The renormalization of quantum fields to absorb the apparent infinities that appear in computations is a fundamental feature of quantum field theories. It is therefore important to understand how to treat this problem in the real time  Hamiltonian description of the evolution of quantum fields and its implementation on a quantum computer. More specifically, we need to understand how to implement the renormalization group for scattering problems in our single-particle framework.

We  begin our discussion with a general overview of the renormalization group procedure in the Hamiltonian formalism. We will  illustrate this picture in perturbation theory. We argue however 
that non-perturbative renormalization is essential to ensure one does not vitiate the reduction in computational complexity presented by quantum computations relative to classical approaches. We will therefore outline a concrete non-perturbative scheme closely paralleling
the corresponding procedure in classical lattice computations in the (Euclidean) path integral formalism.

\subsubsection{Operator formulation}
The renormalization of quantum fields and operators requires finding a Hamiltonian for the effective field theory of interest (defined with an ultraviolet cutoff) concretely through a lattice discretization
as well as the truncation of the Hilbert space imposed by a given digitization scheme. Since renormalization in the Hamiltonian operator formalism has been developed extensively~\cite{wegner1994flow}, as well as its applications to single-particle strategies~\cite{Perry:1993gp,glazek1993renormalization,perry1994renormalization}, we will only outline the relevant ideas in the context of this work. Working in the computational basis (the eigenbasis of the free Hamiltonian $H_0$) introduced in Section~\ref{sec:strategies_for_quantum_simulation}, we can write the Hamiltonian in the block form
\begin{align}
H=\begin{pmatrix}
H_{ll} & H_{lh}\\
H_{hl} & H_{hh}
\end{pmatrix}\,.
\end{align} 
The matrix elements in this representation are between states with energies $E=\sum_{\mathbf{p}} \omega_{\mathbf{p}}  \mathfrak{n}_{\mathbf{p}}$, either below $(l)$  or above $(h)$ a cutoff $\Lambda$.

A renormalization group (RG) transformation consists of the similarity transformation
\begin{align}\label{eq:RGgroupHamiltonian}
H^{\rm eff}\equiv T H T^\dagger\,,
\end{align}
where $T\equiv\exp{(i \eta)}$
 block-diagonalizes $H$, eliminating matrix elements between the low and high energy sectors such that $H^{\rm eff}_{ll}$ in the new basis defines a low energy effective field theory\footnote{Note that a self-consistent formulation of the S-matrix in this picture may provide deeper insight into ambiguities regarding the elementarity of the degrees of freedom included in the EFT~\cite{Dashen:1974yy}. For a recent discussion, see \cite{Beane:2020wjl}.}. The generator $\eta$  of this similarity transformation is not known a priori. It can however be constructed to realize a non-perturbative RG, the so-called similarity RG~\cite{wegner1994flow},  by integrating out one
energy shell at a time in infinitesimal steps. This point is discussed further in  Appendix \ref{app:renormalization}.

 If the coupling is small enough, perturbative renormalization is applicable. This procedure is very familiar to the high energy physicist in its Lorentz covariant path integral formulation; in the Hamiltonian operator picture,
it is best illustrated through a Schrieffer-Wolf transformation, as discussed in~\cite{bravyi2011schrieffer} and worked out in Appendix F. As is shown there, this allows to systematically derive low energy elements of $H_{\rm eff}$ and of any other operator order by order in $\lambda$.

However it is not difficult to see that doing so comes with a factorial increase in the computational complexity, just as the number of Feynman diagrams grows factorially with loop order in a path integral formulation. Moreover such a perturbative computation will break down if there is a phase transition in $\lambda$, as is likely for $D=2,3$ for scalar $\phi^4$ theory; for QCD, this expansion will be problematic for quantum simulations that attempt to treat hadronization of parton single-particle degrees of freedom. 

Therefore to match the quantum advantage of the renormalization procedure with that of the non-perturbative formulation 
of the rest of our treatment of the scattering problem, we will outline below a practical scheme to non-perturbatively renormalize
the theory on a quantum computer.

\subsubsection{Non-perturbative renormalization scheme}
We begin by outlining how exactly renormalization enters our algorithm.
As shown in \Fig{fig:algorithm}, the algorithm includes a turn-on of interactions from a free (but unphysical) theory at $t=0$, where the initial state
can be prepared, to the interacting (physical) theory at $t=\tau_0$ with time-dependent Hamiltonian $H(t)=H(\lambda(t),m(t))$.

It is only the couplings in the physical Hamiltonian at $t\ge \tau_0$,
\begin{align}\label{eq:couplingst0}
\lambda(\tau_0) = \lambda^{\rm ren} \,, \qquad m(\tau_0)=m^{\rm ren}\,,
\end{align}
that are to be determined by a renormalization group procedure which we outline below. The `unphysical'
theories defined by $H(t)=H(\lambda(t),m(t))$ at $t<\tau_0$, including the initial values
\begin{align}\label{eq:couplingszero}
\lambda(0) = 0 \,, \qquad m(0)=m_0\,,
\end{align}
are not renormalized because there is simply no physical renormalization for them.
Instead, one simply works with a linear interpolation 
\begin{align}
&\lambda(t) = \lambda^{\rm ren} \frac{t}{\tau_0}\,,
\qquad m(t)=m_0\big(1-\frac{t}{\tau_0}\big)+m^{\rm ren}\frac{t}{\tau_0}\,, 
\end{align}
for $t\in[0,\tau_0]$ and constant thereafter. From a practical perspective, the unknown parameter $m_0$ may be chosen to represent a relevant energy scale in the weakly coupled regime of the theory such as for example the bare quark mass in QCD.
However if the system undergoes a phase transition during this turn-on procedure, the mass and energy scales of the weakly and strongly coupled regimes of the theory are very different (as is the case in QCD), requiring large lattices to resolve both regimes.

We now turn our attention to determining the renormalized values for the bare parameters  $\lambda(t)$ and $m(t)$ at $t\ge \tau_0$. We will assume form invariance of the Hamiltonian of the form  \Eq{eq:latticeHamiltonian} for all values of lattice spacing $a_s$ and particle number cutoff $M$. In other words, we do not add dimensionful operators that  would be generated by the similarity transformation in \Eq{eq:RGgroupHamiltonian}. (These  could in principle improve the convergence to the continuum limit.)

To renormalize the Hamiltonian operator, it is sufficient to perform the 
computation of a static property and then use the result as the input for the computation of a scattering process\footnote{A caveat here is that since the 
scattering process likely covers a larger range of scales, the continuum extrapolation of
the cross-section is more challenging than that of the low energy spectrum.}. The non-perturbative renormalization strategy consists of the following steps:

\begin{enumerate}
    \item First, one quantum computes a static and dimensionless physical quantity such as the energy ratio of two low lying excitations at a given $a_s$ and $M$. One then repeats the computation adjusting the bare parameters $\lambda, m$ so that the physical value is reproduced for that $a_s$ and $M$.
    We will not discuss the details of such computations here but note that 
    algorithms\footnote{Examples of such algorithms include variational approaches \cite{otterbach2017unsupervised,peruzzo2014variational}, adiabatic state preparation with quantum phase estimation~\cite{kitaev1995quantum,farhi2000quantum}, quantum approximate optimization~\cite{farhi2014quantum,otterbach2017unsupervised}, quantum imaginary time and quantum Lanczos algorithms~\cite{motta2020determining}, and efficient operator averaging techniques~\cite{cleve1998quantum,abrams1999quantum,knill2007optimal,roggero2020short}.} to do so can be applied to our single-particle digitization.
    \item One then repeats the computation at a somewhat different $a_s,M$ along the direction $a_s \rightarrow 0$ and $M\rightarrow\infty$, and adjusts the values of $\lambda, m$ so that the aforementioned physical quantity does not change. \item One repeats this computation for various $a_s,M$ along a line of constant physics. Because there are now two directions $(a_s,M)$, this procedure is in principle ambiguous. In practice however it should be subject to an optimization procedure identifying the most relevant RG direction, such as determined by a steepest decent approach. We will not discuss such a procedure here.
    \item Once the renormalized values $\lambda^{\rm ren}$ and $m^{\rm ren}$ are known for a range of $(a_s,M)$, one performs the scattering experiment outlined in this manuscript with these values as input. This also includes the 
    renormalization of operators $\langle O^{\rm eff} \rangle$ measured in Section \ref{sec:Measurement} such as particle number, momentum and energy density. In the simplest case, one sets $  O  =Z  O^{\rm eff} $ and determines $Z$ in the same way 
    as for the bare $\lambda$ and $m$.
    \item Finally, one performs a continuum extrapolation of the observables obtained in the scattering experiment. This dynamical problem will require determining the $\lambda^{\rm ren}$, $m^{\rm ren}$ and $Z$ over a large range $(a_s,M)$ which is likely  computationally demanding even with a quantum computer.
\end{enumerate}
This procedure is similar to the Luscher formalism that relates energy differences between static long-lived states and S-matrix elements. Extracting the latter from the former is in general an inverse scattering problem and a number of sophisticated techniques have been developed in this regard~\cite{Hansen:2017mnd}. A potential advantage of the quantum computation is that both sides of the Luscher relation can be computed in real time; realizing this in practice is of course very challenging.


\section{Summary and Outlook}\label{sec:outlook}
In this work, we developed a novel single-particle digitization strategy for the quantum simulation of scattering in a relativistic scalar $\phi^4$ field theory in $d$ dimensions. The essence of this picture is a relativistic generalization of a single-particle picture consisting of  $M$ ``particle registers'' whose Hilbert space spans states over a volume $\mathcal{V}$. 
Our approach is non-perturbative and fully general and may offer a quantum  advantage over other digitization strategies for a class of interesting physical problems that are challenging to address with purely classical methods.

 The conceptual elements of this framework are outlined in sections \ref{sec:DIS_Smatrix} and \ref{sec:strategies_for_quantum_simulation}. We developed quantum circuits for the initial state preparation of scattering wavepackets in Section \ref{sec:InitialStatePrep}, their time evolution through the scattering process in Section \ref{sec:TimeEvolution} and the subsequent measurement of final states in Section \ref{sec:Measurement}. We sketched 
in Section \ref{sec:Renormalization} the elements of a non-perturbative renormalization strategy that must  be implemented in the quantum simulation to achieve physically meaningful results. 

The overall gate complexity of the elements of a quantum circuit for a scattering simulation are compactly summarized in Table \ref{tab:overall_cost}. The initial state preparation requires $O(M^\mathfrak{n} \log(\mathcal{V}))$ elementary 
gate operations, where $\mathfrak{n}$ is the initial number of particles (the simplest case being two-particle scattering with $\mathfrak{n}=2$), and
$O(\log(M^\mathfrak{n}))$ ancilla qubits. The algorithm may become probabilistic,  requiring additional measurements for certain choices of $\mathfrak{n}$ and $M$ depending on Bose combinatorics. A Trotter scheme is employed to separate the time evolution operator into free and interaction parts; these are evaluated respectively in momentum and position space representations of the single-particle digitization basis. The change of basis from the former to the latter is achieved through a combination of squeezing and quantum Fourier transform operations.
The dominant cost of the algorithm is from the $O(M^4 \mathcal{V}\, \text{poly}\log(\mathcal{V}))$ gate operations per Trotter step required to compute the
interaction part of the time evolution operator. We believe that one can improve the polynomial cost in the number of registers $M$ by improving the algorithm outlined in section~\ref{sec:interactionpart}. This would open up a broader class of interaction terms and theories that could be efficiently simulated within this approach. The measurement of particle number incurs no additional cost; the estimation of the localized momentum and energy density (in a sub-volume $\mathcal{V}_\mathbf{x}\subset\mathcal{V}$), via the phase estimation algorithm, requires $O(M\text{poly}\log{(\mathcal{V})})$ and $O(M^4 \mathcal{V}_\mathbf{x} \text{poly}\log{(\mathcal{V})})$ operations respectively. We note that some of the unitary operations in our circuit, such as the squeezing operation or the diagonal phase multiplication used in computing the $\phi^4$ interaction term, are available as native gates in certain architectures such as circuit QED~\cite{girvin2011circuit}, potentially improving their resource efficiency and  facilitating a near-term implementation of our strategy.

Apart from the Hilbert space truncation, sources of errors in our algorithm are  from the Trotterization of the time evolution operator, and  imperfect evolution of the qubits on non-error corrected devices. It should be possible to derive rather tight bounds on the Trotter error, using similar techniques as in~\cite{Childs_2021}, and it would be interesting to compare them with~\cite{jordan2011quantum,jordan2012quantum}. Machine errors, such as bit flips, are important because, if they occur in a major bit of the momentum/position of a particle register, they can change a position/momentum eigenstate drastically. Such errors could be protected using linear codes~\cite{steane1996multiple,steane2007tutorial,calderbank1996good}. We also note that, because the momentum/position information is entangled over several registers in our (Bose-symmetric) digitization, such errors will take the state into an unphysical regime and can be detected easily. Whether this symmetry can be used to correct or minimize errors will be explored in future work.

\begin{table}[t]
\begin{tabular}{|c| c c | c| }
\hline
 & \multicolumn{2}{c|}{Elementary gate operations} &  Ancilla qubits  \\ \hline\hline
 \multirow{2}{*}{ Initial State preparation} & \multirow{2}{*}{$O(M^{\mathfrak{n}} \log{(\mathcal{V}}) )$}  & $p_{\rm success}=1$ \quad\;  [$\text{exact}^{*}$] & $\log(M!/(M-\mathfrak{n}!))$ \quad [$\text{exact}^{*}$]  \\
 & &  $p_{\rm success}>1/2$  \, [$\text{probabilistic}^{*}$]   & $O( \log(M^{\mathfrak{n}}))$ \, [$\text{probabilistic}^{*}$] \\\hline
 \multirow{5}{*}{Time Evolution}  & Free part $U_0$ & $O(M  \text{poly}\log{\mathcal{(V)}} \, t)$ & $O(\log{(\mathcal{V})/d})$ \\
  & Squeezing transform $S$ & $O(M^2 \mathcal{V}  \text{poly}\log{\mathcal{(V)}} \, t ) $ & 0 \\
  & quantum Fourier transform & $O(M \text{poly}\log{(\mathcal{V})}\, t)$ & 0 \\
  & Interaction part $U_I$ & $O(M^4 \mathcal{V} \text{poly } \log{(\mathcal{V})} \, t)$ &  $O(\log(\mathcal{V})/d)$ \\ \cline{2-4}
  & {Total} & {$O(M^4 \mathcal{V} \text{poly } \log{(\mathcal{V})} \, t)$} & $O(\log(\mathcal{V})/d)$ \\ \hline
   \multirow{3}{*}{Measurement} & Particle number & 0 & 0\\ 
   & Momentum density &  $O( M\,  \text{poly} \log{(\mathcal{V})})$ (PEA${}^{**}$) & $O(\log(\mathcal{V})/d)$ \\ 
   & Energy density & $O(M^4 \mathcal{V}_\mathbf{x} \, \text{poly} \log{(\mathcal{V})})$ (PEA${}^{**}$) & $O(\log(\mathcal{V})/d)$  \\ \hline
   \end{tabular}
   \caption{Cost of the circuit implementation  discussed in this manuscript, assuming noiseless qubits. We use the following abbreviations:  number of particle registers $M$, volume $\mathcal{V}$, occupied registers in initial state $\mathfrak{n}$, dimension $d$, Trotter time steps $t$. $({}^{*})$ If $\log_2(M!/(M-\mathfrak{n}))$ cannot be chosen integer, the initial state is prepared with  probability $p_{\rm success}> 1/2$, depending on the choice of $M$ and $\mathfrak{n}$. $({}^{**})$ Measurements of (localized) energy and momentum densities are via the phase estimation algorithm (PEA)~\cite{cleve1998quantum,abrams1999quantum,knill2007optimal,roggero2020short}. The cost estimate for the localized energy density includes a factor  $\mathcal{V}_\mathbf{x}\subset \mathcal{V}$ denoting a small sub-volume of the total $\mathcal{V}$, see Section \ref{sec:Measurement}.}
   \label{tab:overall_cost}
\end{table}

Our framework can be compared to the paradigmatic description of scattering on quantum computers by Jordan, Lee and Preskill (JLP)~\cite{jordan2011quantum,jordan2012quantum} which, in contrast, is based on the digitization of field operators\footnote{The implementation of the JLP program for scattering problems has been discussed at length recently~\cite{Klco:2018zqz} and compared to an alternative digitization strategy employing a harmonic oscillator basis in position or momentum space~\cite{Yeter-Aydeniz:2017ubh,Macridin:2018gdw,Macridin:2018oli}.}.
Our digitization strategy differs fundamentally from JLP and other field digitization approaches since the number of degrees of freedom in our approach scales linearly with the particle number (and as a logarithm of the  volume) as opposed to the linear scaling with volume in the field digitization approach. However the logarithmic scaling in our approach only holds if the system is dilute; for dense systems with high occupancy, one recovers linear scaling or greater with the volume and the single-particle strategy is no longer preferred.
This is seen on the algorithmic level when the required Bose-/Fermi- symmetrization creates a large overhead of unphysical/unused states in Hilbert space. Because $M\sim \mathcal{V}$ in such situations, the cost for the time evolution operator would be significantly higher, albeit still polynomial in volume, as can be inferred from Table \ref{tab:overall_cost}.

The situation is analogous to the virial expansion we discussed previously which breaks down for high-density systems. 
Thus just as the virial expansion is very useful for a wide class of many-body problems, our single-particle approach 
   may present a quantum advantage for a number of physical problems. 
   From a purely practical point of view, the logarithmic scaling with volume of our approach will be useful in  benchmark computations for a class of scattering problems with NISQ era quantum hardware, where only few tens to hundreds of noisy qubits will be available. 
A physics application where our strategy may provide a quantum advantage is the Feynman diagram approach to compute scattering amplitudes at weak coupling. As pointed out in~\cite{jordan2011quantum,jordan2012quantum} a quantum computation avoids the combinatorial complexity with increasing precision that burdens classical computations. 
 Another appealing feature of our strategy is the relative simplicity of initial state preparation and of the extraction of inclusive cross-sections; the latter, for instance, requires no additional gate operations. Not least, the single-particle approach, as articulated in Section \ref{sec:Renormalization}, provides a transparent realization of a non-perturbative renormalization scheme that can simultaneously be used to fix lattice masses and couplings from comparisons to static properties of the system and to compute physically meaningful cross-sections.

One can extend our strategy to fermionic theories and theories involving internal symmetries. For a fermionic theory, 
the algorithm in Section \ref{sec:InitialStatePrep} can be modified  to produce antisymmetrized wavefunctions, and may offer a new fermion qubit mapping that is useful in higher dimensions. Internal symmetries such as spin and color can 
also be realized via the strategy discussed in \cite{mueller2020deeply}. For example, to realize a Dirac fermion in $3+1$ dimensions, one only needs to modify \Eq{eq:stateBinaryDecomp} to include two extra fermionic degrees of freedom; these can then be mapped on to two qubits by means of a Jordan-Wigner transformation, thereby realizing the four-dimensional spinor matrix space.  Likewise, for color $SU(3)$ in the fundamental representation, only three extra qubits are required. Details of the spin and extensions to $ SU(N_c)$ in arbitrary representations are given in \cite{mueller2020deeply,mueller2019constructing,Tarasov:2020cwl}.

The theory can be non-pertubatively coupled to gauge fields\footnote{For discussions of first principles quantum simulation of   non-Abelian gauge theories see~\cite{brower1999qcd,banerjee2013atomic,zohar2015quantum,Klco:2019evd,kasper2020jaynes,Davoudi:2020yln,dasgupta2020cold}.}. An important consequence in doing so is that the quadratic term of the Hamiltonian, whose implementation is discussed in Section \ref{sec:freepart},
is not diagonal anymore.  The theory can be non-pertubatively coupled to gauge fields.  An important consequence in doing so is that the quadratic term of the Hamiltonian,  whose implementation is discussed in Section IV B 1,  is not diagonal anymore. Hence  one  would  have  to  develop  an  algorithm  similar  to  that  used  for  the  interaction  term.   A  significant downside would be that one no longer can work with the continuum dispersion relation in momentum space but would instead have to use a lattice discretization of the Laplacian operator in Eq.(11).  This introduces larger discretization errors which are unknown in the strongly coupled regime and are likely more severe than the cost of the squeezing and quantum Fourier transformations that are avoided by working purely in the coordinate space basis.

As a next step, we aim to perform a numerical study focusing
on the simplest case of $d=1$ spatial dimensions. While we work in the eigenbasis of the free Hamiltonian,
we will test,  using exact diagonalization, how well the spectrum of the theory can be reproduced in the interacting theory at finite $\lambda$ for given lattice discretization and $M$. This is similar to what is done in \cite{klco2019digitization} using the digitization of \cite{jordan2011quantum,jordan2012quantum}. At weak coupling, 
the results of such study can be compared with lattice perturbation theory, unlike at strong coupling where the analysis includes varying $M$ and $\mathcal{V}$ over a wider range, hoping for eventual convergence.

Next, one could classically compute our algorithm within the simplest case of $M=2$ in $d=1$ dimensions with $N_s=8$ ($16$) lattice sites. This would correspond to a quantum simulation with $8$ ($10$) qubits, plus an overhead of ancilla qubits. An important
motivation for such a study would be to quantify the consequences of the violation of Bose symmetry by injecting errors into the simulation. 

With this as benchmark, we plan to implement elements of our circuit on available quantum hardware, starting in the simplest case of $M=2$ in $d=1$ dimensions which we assume can be done using lattices up to $N_s\sim O(8)$  sites. 
While negligible for large systems, the overhead from ancilla qubits is a significant part of the computational budget for such small number of sites. Preparing the Bose-symmetrized initial state is already a non-trivial task involving 
entangling the two particle registers. To implement the time evolution algorithm,  a quantum algorithm for the oracle to compute the single particle energy $\omega_\mathbf{q}$ from the momentum $\mathbf{q}$ has to be devised for the free part of the time evolution operator $U_0$. While it is certainly possible to come up with an efficient circuit for $\omega_\mathbf{q}$, a simpler strategy would be to simulate $U_0$ in position space albeit with a lattice discretized Laplacian operator. The resulting complexity of $O(\mathcal{V})$ versus $O(\log (\mathcal{V}))$ would hardly make a difference on lattices this small.

Finally, we should mention that we see important applications of our single particle basis digitization strategy to quantum computing scattering cross-sections,
nuclear structure functions and jet fragmentation functions probed at high energy collider experiments such as the Large Hadron Collider, the Relativistic Heavy Ion Collider and the upcoming Electron-Ion Collider~\cite{accardi2016electron}.
A single particle basis may also be useful to quantify the role of entanglement in
high energy and nuclear physics, for example, between partons probed in DIS experiments~\cite{kharzeev2017deep,hagiwara2018classical,kovner2019entanglement,tu2020einstein}, the evolution of entanglement during the parton fragmentation process~\cite{Berges:2017zws,Berges:2017hne} and its role in the composition of the proton's spin~\cite{beane2020chiral}. We also see novel applications of this approach~\cite{deJong:2020tvx,Zhang:2020uqo} to systems in high energy nuclear and particle physics that can be described by hybrid quantum/classical dynamics such as QCD in the Regge limit~\cite{Gelis:2010nm} and the thermalization dynamics of the quark-gluon plasma in ultrarelativistic heavy ion collisions~\cite{Berges:2020fwq,krasnitz1999non,berges2012overpopulated,mace2020chiral}.

 \section*{Acknowledgments}
 N.M. would like to thank Ning Bao, Zohreh Davoudi, Nikhil Karthik, Alex Shaw and Torsten Zache for discussions. R.V. would like to thank Robert Konik for a useful discussion. We also thank Nathan Wiebe for very useful comments on quantum algorithms. 
 This project was supported by a fellowship to J.B. from ``la Caixa" Foundation (ID 100010434)-- fellowship code LCF/BQ/ DI18/11660057, and by funding from the European Union's Horizon 2020 research and innovation program under the Marie Sklodowska-Curie grant agreement No. 713673. J.B. is supported by Ministerio de Ciencia e Innovacion of Spain under project FPA2017-83814-P; Unidad de Excelencia Maria de Maetzu under project MDM-2016-0692; European research Council project ERC-2018-ADG-835105 YoctoLHC; and Xunta de Galicia (Conselleria de Educacion) and FEDER. J.B. also acknowledges the support from the Fulbright Commission and the hospitality of Brookhaven National Laboratory.
 N.M. acknowledges funding by the Deutsche Forschungsgemeinschaft (DFG, German Research Foundation) - Project 404640738 and by the U.S. Department of Energy, Office of Science, Office of Nuclear Physics, under contract No. DE-SC0012704 whilst at Brookhaven National Lab where a significant part of the work presented was performed, and by the U.S. Department of Energy’s Office of Science, Office of Advanced Scientific Computing Research, Accelerated Research in Quantum Computing program award DE-SC0020312. A.T.’s work is supported by the U.S. Department of Energy, Office of Science, Office of Nuclear Physics under Award No. DE-SC0004286
  and in part by the joint BNL/Stony Brook Center for Frontiers in Nuclear Science (CFNS).  This material is based upon R.V.'s work supported by the U.S. Department of Energy, Office of Science, National Quantum Information Science Research Centers under the ``Co-design Center for Quantum Advantage" award. R.V.'s work is also supported by the U.S. Department of Energy, Office of Science, Office of Nuclear Physics, under contract No. DE-SC0012704.

%
%
%
\bibliographystyle{apsrev4-1} 
\bibliography{references}

\begin{thebibliography}{188}%
\makeatletter
\providecommand \@ifxundefined [1]{%
 \@ifx{#1\undefined}
}%
\providecommand \@ifnum [1]{%
 \ifnum #1\expandafter \@firstoftwo
 \else \expandafter \@secondoftwo
 \fi
}%
\providecommand \@ifx [1]{%
 \ifx #1\expandafter \@firstoftwo
 \else \expandafter \@secondoftwo
 \fi
}%
\providecommand \natexlab [1]{#1}%
\providecommand \enquote  [1]{``#1''}%
\providecommand \bibnamefont  [1]{#1}%
\providecommand \bibfnamefont [1]{#1}%
\providecommand \citenamefont [1]{#1}%
\providecommand \href@noop [0]{\@secondoftwo}%
\providecommand \href [0]{\begingroup \@sanitize@url \@href}%
\providecommand \@href[1]{\@@startlink{#1}\@@href}%
\providecommand \@@href[1]{\endgroup#1\@@endlink}%
\providecommand \@sanitize@url [0]{\catcode `\\12\catcode `\$12\catcode
  `\&12\catcode `\#12\catcode `\^12\catcode `\_12\catcode `\%12\relax}%
\providecommand \@@startlink[1]{}%
\providecommand \@@endlink[0]{}%
\providecommand \url  [0]{\begingroup\@sanitize@url \@url }%
\providecommand \@url [1]{\endgroup\@href {#1}{\urlprefix }}%
\providecommand \urlprefix  [0]{URL }%
\providecommand \Eprint [0]{\href }%
\providecommand \doibase [0]{http://dx.doi.org/}%
\providecommand \selectlanguage [0]{\@gobble}%
\providecommand \bibinfo  [0]{\@secondoftwo}%
\providecommand \bibfield  [0]{\@secondoftwo}%
\providecommand \translation [1]{[#1]}%
\providecommand \BibitemOpen [0]{}%
\providecommand \bibitemStop [0]{}%
\providecommand \bibitemNoStop [0]{.\EOS\space}%
\providecommand \EOS [0]{\spacefactor3000\relax}%
\providecommand \BibitemShut  [1]{\csname bibitem#1\endcsname}%
\let\auto@bib@innerbib\@empty
\bibitem [{\citenamefont {Whitfield}\ \emph {et~al.}(2011)\citenamefont
  {Whitfield}, \citenamefont {Biamonte},\ and\ \citenamefont
  {Aspuru-Guzik}}]{whitfield2011simulation}%
  \BibitemOpen
  \bibfield  {author} {\bibinfo {author} {\bibfnamefont {J.~D.}\ \bibnamefont
  {Whitfield}}, \bibinfo {author} {\bibfnamefont {J.}~\bibnamefont {Biamonte}},
  \ and\ \bibinfo {author} {\bibfnamefont {A.}~\bibnamefont {Aspuru-Guzik}},\
  }\href@noop {} {\bibfield  {journal} {\bibinfo  {journal} {Molecular
  Physics}\ }\textbf {\bibinfo {volume} {109}},\ \bibinfo {pages} {735}
  (\bibinfo {year} {2011})}\BibitemShut {NoStop}%
\bibitem [{\citenamefont {Kassal}\ \emph {et~al.}(2011)\citenamefont {Kassal},
  \citenamefont {Whitfield}, \citenamefont {Perdomo-Ortiz}, \citenamefont
  {Yung},\ and\ \citenamefont {Aspuru-Guzik}}]{kassal2011simulating}%
  \BibitemOpen
  \bibfield  {author} {\bibinfo {author} {\bibfnamefont {I.}~\bibnamefont
  {Kassal}}, \bibinfo {author} {\bibfnamefont {J.~D.}\ \bibnamefont
  {Whitfield}}, \bibinfo {author} {\bibfnamefont {A.}~\bibnamefont
  {Perdomo-Ortiz}}, \bibinfo {author} {\bibfnamefont {M.-H.}\ \bibnamefont
  {Yung}}, \ and\ \bibinfo {author} {\bibfnamefont {A.}~\bibnamefont
  {Aspuru-Guzik}},\ }\href@noop {} {\bibfield  {journal} {\bibinfo  {journal}
  {Annual review of physical chemistry}\ }\textbf {\bibinfo {volume} {62}},\
  \bibinfo {pages} {185} (\bibinfo {year} {2011})}\BibitemShut {NoStop}%
\bibitem [{\citenamefont {O’Malley}\ \emph {et~al.}(2016)\citenamefont
  {O’Malley}, \citenamefont {Babbush}, \citenamefont {Kivlichan},
  \citenamefont {Romero}, \citenamefont {McClean}, \citenamefont {Barends},
  \citenamefont {Kelly}, \citenamefont {Roushan}, \citenamefont {Tranter},
  \citenamefont {Ding} \emph {et~al.}}]{o2016scalable}%
  \BibitemOpen
  \bibfield  {author} {\bibinfo {author} {\bibfnamefont {P.~J.}\ \bibnamefont
  {O’Malley}}, \bibinfo {author} {\bibfnamefont {R.}~\bibnamefont {Babbush}},
  \bibinfo {author} {\bibfnamefont {I.~D.}\ \bibnamefont {Kivlichan}}, \bibinfo
  {author} {\bibfnamefont {J.}~\bibnamefont {Romero}}, \bibinfo {author}
  {\bibfnamefont {J.~R.}\ \bibnamefont {McClean}}, \bibinfo {author}
  {\bibfnamefont {R.}~\bibnamefont {Barends}}, \bibinfo {author} {\bibfnamefont
  {J.}~\bibnamefont {Kelly}}, \bibinfo {author} {\bibfnamefont
  {P.}~\bibnamefont {Roushan}}, \bibinfo {author} {\bibfnamefont
  {A.}~\bibnamefont {Tranter}}, \bibinfo {author} {\bibfnamefont
  {N.}~\bibnamefont {Ding}},  \emph {et~al.},\ }\href@noop {} {\bibfield
  {journal} {\bibinfo  {journal} {Physical Review X}\ }\textbf {\bibinfo
  {volume} {6}},\ \bibinfo {pages} {031007} (\bibinfo {year}
  {2016})}\BibitemShut {NoStop}%
\bibitem [{\citenamefont {Hempel}\ \emph {et~al.}(2018)\citenamefont {Hempel},
  \citenamefont {Maier}, \citenamefont {Romero}, \citenamefont {McClean},
  \citenamefont {Monz}, \citenamefont {Shen}, \citenamefont {Jurcevic},
  \citenamefont {Lanyon}, \citenamefont {Love}, \citenamefont {Babbush} \emph
  {et~al.}}]{hempel2018quantum}%
  \BibitemOpen
  \bibfield  {author} {\bibinfo {author} {\bibfnamefont {C.}~\bibnamefont
  {Hempel}}, \bibinfo {author} {\bibfnamefont {C.}~\bibnamefont {Maier}},
  \bibinfo {author} {\bibfnamefont {J.}~\bibnamefont {Romero}}, \bibinfo
  {author} {\bibfnamefont {J.}~\bibnamefont {McClean}}, \bibinfo {author}
  {\bibfnamefont {T.}~\bibnamefont {Monz}}, \bibinfo {author} {\bibfnamefont
  {H.}~\bibnamefont {Shen}}, \bibinfo {author} {\bibfnamefont {P.}~\bibnamefont
  {Jurcevic}}, \bibinfo {author} {\bibfnamefont {B.~P.}\ \bibnamefont
  {Lanyon}}, \bibinfo {author} {\bibfnamefont {P.}~\bibnamefont {Love}},
  \bibinfo {author} {\bibfnamefont {R.}~\bibnamefont {Babbush}},  \emph
  {et~al.},\ }\href@noop {} {\bibfield  {journal} {\bibinfo  {journal}
  {Physical Review X}\ }\textbf {\bibinfo {volume} {8}},\ \bibinfo {pages}
  {031022} (\bibinfo {year} {2018})}\BibitemShut {NoStop}%
\bibitem [{\citenamefont {Lewenstein}\ \emph {et~al.}(2007)\citenamefont
  {Lewenstein}, \citenamefont {Sanpera}, \citenamefont {Ahufinger},
  \citenamefont {Damski}, \citenamefont {Sen},\ and\ \citenamefont
  {Sen}}]{lewenstein2007ultracold}%
  \BibitemOpen
  \bibfield  {author} {\bibinfo {author} {\bibfnamefont {M.}~\bibnamefont
  {Lewenstein}}, \bibinfo {author} {\bibfnamefont {A.}~\bibnamefont {Sanpera}},
  \bibinfo {author} {\bibfnamefont {V.}~\bibnamefont {Ahufinger}}, \bibinfo
  {author} {\bibfnamefont {B.}~\bibnamefont {Damski}}, \bibinfo {author}
  {\bibfnamefont {A.}~\bibnamefont {Sen}}, \ and\ \bibinfo {author}
  {\bibfnamefont {U.}~\bibnamefont {Sen}},\ }\href@noop {} {\bibfield
  {journal} {\bibinfo  {journal} {Advances in Physics}\ }\textbf {\bibinfo
  {volume} {56}},\ \bibinfo {pages} {243} (\bibinfo {year} {2007})}\BibitemShut
  {NoStop}%
\bibitem [{\citenamefont {Bloch}\ \emph {et~al.}(2012)\citenamefont {Bloch},
  \citenamefont {Dalibard},\ and\ \citenamefont
  {Nascimbene}}]{bloch2012quantum}%
  \BibitemOpen
  \bibfield  {author} {\bibinfo {author} {\bibfnamefont {I.}~\bibnamefont
  {Bloch}}, \bibinfo {author} {\bibfnamefont {J.}~\bibnamefont {Dalibard}}, \
  and\ \bibinfo {author} {\bibfnamefont {S.}~\bibnamefont {Nascimbene}},\
  }\href@noop {} {\bibfield  {journal} {\bibinfo  {journal} {Nature Physics}\
  }\textbf {\bibinfo {volume} {8}},\ \bibinfo {pages} {267} (\bibinfo {year}
  {2012})}\BibitemShut {NoStop}%
\bibitem [{\citenamefont {Bao}\ \emph {et~al.}(2015)\citenamefont {Bao},
  \citenamefont {Hayden}, \citenamefont {Salton},\ and\ \citenamefont
  {Thomas}}]{bao2015universal}%
  \BibitemOpen
  \bibfield  {author} {\bibinfo {author} {\bibfnamefont {N.}~\bibnamefont
  {Bao}}, \bibinfo {author} {\bibfnamefont {P.}~\bibnamefont {Hayden}},
  \bibinfo {author} {\bibfnamefont {G.}~\bibnamefont {Salton}}, \ and\ \bibinfo
  {author} {\bibfnamefont {N.}~\bibnamefont {Thomas}},\ }\href@noop {}
  {\bibfield  {journal} {\bibinfo  {journal} {New Journal of Physics}\ }\textbf
  {\bibinfo {volume} {17}},\ \bibinfo {pages} {093028} (\bibinfo {year}
  {2015})}\BibitemShut {NoStop}%
\bibitem [{\citenamefont {Fischer}\ and\ \citenamefont
  {Sch{\"u}tzhold}(2004)}]{fischer2004quantum}%
  \BibitemOpen
  \bibfield  {author} {\bibinfo {author} {\bibfnamefont {U.~R.}\ \bibnamefont
  {Fischer}}\ and\ \bibinfo {author} {\bibfnamefont {R.}~\bibnamefont
  {Sch{\"u}tzhold}},\ }\href@noop {} {\bibfield  {journal} {\bibinfo  {journal}
  {Physical Review A}\ }\textbf {\bibinfo {volume} {70}},\ \bibinfo {pages}
  {063615} (\bibinfo {year} {2004})}\BibitemShut {NoStop}%
\bibitem [{\citenamefont {Jain}\ \emph {et~al.}(2007)\citenamefont {Jain},
  \citenamefont {Weinfurtner}, \citenamefont {Visser},\ and\ \citenamefont
  {Gardiner}}]{jain2007analog}%
  \BibitemOpen
  \bibfield  {author} {\bibinfo {author} {\bibfnamefont {P.}~\bibnamefont
  {Jain}}, \bibinfo {author} {\bibfnamefont {S.}~\bibnamefont {Weinfurtner}},
  \bibinfo {author} {\bibfnamefont {M.}~\bibnamefont {Visser}}, \ and\ \bibinfo
  {author} {\bibfnamefont {C.}~\bibnamefont {Gardiner}},\ }\href@noop {}
  {\bibfield  {journal} {\bibinfo  {journal} {Physical Review A}\ }\textbf
  {\bibinfo {volume} {76}},\ \bibinfo {pages} {033616} (\bibinfo {year}
  {2007})}\BibitemShut {NoStop}%
\bibitem [{\citenamefont {Chatrchyan}\ \emph {et~al.}(2020)\citenamefont
  {Chatrchyan}, \citenamefont {Geier}, \citenamefont {Oberthaler},
  \citenamefont {Berges},\ and\ \citenamefont {Hauke}}]{chatrchyan2020analog}%
  \BibitemOpen
  \bibfield  {author} {\bibinfo {author} {\bibfnamefont {A.}~\bibnamefont
  {Chatrchyan}}, \bibinfo {author} {\bibfnamefont {K.}~\bibnamefont {Geier}},
  \bibinfo {author} {\bibfnamefont {M.~K.}\ \bibnamefont {Oberthaler}},
  \bibinfo {author} {\bibfnamefont {J.}~\bibnamefont {Berges}}, \ and\ \bibinfo
  {author} {\bibfnamefont {P.}~\bibnamefont {Hauke}},\ }\href@noop {}
  {\bibfield  {journal} {\bibinfo  {journal} {arXiv preprint arXiv:2008.02290}\
  } (\bibinfo {year} {2020})}\BibitemShut {NoStop}%
\bibitem [{\citenamefont {Carlson}\ \emph {et~al.}(2018)\citenamefont
  {Carlson}, \citenamefont {Dean}, \citenamefont {Hjorth-Jensen}, \citenamefont
  {Kaplan}, \citenamefont {Preskill}, \citenamefont {Roche}, \citenamefont
  {Savage},\ and\ \citenamefont {Troyer}}]{carlson2018quantum}%
  \BibitemOpen
  \bibfield  {author} {\bibinfo {author} {\bibfnamefont {J.}~\bibnamefont
  {Carlson}}, \bibinfo {author} {\bibfnamefont {D.~J.}\ \bibnamefont {Dean}},
  \bibinfo {author} {\bibfnamefont {M.}~\bibnamefont {Hjorth-Jensen}}, \bibinfo
  {author} {\bibfnamefont {D.}~\bibnamefont {Kaplan}}, \bibinfo {author}
  {\bibfnamefont {J.}~\bibnamefont {Preskill}}, \bibinfo {author}
  {\bibfnamefont {K.}~\bibnamefont {Roche}}, \bibinfo {author} {\bibfnamefont
  {M.~J.}\ \bibnamefont {Savage}}, \ and\ \bibinfo {author} {\bibfnamefont
  {M.}~\bibnamefont {Troyer}},\ }\href@noop {} {\emph {\bibinfo {title}
  {Quantum Computing for Theoretical Nuclear Physics, A White Paper prepared
  for the US Department of Energy, Office of Science, Office of Nuclear
  Physics}}},\ \bibinfo {type} {Tech. Rep.}\ (\bibinfo  {institution} {USDOE
  Office of Science (SC)(United States)},\ \bibinfo {year} {2018})\BibitemShut
  {NoStop}%
\bibitem [{\citenamefont {McCaskey}\ \emph {et~al.}(2019)\citenamefont
  {McCaskey}, \citenamefont {Parks}, \citenamefont {Jakowski}, \citenamefont
  {Moore}, \citenamefont {Morris}, \citenamefont {Humble},\ and\ \citenamefont
  {Pooser}}]{mccaskey2019quantum}%
  \BibitemOpen
  \bibfield  {author} {\bibinfo {author} {\bibfnamefont {A.~J.}\ \bibnamefont
  {McCaskey}}, \bibinfo {author} {\bibfnamefont {Z.~P.}\ \bibnamefont {Parks}},
  \bibinfo {author} {\bibfnamefont {J.}~\bibnamefont {Jakowski}}, \bibinfo
  {author} {\bibfnamefont {S.~V.}\ \bibnamefont {Moore}}, \bibinfo {author}
  {\bibfnamefont {T.~D.}\ \bibnamefont {Morris}}, \bibinfo {author}
  {\bibfnamefont {T.~S.}\ \bibnamefont {Humble}}, \ and\ \bibinfo {author}
  {\bibfnamefont {R.~C.}\ \bibnamefont {Pooser}},\ }\href@noop {} {\bibfield
  {journal} {\bibinfo  {journal} {npj Quantum Information}\ }\textbf {\bibinfo
  {volume} {5}},\ \bibinfo {pages} {1} (\bibinfo {year} {2019})}\BibitemShut
  {NoStop}%
\bibitem [{\citenamefont {Dumitrescu}\ \emph {et~al.}(2018)\citenamefont
  {Dumitrescu}, \citenamefont {McCaskey}, \citenamefont {Hagen}, \citenamefont
  {Jansen}, \citenamefont {Morris}, \citenamefont {Papenbrock}, \citenamefont
  {Pooser}, \citenamefont {Dean},\ and\ \citenamefont
  {Lougovski}}]{dumitrescu2018cloud}%
  \BibitemOpen
  \bibfield  {author} {\bibinfo {author} {\bibfnamefont {E.~F.}\ \bibnamefont
  {Dumitrescu}}, \bibinfo {author} {\bibfnamefont {A.~J.}\ \bibnamefont
  {McCaskey}}, \bibinfo {author} {\bibfnamefont {G.}~\bibnamefont {Hagen}},
  \bibinfo {author} {\bibfnamefont {G.~R.}\ \bibnamefont {Jansen}}, \bibinfo
  {author} {\bibfnamefont {T.~D.}\ \bibnamefont {Morris}}, \bibinfo {author}
  {\bibfnamefont {T.}~\bibnamefont {Papenbrock}}, \bibinfo {author}
  {\bibfnamefont {R.~C.}\ \bibnamefont {Pooser}}, \bibinfo {author}
  {\bibfnamefont {D.~J.}\ \bibnamefont {Dean}}, \ and\ \bibinfo {author}
  {\bibfnamefont {P.}~\bibnamefont {Lougovski}},\ }\href@noop {} {\bibfield
  {journal} {\bibinfo  {journal} {Physical review letters}\ }\textbf {\bibinfo
  {volume} {120}},\ \bibinfo {pages} {210501} (\bibinfo {year}
  {2018})}\BibitemShut {NoStop}%
\bibitem [{\citenamefont {Rico}\ \emph {et~al.}(2018)\citenamefont {Rico},
  \citenamefont {Dalmonte}, \citenamefont {Zoller}, \citenamefont {Banerjee},
  \citenamefont {B{\"o}gli}, \citenamefont {Stebler},\ and\ \citenamefont
  {Wiese}}]{rico2018so}%
  \BibitemOpen
  \bibfield  {author} {\bibinfo {author} {\bibfnamefont {E.}~\bibnamefont
  {Rico}}, \bibinfo {author} {\bibfnamefont {M.}~\bibnamefont {Dalmonte}},
  \bibinfo {author} {\bibfnamefont {P.}~\bibnamefont {Zoller}}, \bibinfo
  {author} {\bibfnamefont {D.}~\bibnamefont {Banerjee}}, \bibinfo {author}
  {\bibfnamefont {M.}~\bibnamefont {B{\"o}gli}}, \bibinfo {author}
  {\bibfnamefont {P.}~\bibnamefont {Stebler}}, \ and\ \bibinfo {author}
  {\bibfnamefont {U.-J.}\ \bibnamefont {Wiese}},\ }\href@noop {} {\bibfield
  {journal} {\bibinfo  {journal} {Annals of physics}\ }\textbf {\bibinfo
  {volume} {393}},\ \bibinfo {pages} {466} (\bibinfo {year}
  {2018})}\BibitemShut {NoStop}%
\bibitem [{\citenamefont {Clo{\"e}t}\ \emph {et~al.}(2019)\citenamefont
  {Clo{\"e}t}, \citenamefont {Dietrich}, \citenamefont {Arrington},
  \citenamefont {Bazavov}, \citenamefont {Bishof}, \citenamefont {Freese},
  \citenamefont {Gorshkov}, \citenamefont {Grassellino}, \citenamefont
  {Hafidi}, \citenamefont {Jacob} \emph {et~al.}}]{cloet2019opportunities}%
  \BibitemOpen
  \bibfield  {author} {\bibinfo {author} {\bibfnamefont {I.~C.}\ \bibnamefont
  {Clo{\"e}t}}, \bibinfo {author} {\bibfnamefont {M.~R.}\ \bibnamefont
  {Dietrich}}, \bibinfo {author} {\bibfnamefont {J.}~\bibnamefont {Arrington}},
  \bibinfo {author} {\bibfnamefont {A.}~\bibnamefont {Bazavov}}, \bibinfo
  {author} {\bibfnamefont {M.}~\bibnamefont {Bishof}}, \bibinfo {author}
  {\bibfnamefont {A.}~\bibnamefont {Freese}}, \bibinfo {author} {\bibfnamefont
  {A.~V.}\ \bibnamefont {Gorshkov}}, \bibinfo {author} {\bibfnamefont
  {A.}~\bibnamefont {Grassellino}}, \bibinfo {author} {\bibfnamefont
  {K.}~\bibnamefont {Hafidi}}, \bibinfo {author} {\bibfnamefont
  {Z.}~\bibnamefont {Jacob}},  \emph {et~al.},\ }\href@noop {} {\bibfield
  {journal} {\bibinfo  {journal} {arXiv preprint arXiv:1903.05453}\ } (\bibinfo
  {year} {2019})}\BibitemShut {NoStop}%
\bibitem [{\citenamefont {Matchev}\ \emph {et~al.}(2020)\citenamefont
  {Matchev}, \citenamefont {Mrenna}, \citenamefont {Shyamsundar},\ and\
  \citenamefont {Smolinsky}}]{matchev2020quantum}%
  \BibitemOpen
  \bibfield  {author} {\bibinfo {author} {\bibfnamefont {K.}~\bibnamefont
  {Matchev}}, \bibinfo {author} {\bibfnamefont {S.}~\bibnamefont {Mrenna}},
  \bibinfo {author} {\bibfnamefont {P.}~\bibnamefont {Shyamsundar}}, \ and\
  \bibinfo {author} {\bibfnamefont {J.}~\bibnamefont {Smolinsky}},\ }\href@noop
  {} {\bibfield  {journal} {\bibinfo  {journal} {Quantum}\ } (\bibinfo {year}
  {2020})}\BibitemShut {NoStop}%
\bibitem [{\citenamefont {Kielpinski}\ \emph {et~al.}(2002)\citenamefont
  {Kielpinski}, \citenamefont {Monroe},\ and\ \citenamefont
  {Wineland}}]{kielpinski2002architecture}%
  \BibitemOpen
  \bibfield  {author} {\bibinfo {author} {\bibfnamefont {D.}~\bibnamefont
  {Kielpinski}}, \bibinfo {author} {\bibfnamefont {C.}~\bibnamefont {Monroe}},
  \ and\ \bibinfo {author} {\bibfnamefont {D.~J.}\ \bibnamefont {Wineland}},\
  }\href@noop {} {\bibfield  {journal} {\bibinfo  {journal} {Nature}\ }\textbf
  {\bibinfo {volume} {417}},\ \bibinfo {pages} {709} (\bibinfo {year}
  {2002})}\BibitemShut {NoStop}%
\bibitem [{\citenamefont {Monroe}(2002)}]{monroe2002quantum}%
  \BibitemOpen
  \bibfield  {author} {\bibinfo {author} {\bibfnamefont {C.}~\bibnamefont
  {Monroe}},\ }\href@noop {} {\bibfield  {journal} {\bibinfo  {journal}
  {Nature}\ }\textbf {\bibinfo {volume} {416}},\ \bibinfo {pages} {238}
  (\bibinfo {year} {2002})}\BibitemShut {NoStop}%
\bibitem [{\citenamefont {Blais}\ \emph {et~al.}(2004)\citenamefont {Blais},
  \citenamefont {Huang}, \citenamefont {Wallraff}, \citenamefont {Girvin},\
  and\ \citenamefont {Schoelkopf}}]{blais2004cavity}%
  \BibitemOpen
  \bibfield  {author} {\bibinfo {author} {\bibfnamefont {A.}~\bibnamefont
  {Blais}}, \bibinfo {author} {\bibfnamefont {R.-S.}\ \bibnamefont {Huang}},
  \bibinfo {author} {\bibfnamefont {A.}~\bibnamefont {Wallraff}}, \bibinfo
  {author} {\bibfnamefont {S.~M.}\ \bibnamefont {Girvin}}, \ and\ \bibinfo
  {author} {\bibfnamefont {R.~J.}\ \bibnamefont {Schoelkopf}},\ }\href@noop {}
  {\bibfield  {journal} {\bibinfo  {journal} {Physical Review A}\ }\textbf
  {\bibinfo {volume} {69}},\ \bibinfo {pages} {062320} (\bibinfo {year}
  {2004})}\BibitemShut {NoStop}%
\bibitem [{\citenamefont {Cirac}\ and\ \citenamefont
  {Zoller}(2012)}]{cirac2012goals}%
  \BibitemOpen
  \bibfield  {author} {\bibinfo {author} {\bibfnamefont {J.~I.}\ \bibnamefont
  {Cirac}}\ and\ \bibinfo {author} {\bibfnamefont {P.}~\bibnamefont {Zoller}},\
  }\href@noop {} {\bibfield  {journal} {\bibinfo  {journal} {Nature Physics}\
  }\textbf {\bibinfo {volume} {8}},\ \bibinfo {pages} {264} (\bibinfo {year}
  {2012})}\BibitemShut {NoStop}%
\bibitem [{\citenamefont {Hauke}\ \emph {et~al.}(2012)\citenamefont {Hauke},
  \citenamefont {Cucchietti}, \citenamefont {Tagliacozzo}, \citenamefont
  {Deutsch},\ and\ \citenamefont {Lewenstein}}]{hauke2012can}%
  \BibitemOpen
  \bibfield  {author} {\bibinfo {author} {\bibfnamefont {P.}~\bibnamefont
  {Hauke}}, \bibinfo {author} {\bibfnamefont {F.~M.}\ \bibnamefont
  {Cucchietti}}, \bibinfo {author} {\bibfnamefont {L.}~\bibnamefont
  {Tagliacozzo}}, \bibinfo {author} {\bibfnamefont {I.}~\bibnamefont
  {Deutsch}}, \ and\ \bibinfo {author} {\bibfnamefont {M.}~\bibnamefont
  {Lewenstein}},\ }\href@noop {} {\bibfield  {journal} {\bibinfo  {journal}
  {Reports on Progress in Physics}\ }\textbf {\bibinfo {volume} {75}},\
  \bibinfo {pages} {082401} (\bibinfo {year} {2012})}\BibitemShut {NoStop}%
\bibitem [{\citenamefont {Preskill}(2018)}]{preskill2018quantum}%
  \BibitemOpen
  \bibfield  {author} {\bibinfo {author} {\bibfnamefont {J.}~\bibnamefont
  {Preskill}},\ }\href@noop {} {\bibfield  {journal} {\bibinfo  {journal}
  {Quantum}\ }\textbf {\bibinfo {volume} {2}},\ \bibinfo {pages} {79} (\bibinfo
  {year} {2018})}\BibitemShut {NoStop}%
\bibitem [{\citenamefont {Roggero}\ and\ \citenamefont
  {Carlson}(2018)}]{roggero2018linear}%
  \BibitemOpen
  \bibfield  {author} {\bibinfo {author} {\bibfnamefont {A.}~\bibnamefont
  {Roggero}}\ and\ \bibinfo {author} {\bibfnamefont {J.}~\bibnamefont
  {Carlson}},\ }\href@noop {} {\bibfield  {journal} {\bibinfo  {journal} {arXiv
  preprint arXiv:1804.01505}\ } (\bibinfo {year} {2018})}\BibitemShut {NoStop}%
\bibitem [{\citenamefont {Roggero}\ \emph {et~al.}(2020)\citenamefont
  {Roggero}, \citenamefont {Li}, \citenamefont {Carlson}, \citenamefont
  {Gupta},\ and\ \citenamefont {Perdue}}]{roggero2020quantum}%
  \BibitemOpen
  \bibfield  {author} {\bibinfo {author} {\bibfnamefont {A.}~\bibnamefont
  {Roggero}}, \bibinfo {author} {\bibfnamefont {A.~C.}\ \bibnamefont {Li}},
  \bibinfo {author} {\bibfnamefont {J.}~\bibnamefont {Carlson}}, \bibinfo
  {author} {\bibfnamefont {R.}~\bibnamefont {Gupta}}, \ and\ \bibinfo {author}
  {\bibfnamefont {G.~N.}\ \bibnamefont {Perdue}},\ }\href@noop {} {\bibfield
  {journal} {\bibinfo  {journal} {Physical Review D}\ }\textbf {\bibinfo
  {volume} {101}},\ \bibinfo {pages} {074038} (\bibinfo {year}
  {2020})}\BibitemShut {NoStop}%
\bibitem [{\citenamefont {Berges}\ \emph {et~al.}(2020)\citenamefont {Berges},
  \citenamefont {Heller}, \citenamefont {Mazeliauskas},\ and\ \citenamefont
  {Venugopalan}}]{Berges:2020fwq}%
  \BibitemOpen
  \bibfield  {author} {\bibinfo {author} {\bibfnamefont {J.}~\bibnamefont
  {Berges}}, \bibinfo {author} {\bibfnamefont {M.~P.}\ \bibnamefont {Heller}},
  \bibinfo {author} {\bibfnamefont {A.}~\bibnamefont {Mazeliauskas}}, \ and\
  \bibinfo {author} {\bibfnamefont {R.}~\bibnamefont {Venugopalan}},\
  }\href@noop {} {\  (\bibinfo {year} {2020})},\ \Eprint
  {http://arxiv.org/abs/2005.12299} {arXiv:2005.12299 [hep-th]} \BibitemShut
  {NoStop}%
\bibitem [{\citenamefont {Breidenbach}\ \emph {et~al.}(1969)\citenamefont
  {Breidenbach}, \citenamefont {Friedman}, \citenamefont {Kendall},
  \citenamefont {Bloom}, \citenamefont {Coward}, \citenamefont {DeStaebler},
  \citenamefont {Drees}, \citenamefont {Mo},\ and\ \citenamefont
  {Taylor}}]{Breidenbach:1969kd}%
  \BibitemOpen
  \bibfield  {author} {\bibinfo {author} {\bibfnamefont {M.}~\bibnamefont
  {Breidenbach}}, \bibinfo {author} {\bibfnamefont {J.~I.}\ \bibnamefont
  {Friedman}}, \bibinfo {author} {\bibfnamefont {H.~W.}\ \bibnamefont
  {Kendall}}, \bibinfo {author} {\bibfnamefont {E.~D.}\ \bibnamefont {Bloom}},
  \bibinfo {author} {\bibfnamefont {D.~H.}\ \bibnamefont {Coward}}, \bibinfo
  {author} {\bibfnamefont {H.~C.}\ \bibnamefont {DeStaebler}}, \bibinfo
  {author} {\bibfnamefont {J.}~\bibnamefont {Drees}}, \bibinfo {author}
  {\bibfnamefont {L.~W.}\ \bibnamefont {Mo}}, \ and\ \bibinfo {author}
  {\bibfnamefont {R.~E.}\ \bibnamefont {Taylor}},\ }\href {\doibase
  10.1103/PhysRevLett.23.935} {\bibfield  {journal} {\bibinfo  {journal} {Phys.
  Rev. Lett.}\ }\textbf {\bibinfo {volume} {23}},\ \bibinfo {pages} {935}
  (\bibinfo {year} {1969})}\BibitemShut {NoStop}%
\bibitem [{\citenamefont {Bjorken}(1969)}]{Bjorken:1968dy}%
  \BibitemOpen
  \bibfield  {author} {\bibinfo {author} {\bibfnamefont {J.~D.}\ \bibnamefont
  {Bjorken}},\ }\href {\doibase 10.1103/PhysRev.179.1547} {\bibfield  {journal}
  {\bibinfo  {journal} {Phys. Rev.}\ }\textbf {\bibinfo {volume} {179}},\
  \bibinfo {pages} {1547} (\bibinfo {year} {1969})}\BibitemShut {NoStop}%
\bibitem [{\citenamefont {Bjorken}\ and\ \citenamefont
  {Paschos}(1969)}]{Bjorken:1969ja}%
  \BibitemOpen
  \bibfield  {author} {\bibinfo {author} {\bibfnamefont {J.~D.}\ \bibnamefont
  {Bjorken}}\ and\ \bibinfo {author} {\bibfnamefont {E.~A.}\ \bibnamefont
  {Paschos}},\ }\href {\doibase 10.1103/PhysRev.185.1975} {\bibfield  {journal}
  {\bibinfo  {journal} {Phys. Rev.}\ }\textbf {\bibinfo {volume} {185}},\
  \bibinfo {pages} {1975} (\bibinfo {year} {1969})}\BibitemShut {NoStop}%
\bibitem [{\citenamefont {Gross}\ and\ \citenamefont
  {Wilczek}(1973)}]{Gross:1973id}%
  \BibitemOpen
  \bibfield  {author} {\bibinfo {author} {\bibfnamefont {D.~J.}\ \bibnamefont
  {Gross}}\ and\ \bibinfo {author} {\bibfnamefont {F.}~\bibnamefont
  {Wilczek}},\ }\href {\doibase 10.1103/PhysRevLett.30.1343} {\bibfield
  {journal} {\bibinfo  {journal} {Phys. Rev. Lett.}\ }\textbf {\bibinfo
  {volume} {30}},\ \bibinfo {pages} {1343} (\bibinfo {year} {1973})},\ \bibinfo
  {note} {[,271(1973)]}\BibitemShut {NoStop}%
\bibitem [{\citenamefont {Blumlein}(2013)}]{Blumlein:2012bf}%
  \BibitemOpen
  \bibfield  {author} {\bibinfo {author} {\bibfnamefont {J.}~\bibnamefont
  {Blumlein}},\ }\href {\doibase 10.1016/j.ppnp.2012.09.006} {\bibfield
  {journal} {\bibinfo  {journal} {Prog. Part. Nucl. Phys.}\ }\textbf {\bibinfo
  {volume} {69}},\ \bibinfo {pages} {28} (\bibinfo {year} {2013})},\ \Eprint
  {http://arxiv.org/abs/1208.6087} {arXiv:1208.6087 [hep-ph]} \BibitemShut
  {NoStop}%
\bibitem [{\citenamefont {Mueller}\ \emph {et~al.}(2020)\citenamefont
  {Mueller}, \citenamefont {Tarasov},\ and\ \citenamefont
  {Venugopalan}}]{mueller2020deeply}%
  \BibitemOpen
  \bibfield  {author} {\bibinfo {author} {\bibfnamefont {N.}~\bibnamefont
  {Mueller}}, \bibinfo {author} {\bibfnamefont {A.}~\bibnamefont {Tarasov}}, \
  and\ \bibinfo {author} {\bibfnamefont {R.}~\bibnamefont {Venugopalan}},\
  }\href@noop {} {\bibfield  {journal} {\bibinfo  {journal} {Physical Review
  D}\ }\textbf {\bibinfo {volume} {102}},\ \bibinfo {pages} {016007} (\bibinfo
  {year} {2020})}\BibitemShut {NoStop}%
\bibitem [{\citenamefont {Lamm}\ \emph
  {et~al.}(2020{\natexlab{a}})\citenamefont {Lamm}, \citenamefont {Lawrence},
  \citenamefont {Yamauchi}, \citenamefont {Collaboration} \emph
  {et~al.}}]{lamm2020parton}%
  \BibitemOpen
  \bibfield  {author} {\bibinfo {author} {\bibfnamefont {H.}~\bibnamefont
  {Lamm}}, \bibinfo {author} {\bibfnamefont {S.}~\bibnamefont {Lawrence}},
  \bibinfo {author} {\bibfnamefont {Y.}~\bibnamefont {Yamauchi}}, \bibinfo
  {author} {\bibfnamefont {N.}~\bibnamefont {Collaboration}},  \emph {et~al.},\
  }\href@noop {} {\bibfield  {journal} {\bibinfo  {journal} {Physical Review
  Research}\ }\textbf {\bibinfo {volume} {2}},\ \bibinfo {pages} {013272}
  (\bibinfo {year} {2020}{\natexlab{a}})}\BibitemShut {NoStop}%
\bibitem [{\citenamefont {Kreshchuk}\ \emph
  {et~al.}(2020{\natexlab{a}})\citenamefont {Kreshchuk}, \citenamefont {Kirby},
  \citenamefont {Goldstein}, \citenamefont {Beauchemin},\ and\ \citenamefont
  {Love}}]{kreshchuk2020quantum}%
  \BibitemOpen
  \bibfield  {author} {\bibinfo {author} {\bibfnamefont {M.}~\bibnamefont
  {Kreshchuk}}, \bibinfo {author} {\bibfnamefont {W.~M.}\ \bibnamefont
  {Kirby}}, \bibinfo {author} {\bibfnamefont {G.}~\bibnamefont {Goldstein}},
  \bibinfo {author} {\bibfnamefont {H.}~\bibnamefont {Beauchemin}}, \ and\
  \bibinfo {author} {\bibfnamefont {P.~J.}\ \bibnamefont {Love}},\ }\href@noop
  {} {\bibfield  {journal} {\bibinfo  {journal} {arXiv preprint
  arXiv:2002.04016}\ } (\bibinfo {year} {2020}{\natexlab{a}})}\BibitemShut
  {NoStop}%
\bibitem [{\citenamefont {Bassetto}\ \emph {et~al.}(1983)\citenamefont
  {Bassetto}, \citenamefont {Ciafaloni},\ and\ \citenamefont
  {Marchesini}}]{Bassetto:1984ik}%
  \BibitemOpen
  \bibfield  {author} {\bibinfo {author} {\bibfnamefont {A.}~\bibnamefont
  {Bassetto}}, \bibinfo {author} {\bibfnamefont {M.}~\bibnamefont {Ciafaloni}},
  \ and\ \bibinfo {author} {\bibfnamefont {G.}~\bibnamefont {Marchesini}},\
  }\href {\doibase 10.1016/0370-1573(83)90083-2} {\bibfield  {journal}
  {\bibinfo  {journal} {Phys. Rept.}\ }\textbf {\bibinfo {volume} {100}},\
  \bibinfo {pages} {201} (\bibinfo {year} {1983})}\BibitemShut {NoStop}%
\bibitem [{\citenamefont {Dokshitzer}\ \emph {et~al.}(1988)\citenamefont
  {Dokshitzer}, \citenamefont {Khoze}, \citenamefont {Troian},\ and\
  \citenamefont {Mueller}}]{Dokshitzer:1987nm}%
  \BibitemOpen
  \bibfield  {author} {\bibinfo {author} {\bibfnamefont {Y.~L.}\ \bibnamefont
  {Dokshitzer}}, \bibinfo {author} {\bibfnamefont {V.~A.}\ \bibnamefont
  {Khoze}}, \bibinfo {author} {\bibfnamefont {S.}~\bibnamefont {Troian}}, \
  and\ \bibinfo {author} {\bibfnamefont {A.~H.}\ \bibnamefont {Mueller}},\
  }\href {\doibase 10.1103/RevModPhys.60.373} {\bibfield  {journal} {\bibinfo
  {journal} {Rev. Mod. Phys.}\ }\textbf {\bibinfo {volume} {60}},\ \bibinfo
  {pages} {373} (\bibinfo {year} {1988})}\BibitemShut {NoStop}%
\bibitem [{\citenamefont {Metz}\ and\ \citenamefont
  {Vossen}(2016)}]{Metz:2016swz}%
  \BibitemOpen
  \bibfield  {author} {\bibinfo {author} {\bibfnamefont {A.}~\bibnamefont
  {Metz}}\ and\ \bibinfo {author} {\bibfnamefont {A.}~\bibnamefont {Vossen}},\
  }\href {\doibase 10.1016/j.ppnp.2016.08.003} {\bibfield  {journal} {\bibinfo
  {journal} {Prog. Part. Nucl. Phys.}\ }\textbf {\bibinfo {volume} {91}},\
  \bibinfo {pages} {136} (\bibinfo {year} {2016})},\ \Eprint
  {http://arxiv.org/abs/1607.02521} {arXiv:1607.02521 [hep-ex]} \BibitemShut
  {NoStop}%
\bibitem [{\citenamefont {Winter}\ \emph {et~al.}(2017)\citenamefont {Winter},
  \citenamefont {Detmold}, \citenamefont {Gambhir}, \citenamefont {Orginos},
  \citenamefont {Savage}, \citenamefont {Shanahan},\ and\ \citenamefont
  {Wagman}}]{Winter:2017bfs}%
  \BibitemOpen
  \bibfield  {author} {\bibinfo {author} {\bibfnamefont {F.}~\bibnamefont
  {Winter}}, \bibinfo {author} {\bibfnamefont {W.}~\bibnamefont {Detmold}},
  \bibinfo {author} {\bibfnamefont {A.~S.}\ \bibnamefont {Gambhir}}, \bibinfo
  {author} {\bibfnamefont {K.}~\bibnamefont {Orginos}}, \bibinfo {author}
  {\bibfnamefont {M.~J.}\ \bibnamefont {Savage}}, \bibinfo {author}
  {\bibfnamefont {P.~E.}\ \bibnamefont {Shanahan}}, \ and\ \bibinfo {author}
  {\bibfnamefont {M.~L.}\ \bibnamefont {Wagman}},\ }\href {\doibase
  10.1103/PhysRevD.96.094512} {\bibfield  {journal} {\bibinfo  {journal} {Phys.
  Rev.}\ }\textbf {\bibinfo {volume} {D96}},\ \bibinfo {pages} {094512}
  (\bibinfo {year} {2017})},\ \Eprint {http://arxiv.org/abs/1709.00395}
  {arXiv:1709.00395 [hep-lat]} \BibitemShut {NoStop}%
\bibitem [{\citenamefont {Ji}(2013)}]{Ji:2013dva}%
  \BibitemOpen
  \bibfield  {author} {\bibinfo {author} {\bibfnamefont {X.}~\bibnamefont
  {Ji}},\ }\href {\doibase 10.1103/PhysRevLett.110.262002} {\bibfield
  {journal} {\bibinfo  {journal} {Phys. Rev. Lett.}\ }\textbf {\bibinfo
  {volume} {110}},\ \bibinfo {pages} {262002} (\bibinfo {year} {2013})},\
  \Eprint {http://arxiv.org/abs/1305.1539} {arXiv:1305.1539 [hep-ph]}
  \BibitemShut {NoStop}%
\bibitem [{\citenamefont {Alexandrou}\ \emph {et~al.}(2015)\citenamefont
  {Alexandrou}, \citenamefont {Cichy}, \citenamefont {Drach}, \citenamefont
  {Garcia-Ramos}, \citenamefont {Hadjiyiannakou}, \citenamefont {Jansen},
  \citenamefont {Steffens},\ and\ \citenamefont {Wiese}}]{Alexandrou:2015rja}%
  \BibitemOpen
  \bibfield  {author} {\bibinfo {author} {\bibfnamefont {C.}~\bibnamefont
  {Alexandrou}}, \bibinfo {author} {\bibfnamefont {K.}~\bibnamefont {Cichy}},
  \bibinfo {author} {\bibfnamefont {V.}~\bibnamefont {Drach}}, \bibinfo
  {author} {\bibfnamefont {E.}~\bibnamefont {Garcia-Ramos}}, \bibinfo {author}
  {\bibfnamefont {K.}~\bibnamefont {Hadjiyiannakou}}, \bibinfo {author}
  {\bibfnamefont {K.}~\bibnamefont {Jansen}}, \bibinfo {author} {\bibfnamefont
  {F.}~\bibnamefont {Steffens}}, \ and\ \bibinfo {author} {\bibfnamefont
  {C.}~\bibnamefont {Wiese}},\ }\href {\doibase 10.1103/PhysRevD.92.014502}
  {\bibfield  {journal} {\bibinfo  {journal} {Phys. Rev.}\ }\textbf {\bibinfo
  {volume} {D92}},\ \bibinfo {pages} {014502} (\bibinfo {year} {2015})},\
  \Eprint {http://arxiv.org/abs/1504.07455} {arXiv:1504.07455 [hep-lat]}
  \BibitemShut {NoStop}%
\bibitem [{\citenamefont {Chen}\ \emph {et~al.}(2016)\citenamefont {Chen},
  \citenamefont {Cohen}, \citenamefont {Ji}, \citenamefont {Lin},\ and\
  \citenamefont {Zhang}}]{Chen:2016utp}%
  \BibitemOpen
  \bibfield  {author} {\bibinfo {author} {\bibfnamefont {J.-W.}\ \bibnamefont
  {Chen}}, \bibinfo {author} {\bibfnamefont {S.~D.}\ \bibnamefont {Cohen}},
  \bibinfo {author} {\bibfnamefont {X.}~\bibnamefont {Ji}}, \bibinfo {author}
  {\bibfnamefont {H.-W.}\ \bibnamefont {Lin}}, \ and\ \bibinfo {author}
  {\bibfnamefont {J.-H.}\ \bibnamefont {Zhang}},\ }\href {\doibase
  10.1016/j.nuclphysb.2016.07.033} {\bibfield  {journal} {\bibinfo  {journal}
  {Nucl. Phys.}\ }\textbf {\bibinfo {volume} {B911}},\ \bibinfo {pages} {246}
  (\bibinfo {year} {2016})},\ \Eprint {http://arxiv.org/abs/1603.06664}
  {arXiv:1603.06664 [hep-ph]} \BibitemShut {NoStop}%
\bibitem [{\citenamefont {Radyushkin}(2017)}]{Radyushkin:2016hsy}%
  \BibitemOpen
  \bibfield  {author} {\bibinfo {author} {\bibfnamefont {A.}~\bibnamefont
  {Radyushkin}},\ }\href {\doibase 10.1016/j.physletb.2017.02.019} {\bibfield
  {journal} {\bibinfo  {journal} {Phys. Lett.}\ }\textbf {\bibinfo {volume}
  {B767}},\ \bibinfo {pages} {314} (\bibinfo {year} {2017})},\ \Eprint
  {http://arxiv.org/abs/1612.05170} {arXiv:1612.05170 [hep-ph]} \BibitemShut
  {NoStop}%
\bibitem [{\citenamefont {Lin}\ \emph {et~al.}(2018)\citenamefont {Lin} \emph
  {et~al.}}]{Lin:2017snn}%
  \BibitemOpen
  \bibfield  {author} {\bibinfo {author} {\bibfnamefont {H.-W.}\ \bibnamefont
  {Lin}} \emph {et~al.},\ }\href {\doibase 10.1016/j.ppnp.2018.01.007}
  {\bibfield  {journal} {\bibinfo  {journal} {Prog. Part. Nucl. Phys.}\
  }\textbf {\bibinfo {volume} {100}},\ \bibinfo {pages} {107} (\bibinfo {year}
  {2018})},\ \Eprint {http://arxiv.org/abs/1711.07916} {arXiv:1711.07916
  [hep-ph]} \BibitemShut {NoStop}%
\bibitem [{\citenamefont {Detmold}\ \emph {et~al.}(2019)\citenamefont
  {Detmold}, \citenamefont {Edwards}, \citenamefont {Dudek}, \citenamefont
  {Engelhardt}, \citenamefont {Lin}, \citenamefont {Meinel}, \citenamefont
  {Orginos},\ and\ \citenamefont {Shanahan}}]{Detmold:2019ghl}%
  \BibitemOpen
  \bibfield  {author} {\bibinfo {author} {\bibfnamefont {W.}~\bibnamefont
  {Detmold}}, \bibinfo {author} {\bibfnamefont {R.~G.}\ \bibnamefont
  {Edwards}}, \bibinfo {author} {\bibfnamefont {J.~J.}\ \bibnamefont {Dudek}},
  \bibinfo {author} {\bibfnamefont {M.}~\bibnamefont {Engelhardt}}, \bibinfo
  {author} {\bibfnamefont {H.-W.}\ \bibnamefont {Lin}}, \bibinfo {author}
  {\bibfnamefont {S.}~\bibnamefont {Meinel}}, \bibinfo {author} {\bibfnamefont
  {K.}~\bibnamefont {Orginos}}, \ and\ \bibinfo {author} {\bibfnamefont
  {P.}~\bibnamefont {Shanahan}} (\bibinfo {collaboration} {USQCD}),\
  }\href@noop {} {\  (\bibinfo {year} {2019})},\ \Eprint
  {http://arxiv.org/abs/1904.09512} {arXiv:1904.09512 [hep-lat]} \BibitemShut
  {NoStop}%
\bibitem [{\citenamefont {Jordan}\ \emph
  {et~al.}(2011{\natexlab{a}})\citenamefont {Jordan}, \citenamefont {Lee},\
  and\ \citenamefont {Preskill}}]{jordan2011quantum}%
  \BibitemOpen
  \bibfield  {author} {\bibinfo {author} {\bibfnamefont {S.~P.}\ \bibnamefont
  {Jordan}}, \bibinfo {author} {\bibfnamefont {K.~S.}\ \bibnamefont {Lee}}, \
  and\ \bibinfo {author} {\bibfnamefont {J.}~\bibnamefont {Preskill}},\
  }\href@noop {} {\bibfield  {journal} {\bibinfo  {journal} {arXiv preprint
  arXiv:1112.4833}\ } (\bibinfo {year} {2011}{\natexlab{a}})}\BibitemShut
  {NoStop}%
\bibitem [{\citenamefont {Jordan}\ \emph
  {et~al.}(2012{\natexlab{a}})\citenamefont {Jordan}, \citenamefont {Lee},\
  and\ \citenamefont {Preskill}}]{jordan2012quantum}%
  \BibitemOpen
  \bibfield  {author} {\bibinfo {author} {\bibfnamefont {S.~P.}\ \bibnamefont
  {Jordan}}, \bibinfo {author} {\bibfnamefont {K.~S.}\ \bibnamefont {Lee}}, \
  and\ \bibinfo {author} {\bibfnamefont {J.}~\bibnamefont {Preskill}},\
  }\href@noop {} {\bibfield  {journal} {\bibinfo  {journal} {Science}\ }\textbf
  {\bibinfo {volume} {336}},\ \bibinfo {pages} {1130} (\bibinfo {year}
  {2012}{\natexlab{a}})}\BibitemShut {NoStop}%
\bibitem [{\citenamefont {Klco}\ and\ \citenamefont
  {Savage}(2020)}]{klco2020fixed}%
  \BibitemOpen
  \bibfield  {author} {\bibinfo {author} {\bibfnamefont {N.}~\bibnamefont
  {Klco}}\ and\ \bibinfo {author} {\bibfnamefont {M.~J.}\ \bibnamefont
  {Savage}},\ }\href@noop {} {\bibfield  {journal} {\bibinfo  {journal} {arXiv
  preprint arXiv:2002.02018}\ } (\bibinfo {year} {2020})}\BibitemShut {NoStop}%
\bibitem [{\citenamefont {Roggero}\ and\ \citenamefont
  {Carlson}(2019)}]{roggero2019dynamic}%
  \BibitemOpen
  \bibfield  {author} {\bibinfo {author} {\bibfnamefont {A.}~\bibnamefont
  {Roggero}}\ and\ \bibinfo {author} {\bibfnamefont {J.}~\bibnamefont
  {Carlson}},\ }\href@noop {} {\bibfield  {journal} {\bibinfo  {journal}
  {Physical Review C}\ }\textbf {\bibinfo {volume} {100}},\ \bibinfo {pages}
  {034610} (\bibinfo {year} {2019})}\BibitemShut {NoStop}%
\bibitem [{\citenamefont {Strassler}(1992)}]{Strassler:1992zr}%
  \BibitemOpen
  \bibfield  {author} {\bibinfo {author} {\bibfnamefont {M.~J.}\ \bibnamefont
  {Strassler}},\ }\href {\doibase 10.1016/0550-3213(92)90098-V} {\bibfield
  {journal} {\bibinfo  {journal} {Nucl. Phys.}\ }\textbf {\bibinfo {volume}
  {B385}},\ \bibinfo {pages} {145} (\bibinfo {year} {1992})},\ \Eprint
  {http://arxiv.org/abs/hep-ph/9205205} {arXiv:hep-ph/9205205 [hep-ph]}
  \BibitemShut {NoStop}%
\bibitem [{\citenamefont {Bauer}\ \emph {et~al.}(2016)\citenamefont {Bauer},
  \citenamefont {Wecker}, \citenamefont {Millis}, \citenamefont {Hastings},\
  and\ \citenamefont {Troyer}}]{PhysRevX.6.031045}%
  \BibitemOpen
  \bibfield  {author} {\bibinfo {author} {\bibfnamefont {B.}~\bibnamefont
  {Bauer}}, \bibinfo {author} {\bibfnamefont {D.}~\bibnamefont {Wecker}},
  \bibinfo {author} {\bibfnamefont {A.~J.}\ \bibnamefont {Millis}}, \bibinfo
  {author} {\bibfnamefont {M.~B.}\ \bibnamefont {Hastings}}, \ and\ \bibinfo
  {author} {\bibfnamefont {M.}~\bibnamefont {Troyer}},\ }\href {\doibase
  10.1103/PhysRevX.6.031045} {\bibfield  {journal} {\bibinfo  {journal} {Phys.
  Rev. X}\ }\textbf {\bibinfo {volume} {6}},\ \bibinfo {pages} {031045}
  (\bibinfo {year} {2016})}\BibitemShut {NoStop}%
\bibitem [{\citenamefont {de~Jong}\ \emph {et~al.}(2020)\citenamefont
  {de~Jong}, \citenamefont {Metcalf}, \citenamefont {Mulligan}, \citenamefont
  {P\l{}osko\'n}, \citenamefont {Ringer},\ and\ \citenamefont
  {Yao}}]{deJong:2020tvx}%
  \BibitemOpen
  \bibfield  {author} {\bibinfo {author} {\bibfnamefont {W.~A.}\ \bibnamefont
  {de~Jong}}, \bibinfo {author} {\bibfnamefont {M.}~\bibnamefont {Metcalf}},
  \bibinfo {author} {\bibfnamefont {J.}~\bibnamefont {Mulligan}}, \bibinfo
  {author} {\bibfnamefont {M.}~\bibnamefont {P\l{}osko\'n}}, \bibinfo {author}
  {\bibfnamefont {F.}~\bibnamefont {Ringer}}, \ and\ \bibinfo {author}
  {\bibfnamefont {X.}~\bibnamefont {Yao}},\ }\href@noop {} {\  (\bibinfo {year}
  {2020})},\ \Eprint {http://arxiv.org/abs/2010.03571} {arXiv:2010.03571
  [hep-ph]} \BibitemShut {NoStop}%
\bibitem [{\citenamefont {Lippmann}\ and\ \citenamefont
  {Schwinger}(1950)}]{lippmann1950variational}%
  \BibitemOpen
  \bibfield  {author} {\bibinfo {author} {\bibfnamefont {B.~A.}\ \bibnamefont
  {Lippmann}}\ and\ \bibinfo {author} {\bibfnamefont {J.}~\bibnamefont
  {Schwinger}},\ }\href@noop {} {\bibfield  {journal} {\bibinfo  {journal}
  {Physical Review}\ }\textbf {\bibinfo {volume} {79}},\ \bibinfo {pages} {469}
  (\bibinfo {year} {1950})}\BibitemShut {NoStop}%
\bibitem [{\citenamefont {Newton}(2013)}]{newton2013scattering}%
  \BibitemOpen
  \bibfield  {author} {\bibinfo {author} {\bibfnamefont {R.~G.}\ \bibnamefont
  {Newton}},\ }\href@noop {} {\emph {\bibinfo {title} {Scattering theory of
  waves and particles}}}\ (\bibinfo  {publisher} {Springer Science \& Business
  Media},\ \bibinfo {year} {2013})\BibitemShut {NoStop}%
\bibitem [{\citenamefont {Meyer}\ \emph {et~al.}(1991)\citenamefont {Meyer},
  \citenamefont {Hor{\'a}ek},\ and\ \citenamefont
  {Cederbaum}}]{meyer1991schwinger}%
  \BibitemOpen
  \bibfield  {author} {\bibinfo {author} {\bibfnamefont {H.-D.}\ \bibnamefont
  {Meyer}}, \bibinfo {author} {\bibfnamefont {J.}~\bibnamefont {Hor{\'a}ek}}, \
  and\ \bibinfo {author} {\bibfnamefont {L.}~\bibnamefont {Cederbaum}},\
  }\href@noop {} {\bibfield  {journal} {\bibinfo  {journal} {Physical Review
  A}\ }\textbf {\bibinfo {volume} {43}},\ \bibinfo {pages} {3587} (\bibinfo
  {year} {1991})}\BibitemShut {NoStop}%
\bibitem [{\citenamefont {Wigner}\ and\ \citenamefont
  {Eisenbud}(1947)}]{wigner1947higher}%
  \BibitemOpen
  \bibfield  {author} {\bibinfo {author} {\bibfnamefont {E.~P.}\ \bibnamefont
  {Wigner}}\ and\ \bibinfo {author} {\bibfnamefont {L.}~\bibnamefont
  {Eisenbud}},\ }\href@noop {} {\bibfield  {journal} {\bibinfo  {journal}
  {Physical Review}\ }\textbf {\bibinfo {volume} {72}},\ \bibinfo {pages} {29}
  (\bibinfo {year} {1947})}\BibitemShut {NoStop}%
\bibitem [{\citenamefont {Yeter-Aydeniz}\ \emph {et~al.}(2020)\citenamefont
  {Yeter-Aydeniz}, \citenamefont {Siopsis},\ and\ \citenamefont
  {Pooser}}]{Yeter-Aydeniz:2020jte}%
  \BibitemOpen
  \bibfield  {author} {\bibinfo {author} {\bibfnamefont {K.}~\bibnamefont
  {Yeter-Aydeniz}}, \bibinfo {author} {\bibfnamefont {G.}~\bibnamefont
  {Siopsis}}, \ and\ \bibinfo {author} {\bibfnamefont {R.~C.}\ \bibnamefont
  {Pooser}},\ }\href@noop {} {\  (\bibinfo {year} {2020})},\ \Eprint
  {http://arxiv.org/abs/2008.08763} {arXiv:2008.08763 [quant-ph]} \BibitemShut
  {NoStop}%
\bibitem [{\citenamefont {Jordan}\ \emph
  {et~al.}(2011{\natexlab{b}})\citenamefont {Jordan}, \citenamefont {Lee},\
  and\ \citenamefont {Preskill}}]{Jordan:2011ci}%
  \BibitemOpen
  \bibfield  {author} {\bibinfo {author} {\bibfnamefont {S.~P.}\ \bibnamefont
  {Jordan}}, \bibinfo {author} {\bibfnamefont {K.~S.~M.}\ \bibnamefont {Lee}},
  \ and\ \bibinfo {author} {\bibfnamefont {J.}~\bibnamefont {Preskill}},\
  }\href@noop {} {\bibfield  {journal} {\bibinfo  {journal} {Quant. Inf.
  Comput.14,1014(2014)}\ } (\bibinfo {year} {2011}{\natexlab{b}})},\ \Eprint
  {http://arxiv.org/abs/1112.4833} {arXiv:1112.4833 [hep-th]} \BibitemShut
  {NoStop}%
\bibitem [{\citenamefont {Jordan}\ \emph
  {et~al.}(2012{\natexlab{b}})\citenamefont {Jordan}, \citenamefont {Lee},\
  and\ \citenamefont {Preskill}}]{Jordan:2011ne}%
  \BibitemOpen
  \bibfield  {author} {\bibinfo {author} {\bibfnamefont {S.~P.}\ \bibnamefont
  {Jordan}}, \bibinfo {author} {\bibfnamefont {K.~S.~M.}\ \bibnamefont {Lee}},
  \ and\ \bibinfo {author} {\bibfnamefont {J.}~\bibnamefont {Preskill}},\
  }\href {\doibase 10.1126/science.1217069} {\bibfield  {journal} {\bibinfo
  {journal} {Science}\ }\textbf {\bibinfo {volume} {336}},\ \bibinfo {pages}
  {1130} (\bibinfo {year} {2012}{\natexlab{b}})},\ \Eprint
  {http://arxiv.org/abs/1111.3633} {arXiv:1111.3633 [quant-ph]} \BibitemShut
  {NoStop}%
\bibitem [{\citenamefont {Dorfman}\ \emph {et~al.}(1994)\citenamefont
  {Dorfman}, \citenamefont {Kirkpatrick},\ and\ \citenamefont
  {Sengers}}]{Dorfman}%
  \BibitemOpen
  \bibfield  {author} {\bibinfo {author} {\bibfnamefont {J.}~\bibnamefont
  {Dorfman}}, \bibinfo {author} {\bibfnamefont {T.~R.}\ \bibnamefont
  {Kirkpatrick}}, \ and\ \bibinfo {author} {\bibfnamefont {J.~V.}\ \bibnamefont
  {Sengers}},\ }\href@noop {} {\bibfield  {journal} {\bibinfo  {journal} {Annu.
  Rev. Phys. Chem.}\ }\textbf {\bibinfo {volume} {45}},\ \bibinfo {pages} {213}
  (\bibinfo {year} {1994})}\BibitemShut {NoStop}%
\bibitem [{\citenamefont {Dorfman}\ \emph {et~al.}()\citenamefont {Dorfman},
  \citenamefont {Kirkpatrick},\ and\ \citenamefont {Sengers}}]{Dorfman2}%
  \BibitemOpen
  \bibfield  {author} {\bibinfo {author} {\bibfnamefont {J.}~\bibnamefont
  {Dorfman}}, \bibinfo {author} {\bibfnamefont {T.~R.}\ \bibnamefont
  {Kirkpatrick}}, \ and\ \bibinfo {author} {\bibfnamefont {J.~V.}\ \bibnamefont
  {Sengers}},\ }\href@noop {} {\ }\Eprint {http://arxiv.org/abs/1512.02679}
  {arXiv:1512.02679 [cond-mat.stat-mech]} \BibitemShut {NoStop}%
\bibitem [{\citenamefont {Huang}\ and\ \citenamefont
  {Yang}(1957)}]{Huang:1957im}%
  \BibitemOpen
  \bibfield  {author} {\bibinfo {author} {\bibfnamefont {K.}~\bibnamefont
  {Huang}}\ and\ \bibinfo {author} {\bibfnamefont {C.~N.}\ \bibnamefont
  {Yang}},\ }\href {\doibase 10.1103/PhysRev.105.767} {\bibfield  {journal}
  {\bibinfo  {journal} {Phys. Rev.}\ }\textbf {\bibinfo {volume} {105}},\
  \bibinfo {pages} {767} (\bibinfo {year} {1957})}\BibitemShut {NoStop}%
\bibitem [{\citenamefont {Huang}\ \emph {et~al.}(1957)\citenamefont {Huang},
  \citenamefont {Yang},\ and\ \citenamefont {Luttinger}}]{Huang:1957zz}%
  \BibitemOpen
  \bibfield  {author} {\bibinfo {author} {\bibfnamefont {K.}~\bibnamefont
  {Huang}}, \bibinfo {author} {\bibfnamefont {C.~N.}\ \bibnamefont {Yang}}, \
  and\ \bibinfo {author} {\bibfnamefont {J.~M.}\ \bibnamefont {Luttinger}},\
  }\href {\doibase 10.1103/PhysRev.105.776} {\bibfield  {journal} {\bibinfo
  {journal} {Phys. Rev.}\ }\textbf {\bibinfo {volume} {105}},\ \bibinfo {pages}
  {776} (\bibinfo {year} {1957})}\BibitemShut {NoStop}%
\bibitem [{\citenamefont {Dashen}\ \emph {et~al.}(1969)\citenamefont {Dashen},
  \citenamefont {Ma},\ and\ \citenamefont {Bernstein}}]{Dashen:1969ep}%
  \BibitemOpen
  \bibfield  {author} {\bibinfo {author} {\bibfnamefont {R.}~\bibnamefont
  {Dashen}}, \bibinfo {author} {\bibfnamefont {S.-K.}\ \bibnamefont {Ma}}, \
  and\ \bibinfo {author} {\bibfnamefont {H.~J.}\ \bibnamefont {Bernstein}},\
  }\href {\doibase 10.1103/PhysRev.187.345} {\bibfield  {journal} {\bibinfo
  {journal} {Phys. Rev.}\ }\textbf {\bibinfo {volume} {187}},\ \bibinfo {pages}
  {345} (\bibinfo {year} {1969})}\BibitemShut {NoStop}%
\bibitem [{\citenamefont {Dashen}\ and\ \citenamefont
  {Rajaraman}(1974)}]{Dashen:1974yy}%
  \BibitemOpen
  \bibfield  {author} {\bibinfo {author} {\bibfnamefont {R.}~\bibnamefont
  {Dashen}}\ and\ \bibinfo {author} {\bibfnamefont {R.}~\bibnamefont
  {Rajaraman}},\ }\href {\doibase 10.1103/PhysRevD.10.708} {\bibfield
  {journal} {\bibinfo  {journal} {Phys. Rev. D}\ }\textbf {\bibinfo {volume}
  {10}},\ \bibinfo {pages} {708} (\bibinfo {year} {1974})}\BibitemShut
  {NoStop}%
\bibitem [{\citenamefont {Feynman}(2018)}]{feynman2018photon}%
  \BibitemOpen
  \bibfield  {author} {\bibinfo {author} {\bibfnamefont {R.~P.}\ \bibnamefont
  {Feynman}},\ }\href@noop {} {\emph {\bibinfo {title} {Photon-hadron
  interactions}}}\ (\bibinfo  {publisher} {CRC Press},\ \bibinfo {year}
  {2018})\BibitemShut {NoStop}%
\bibitem [{\citenamefont {Ioffe}(1969)}]{Ioffe:1969kf}%
  \BibitemOpen
  \bibfield  {author} {\bibinfo {author} {\bibfnamefont {B.}~\bibnamefont
  {Ioffe}},\ }\href {\doibase 10.1016/0370-2693(69)90415-8} {\bibfield
  {journal} {\bibinfo  {journal} {Phys. Lett. B}\ }\textbf {\bibinfo {volume}
  {30}},\ \bibinfo {pages} {123} (\bibinfo {year} {1969})}\BibitemShut
  {NoStop}%
\bibitem [{\citenamefont {Kovchegov}\ and\ \citenamefont
  {Strikman}(2001)}]{Kovchegov:2001dh}%
  \BibitemOpen
  \bibfield  {author} {\bibinfo {author} {\bibfnamefont {Y.~V.}\ \bibnamefont
  {Kovchegov}}\ and\ \bibinfo {author} {\bibfnamefont {M.}~\bibnamefont
  {Strikman}},\ }\href {\doibase 10.1016/S0370-2693(01)00953-4} {\bibfield
  {journal} {\bibinfo  {journal} {Phys. Lett. B}\ }\textbf {\bibinfo {volume}
  {516}},\ \bibinfo {pages} {314} (\bibinfo {year} {2001})},\ \Eprint
  {http://arxiv.org/abs/hep-ph/0107015} {arXiv:hep-ph/0107015} \BibitemShut
  {NoStop}%
\bibitem [{\citenamefont {Wigner}(1955)}]{Wigner:1955zz}%
  \BibitemOpen
  \bibfield  {author} {\bibinfo {author} {\bibfnamefont {E.~P.}\ \bibnamefont
  {Wigner}},\ }\href {\doibase 10.1103/PhysRev.98.145} {\bibfield  {journal}
  {\bibinfo  {journal} {Phys. Rev.}\ }\textbf {\bibinfo {volume} {98}},\
  \bibinfo {pages} {145} (\bibinfo {year} {1955})}\BibitemShut {NoStop}%
\bibitem [{\citenamefont {Luscher}(1986)}]{Luscher:1986pf}%
  \BibitemOpen
  \bibfield  {author} {\bibinfo {author} {\bibfnamefont {M.}~\bibnamefont
  {Luscher}},\ }\href {\doibase 10.1007/BF01211097} {\bibfield  {journal}
  {\bibinfo  {journal} {Commun. Math. Phys.}\ }\textbf {\bibinfo {volume}
  {105}},\ \bibinfo {pages} {153} (\bibinfo {year} {1986})}\BibitemShut
  {NoStop}%
\bibitem [{\citenamefont {Luscher}(1991)}]{Luscher:1990ux}%
  \BibitemOpen
  \bibfield  {author} {\bibinfo {author} {\bibfnamefont {M.}~\bibnamefont
  {Luscher}},\ }\href {\doibase 10.1016/0550-3213(91)90366-6} {\bibfield
  {journal} {\bibinfo  {journal} {Nucl. Phys. B}\ }\textbf {\bibinfo {volume}
  {354}},\ \bibinfo {pages} {531} (\bibinfo {year} {1991})}\BibitemShut
  {NoStop}%
\bibitem [{\citenamefont {Hansen}\ and\ \citenamefont
  {Sharpe}(2019)}]{Hansen:2019nir}%
  \BibitemOpen
  \bibfield  {author} {\bibinfo {author} {\bibfnamefont {M.~T.}\ \bibnamefont
  {Hansen}}\ and\ \bibinfo {author} {\bibfnamefont {S.~R.}\ \bibnamefont
  {Sharpe}},\ }\href {\doibase 10.1146/annurev-nucl-101918-023723} {\bibfield
  {journal} {\bibinfo  {journal} {Ann. Rev. Nucl. Part. Sci.}\ }\textbf
  {\bibinfo {volume} {69}},\ \bibinfo {pages} {65} (\bibinfo {year} {2019})},\
  \Eprint {http://arxiv.org/abs/1901.00483} {arXiv:1901.00483 [hep-lat]}
  \BibitemShut {NoStop}%
\bibitem [{\citenamefont {Briceno}\ \emph {et~al.}(2018)\citenamefont
  {Briceno}, \citenamefont {Dudek},\ and\ \citenamefont
  {Young}}]{Briceno:2017max}%
  \BibitemOpen
  \bibfield  {author} {\bibinfo {author} {\bibfnamefont {R.~A.}\ \bibnamefont
  {Briceno}}, \bibinfo {author} {\bibfnamefont {J.~J.}\ \bibnamefont {Dudek}},
  \ and\ \bibinfo {author} {\bibfnamefont {R.~D.}\ \bibnamefont {Young}},\
  }\href {\doibase 10.1103/RevModPhys.90.025001} {\bibfield  {journal}
  {\bibinfo  {journal} {Rev. Mod. Phys.}\ }\textbf {\bibinfo {volume} {90}},\
  \bibinfo {pages} {025001} (\bibinfo {year} {2018})},\ \Eprint
  {http://arxiv.org/abs/1706.06223} {arXiv:1706.06223 [hep-lat]} \BibitemShut
  {NoStop}%
\bibitem [{\citenamefont {Brice\~no}\ \emph {et~al.}(2020)\citenamefont
  {Brice\~no}, \citenamefont {Guerrero}, \citenamefont {Hansen},\ and\
  \citenamefont {Sturzu}}]{Briceno:2020rar}%
  \BibitemOpen
  \bibfield  {author} {\bibinfo {author} {\bibfnamefont {R.~A.}\ \bibnamefont
  {Brice\~no}}, \bibinfo {author} {\bibfnamefont {J.~V.}\ \bibnamefont
  {Guerrero}}, \bibinfo {author} {\bibfnamefont {M.~T.}\ \bibnamefont
  {Hansen}}, \ and\ \bibinfo {author} {\bibfnamefont {A.}~\bibnamefont
  {Sturzu}},\ }\href@noop {} {\  (\bibinfo {year} {2020})},\ \Eprint
  {http://arxiv.org/abs/2007.01155} {arXiv:2007.01155 [hep-lat]} \BibitemShut
  {NoStop}%
\bibitem [{\citenamefont {Brodsky}\ \emph {et~al.}(1998)\citenamefont
  {Brodsky}, \citenamefont {Pauli},\ and\ \citenamefont
  {Pinsky}}]{Brodsky:1997de}%
  \BibitemOpen
  \bibfield  {author} {\bibinfo {author} {\bibfnamefont {S.~J.}\ \bibnamefont
  {Brodsky}}, \bibinfo {author} {\bibfnamefont {H.-C.}\ \bibnamefont {Pauli}},
  \ and\ \bibinfo {author} {\bibfnamefont {S.~S.}\ \bibnamefont {Pinsky}},\
  }\href {\doibase 10.1016/S0370-1573(97)00089-6} {\bibfield  {journal}
  {\bibinfo  {journal} {Phys. Rept.}\ }\textbf {\bibinfo {volume} {301}},\
  \bibinfo {pages} {299} (\bibinfo {year} {1998})},\ \Eprint
  {http://arxiv.org/abs/hep-ph/9705477} {arXiv:hep-ph/9705477} \BibitemShut
  {NoStop}%
\bibitem [{\citenamefont {Susskind}(1968)}]{Susskind:1967rg}%
  \BibitemOpen
  \bibfield  {author} {\bibinfo {author} {\bibfnamefont {L.}~\bibnamefont
  {Susskind}},\ }\href {\doibase 10.1103/PhysRev.165.1535} {\bibfield
  {journal} {\bibinfo  {journal} {Phys. Rev.}\ }\textbf {\bibinfo {volume}
  {165}},\ \bibinfo {pages} {1535} (\bibinfo {year} {1968})}\BibitemShut
  {NoStop}%
\bibitem [{\citenamefont {Bjorken}\ \emph {et~al.}(1971)\citenamefont
  {Bjorken}, \citenamefont {Kogut},\ and\ \citenamefont
  {Soper}}]{Bjorken:1970ah}%
  \BibitemOpen
  \bibfield  {author} {\bibinfo {author} {\bibfnamefont {J.}~\bibnamefont
  {Bjorken}}, \bibinfo {author} {\bibfnamefont {J.~B.}\ \bibnamefont {Kogut}},
  \ and\ \bibinfo {author} {\bibfnamefont {D.~E.}\ \bibnamefont {Soper}},\
  }\href {\doibase 10.1103/PhysRevD.3.1382} {\bibfield  {journal} {\bibinfo
  {journal} {Phys. Rev. D}\ }\textbf {\bibinfo {volume} {3}},\ \bibinfo {pages}
  {1382} (\bibinfo {year} {1971})}\BibitemShut {NoStop}%
\bibitem [{\citenamefont {Weinberg}(1967)}]{Weinberg:1966fm}%
  \BibitemOpen
  \bibfield  {author} {\bibinfo {author} {\bibfnamefont {S.}~\bibnamefont
  {Weinberg}},\ }\href {\doibase 10.1103/PhysRevLett.18.188} {\bibfield
  {journal} {\bibinfo  {journal} {Phys. Rev. Lett.}\ }\textbf {\bibinfo
  {volume} {18}},\ \bibinfo {pages} {188} (\bibinfo {year} {1967})}\BibitemShut
  {NoStop}%
\bibitem [{\citenamefont {Fitzpatrick}\ \emph {et~al.}(2020)\citenamefont
  {Fitzpatrick}, \citenamefont {Katz},\ and\ \citenamefont
  {Walters}}]{Fitzpatrick:2018xlz}%
  \BibitemOpen
  \bibfield  {author} {\bibinfo {author} {\bibfnamefont {A.~L.}\ \bibnamefont
  {Fitzpatrick}}, \bibinfo {author} {\bibfnamefont {E.}~\bibnamefont {Katz}}, \
  and\ \bibinfo {author} {\bibfnamefont {M.~T.}\ \bibnamefont {Walters}},\
  }\href {\doibase 10.1007/JHEP10(2020)092} {\bibfield  {journal} {\bibinfo
  {journal} {JHEP}\ }\textbf {\bibinfo {volume} {10}},\ \bibinfo {pages} {092}
  (\bibinfo {year} {2020})},\ \Eprint {http://arxiv.org/abs/1812.08177}
  {arXiv:1812.08177 [hep-th]} \BibitemShut {NoStop}%
\bibitem [{\citenamefont {Anand}\ \emph {et~al.}(2020)\citenamefont {Anand},
  \citenamefont {Fitzpatrick}, \citenamefont {Katz}, \citenamefont {Khandker},
  \citenamefont {Walters},\ and\ \citenamefont {Xin}}]{Anand:2020gnn}%
  \BibitemOpen
  \bibfield  {author} {\bibinfo {author} {\bibfnamefont {N.}~\bibnamefont
  {Anand}}, \bibinfo {author} {\bibfnamefont {A.~L.}\ \bibnamefont
  {Fitzpatrick}}, \bibinfo {author} {\bibfnamefont {E.}~\bibnamefont {Katz}},
  \bibinfo {author} {\bibfnamefont {Z.~U.}\ \bibnamefont {Khandker}}, \bibinfo
  {author} {\bibfnamefont {M.~T.}\ \bibnamefont {Walters}}, \ and\ \bibinfo
  {author} {\bibfnamefont {Y.}~\bibnamefont {Xin}},\ }\href@noop {} {\
  (\bibinfo {year} {2020})},\ \Eprint {http://arxiv.org/abs/2005.13544}
  {arXiv:2005.13544 [hep-th]} \BibitemShut {NoStop}%
\bibitem [{\citenamefont {Liu}\ and\ \citenamefont {Xin}(2020)}]{Liu:2020eoa}%
  \BibitemOpen
  \bibfield  {author} {\bibinfo {author} {\bibfnamefont {J.}~\bibnamefont
  {Liu}}\ and\ \bibinfo {author} {\bibfnamefont {Y.}~\bibnamefont {Xin}},\
  }\href@noop {} {\  (\bibinfo {year} {2020})},\ \Eprint
  {http://arxiv.org/abs/2004.13234} {arXiv:2004.13234 [hep-th]} \BibitemShut
  {NoStop}%
\bibitem [{\citenamefont {James}\ \emph {et~al.}(2018)\citenamefont {James},
  \citenamefont {Konik}, \citenamefont {Lecheminant}, \citenamefont
  {Robinson},\ and\ \citenamefont {Tsvelik}}]{James_2018}%
  \BibitemOpen
  \bibfield  {author} {\bibinfo {author} {\bibfnamefont {A.~J.~A.}\
  \bibnamefont {James}}, \bibinfo {author} {\bibfnamefont {R.~M.}\ \bibnamefont
  {Konik}}, \bibinfo {author} {\bibfnamefont {P.}~\bibnamefont {Lecheminant}},
  \bibinfo {author} {\bibfnamefont {N.~J.}\ \bibnamefont {Robinson}}, \ and\
  \bibinfo {author} {\bibfnamefont {A.~M.}\ \bibnamefont {Tsvelik}},\ }\href
  {\doibase 10.1088/1361-6633/aa91ea} {\bibfield  {journal} {\bibinfo
  {journal} {Prog. Part. Nucl. Phys.}\ }\textbf {\bibinfo {volume} {81}},\
  \bibinfo {pages} {046002} (\bibinfo {year} {2018})},\ \Eprint
  {http://arxiv.org/abs/1703.08421v1} {arXiv:1703.08421v1 [cond-mat.str-el]}
  \BibitemShut {NoStop}%
\bibitem [{\citenamefont {Kreshchuk}\ \emph
  {et~al.}(2020{\natexlab{b}})\citenamefont {Kreshchuk}, \citenamefont {Kirby},
  \citenamefont {Goldstein}, \citenamefont {Beauchemin},\ and\ \citenamefont
  {Love}}]{Kreshchuk:2020dla}%
  \BibitemOpen
  \bibfield  {author} {\bibinfo {author} {\bibfnamefont {M.}~\bibnamefont
  {Kreshchuk}}, \bibinfo {author} {\bibfnamefont {W.~M.}\ \bibnamefont
  {Kirby}}, \bibinfo {author} {\bibfnamefont {G.}~\bibnamefont {Goldstein}},
  \bibinfo {author} {\bibfnamefont {H.}~\bibnamefont {Beauchemin}}, \ and\
  \bibinfo {author} {\bibfnamefont {P.~J.}\ \bibnamefont {Love}},\ }\href@noop
  {} {\  (\bibinfo {year} {2020}{\natexlab{b}})},\ \Eprint
  {http://arxiv.org/abs/2002.04016} {arXiv:2002.04016 [quant-ph]} \BibitemShut
  {NoStop}%
\bibitem [{\citenamefont {Kreshchuk}\ \emph
  {et~al.}(2020{\natexlab{c}})\citenamefont {Kreshchuk}, \citenamefont {Jia},
  \citenamefont {Kirby}, \citenamefont {Goldstein}, \citenamefont {Vary},\ and\
  \citenamefont {Love}}]{Kreshchuk:2020kcz}%
  \BibitemOpen
  \bibfield  {author} {\bibinfo {author} {\bibfnamefont {M.}~\bibnamefont
  {Kreshchuk}}, \bibinfo {author} {\bibfnamefont {S.}~\bibnamefont {Jia}},
  \bibinfo {author} {\bibfnamefont {W.~M.}\ \bibnamefont {Kirby}}, \bibinfo
  {author} {\bibfnamefont {G.}~\bibnamefont {Goldstein}}, \bibinfo {author}
  {\bibfnamefont {J.~P.}\ \bibnamefont {Vary}}, \ and\ \bibinfo {author}
  {\bibfnamefont {P.~J.}\ \bibnamefont {Love}},\ }\href@noop {} {\  (\bibinfo
  {year} {2020}{\natexlab{c}})},\ \Eprint {http://arxiv.org/abs/2009.07885}
  {arXiv:2009.07885 [quant-ph]} \BibitemShut {NoStop}%
\bibitem [{\citenamefont {Bauer}\ \emph {et~al.}(2001)\citenamefont {Bauer},
  \citenamefont {Fleming}, \citenamefont {Pirjol},\ and\ \citenamefont
  {Stewart}}]{Bauer:2000yr}%
  \BibitemOpen
  \bibfield  {author} {\bibinfo {author} {\bibfnamefont {C.~W.}\ \bibnamefont
  {Bauer}}, \bibinfo {author} {\bibfnamefont {S.}~\bibnamefont {Fleming}},
  \bibinfo {author} {\bibfnamefont {D.}~\bibnamefont {Pirjol}}, \ and\ \bibinfo
  {author} {\bibfnamefont {I.~W.}\ \bibnamefont {Stewart}},\ }\href {\doibase
  10.1103/PhysRevD.63.114020} {\bibfield  {journal} {\bibinfo  {journal} {Phys.
  Rev. D}\ }\textbf {\bibinfo {volume} {63}},\ \bibinfo {pages} {114020}
  (\bibinfo {year} {2001})},\ \Eprint {http://arxiv.org/abs/hep-ph/0011336}
  {arXiv:hep-ph/0011336} \BibitemShut {NoStop}%
\bibitem [{\citenamefont {Gelis}\ \emph {et~al.}(2010)\citenamefont {Gelis},
  \citenamefont {Iancu}, \citenamefont {Jalilian-Marian},\ and\ \citenamefont
  {Venugopalan}}]{Gelis:2010nm}%
  \BibitemOpen
  \bibfield  {author} {\bibinfo {author} {\bibfnamefont {F.}~\bibnamefont
  {Gelis}}, \bibinfo {author} {\bibfnamefont {E.}~\bibnamefont {Iancu}},
  \bibinfo {author} {\bibfnamefont {J.}~\bibnamefont {Jalilian-Marian}}, \ and\
  \bibinfo {author} {\bibfnamefont {R.}~\bibnamefont {Venugopalan}},\ }\href
  {\doibase 10.1146/annurev.nucl.010909.083629} {\bibfield  {journal} {\bibinfo
   {journal} {Ann. Rev. Nucl. Part. Sci.}\ }\textbf {\bibinfo {volume} {60}},\
  \bibinfo {pages} {463} (\bibinfo {year} {2010})},\ \Eprint
  {http://arxiv.org/abs/1002.0333} {arXiv:1002.0333 [hep-ph]} \BibitemShut
  {NoStop}%
\bibitem [{\citenamefont {Braaten}\ and\ \citenamefont
  {Pisarski}(1990)}]{Braaten:1989mz}%
  \BibitemOpen
  \bibfield  {author} {\bibinfo {author} {\bibfnamefont {E.}~\bibnamefont
  {Braaten}}\ and\ \bibinfo {author} {\bibfnamefont {R.~D.}\ \bibnamefont
  {Pisarski}},\ }\href {\doibase 10.1016/0550-3213(90)90508-B} {\bibfield
  {journal} {\bibinfo  {journal} {Nucl. Phys. B}\ }\textbf {\bibinfo {volume}
  {337}},\ \bibinfo {pages} {569} (\bibinfo {year} {1990})}\BibitemShut
  {NoStop}%
\bibitem [{\citenamefont {Klco}\ and\ \citenamefont
  {Savage}(2019{\natexlab{a}})}]{klco2019digitization}%
  \BibitemOpen
  \bibfield  {author} {\bibinfo {author} {\bibfnamefont {N.}~\bibnamefont
  {Klco}}\ and\ \bibinfo {author} {\bibfnamefont {M.~J.}\ \bibnamefont
  {Savage}},\ }\href@noop {} {\bibfield  {journal} {\bibinfo  {journal}
  {Physical Review A}\ }\textbf {\bibinfo {volume} {99}},\ \bibinfo {pages}
  {052335} (\bibinfo {year} {2019}{\natexlab{a}})}\BibitemShut {NoStop}%
\bibitem [{\citenamefont {Childs}\ \emph {et~al.}(2017)\citenamefont {Childs},
  \citenamefont {Kothari},\ and\ \citenamefont {Somma}}]{childs2017quantum}%
  \BibitemOpen
  \bibfield  {author} {\bibinfo {author} {\bibfnamefont {A.~M.}\ \bibnamefont
  {Childs}}, \bibinfo {author} {\bibfnamefont {R.}~\bibnamefont {Kothari}}, \
  and\ \bibinfo {author} {\bibfnamefont {R.~D.}\ \bibnamefont {Somma}},\
  }\href@noop {} {\bibfield  {journal} {\bibinfo  {journal} {SIAM Journal on
  Computing}\ }\textbf {\bibinfo {volume} {46}},\ \bibinfo {pages} {1920}
  (\bibinfo {year} {2017})}\BibitemShut {NoStop}%
\bibitem [{\citenamefont {Gily{\'e}n}\ \emph {et~al.}(2019)\citenamefont
  {Gily{\'e}n}, \citenamefont {Su}, \citenamefont {Low},\ and\ \citenamefont
  {Wiebe}}]{gilyen2019quantum}%
  \BibitemOpen
  \bibfield  {author} {\bibinfo {author} {\bibfnamefont {A.}~\bibnamefont
  {Gily{\'e}n}}, \bibinfo {author} {\bibfnamefont {Y.}~\bibnamefont {Su}},
  \bibinfo {author} {\bibfnamefont {G.~H.}\ \bibnamefont {Low}}, \ and\
  \bibinfo {author} {\bibfnamefont {N.}~\bibnamefont {Wiebe}},\ }in\ \href@noop
  {} {\emph {\bibinfo {booktitle} {Proceedings of the 51st Annual ACM SIGACT
  Symposium on Theory of Computing}}}\ (\bibinfo {year} {2019})\ pp.\ \bibinfo
  {pages} {193--204}\BibitemShut {NoStop}%
\bibitem [{\citenamefont {Trotter}(1959)}]{Trotter:1959}%
  \BibitemOpen
  \bibfield  {author} {\bibinfo {author} {\bibfnamefont {H.~F.}\ \bibnamefont
  {Trotter}},\ }\href@noop {} {\bibfield  {journal} {\bibinfo  {journal} {Proc.
  Am. Math. Soc.}\ }\textbf {\bibinfo {volume} {10}},\ \bibinfo {pages} {545}
  (\bibinfo {year} {1959})}\BibitemShut {NoStop}%
\bibitem [{\citenamefont {Suzuki}(1976)}]{Suzuki:1976}%
  \BibitemOpen
  \bibfield  {author} {\bibinfo {author} {\bibfnamefont {M.}~\bibnamefont
  {Suzuki}},\ }\href@noop {} {\bibfield  {journal} {\bibinfo  {journal}
  {Commun. Math. Phys.}\ }\textbf {\bibinfo {volume} {51}},\ \bibinfo {pages}
  {183} (\bibinfo {year} {1976})}\BibitemShut {NoStop}%
\bibitem [{\citenamefont {Gerry}\ and\ \citenamefont
  {Knight}(2000)}]{gerry2000quantum}%
  \BibitemOpen
  \bibfield  {author} {\bibinfo {author} {\bibfnamefont {C.}~\bibnamefont
  {Gerry}}\ and\ \bibinfo {author} {\bibfnamefont {P.}~\bibnamefont {Knight}},\
  }\href@noop {} {\enquote {\bibinfo {title} {Quantum optics},}\ } (\bibinfo
  {year} {2000})\BibitemShut {NoStop}%
\bibitem [{\citenamefont {Yeter-Aydeniz}\ and\ \citenamefont
  {Siopsis}(2018{\natexlab{a}})}]{yeter2018quantum}%
  \BibitemOpen
  \bibfield  {author} {\bibinfo {author} {\bibfnamefont {K.}~\bibnamefont
  {Yeter-Aydeniz}}\ and\ \bibinfo {author} {\bibfnamefont {G.}~\bibnamefont
  {Siopsis}},\ }\href@noop {} {\bibfield  {journal} {\bibinfo  {journal}
  {Physical Review D}\ }\textbf {\bibinfo {volume} {97}},\ \bibinfo {pages}
  {036004} (\bibinfo {year} {2018}{\natexlab{a}})}\BibitemShut {NoStop}%
\bibitem [{\citenamefont {Nielsen}\ and\ \citenamefont
  {Chuang}(2010)}]{NielsenChuang}%
  \BibitemOpen
  \bibfield  {author} {\bibinfo {author} {\bibfnamefont {M.~A.}\ \bibnamefont
  {Nielsen}}\ and\ \bibinfo {author} {\bibfnamefont {I.~L.}\ \bibnamefont
  {Chuang}},\ }\href@noop {} {\emph {\bibinfo {title} {{Quantum Computation and
  Quantum Information}}}}\ (\bibinfo  {publisher} {Cambridge University
  Press},\ \bibinfo {year} {2010})\BibitemShut {NoStop}%
\bibitem [{\citenamefont {Su}\ \emph {et~al.}(1997)\citenamefont {Su},
  \citenamefont {Smetanko},\ and\ \citenamefont {Grobe}}]{Su}%
  \BibitemOpen
  \bibfield  {author} {\bibinfo {author} {\bibfnamefont {Q.}~\bibnamefont
  {Su}}, \bibinfo {author} {\bibfnamefont {B.~A.}\ \bibnamefont {Smetanko}}, \
  and\ \bibinfo {author} {\bibfnamefont {B.}~\bibnamefont {Grobe}},\
  }\href@noop {} {\bibfield  {journal} {\bibinfo  {journal} {Opt. Express}\
  }\textbf {\bibinfo {volume} {2}},\ \bibinfo {pages} {277} (\bibinfo {year}
  {1997})}\BibitemShut {NoStop}%
\bibitem [{\citenamefont {Lamm}\ and\ \citenamefont
  {Lawrence}(2018)}]{lamm2018simulation}%
  \BibitemOpen
  \bibfield  {author} {\bibinfo {author} {\bibfnamefont {H.}~\bibnamefont
  {Lamm}}\ and\ \bibinfo {author} {\bibfnamefont {S.}~\bibnamefont
  {Lawrence}},\ }\href@noop {} {\bibfield  {journal} {\bibinfo  {journal}
  {Physical review letters}\ }\textbf {\bibinfo {volume} {121}},\ \bibinfo
  {pages} {170501} (\bibinfo {year} {2018})}\BibitemShut {NoStop}%
\bibitem [{\citenamefont {Kokail}\ \emph {et~al.}(2019)\citenamefont {Kokail},
  \citenamefont {Maier}, \citenamefont {van Bijnen}, \citenamefont {Brydges},
  \citenamefont {Joshi}, \citenamefont {Jurcevic}, \citenamefont {Muschik},
  \citenamefont {Silvi}, \citenamefont {Blatt}, \citenamefont {Roos} \emph
  {et~al.}}]{kokail2019self}%
  \BibitemOpen
  \bibfield  {author} {\bibinfo {author} {\bibfnamefont {C.}~\bibnamefont
  {Kokail}}, \bibinfo {author} {\bibfnamefont {C.}~\bibnamefont {Maier}},
  \bibinfo {author} {\bibfnamefont {R.}~\bibnamefont {van Bijnen}}, \bibinfo
  {author} {\bibfnamefont {T.}~\bibnamefont {Brydges}}, \bibinfo {author}
  {\bibfnamefont {M.~K.}\ \bibnamefont {Joshi}}, \bibinfo {author}
  {\bibfnamefont {P.}~\bibnamefont {Jurcevic}}, \bibinfo {author}
  {\bibfnamefont {C.~A.}\ \bibnamefont {Muschik}}, \bibinfo {author}
  {\bibfnamefont {P.}~\bibnamefont {Silvi}}, \bibinfo {author} {\bibfnamefont
  {R.}~\bibnamefont {Blatt}}, \bibinfo {author} {\bibfnamefont {C.~F.}\
  \bibnamefont {Roos}},  \emph {et~al.},\ }\href@noop {} {\bibfield  {journal}
  {\bibinfo  {journal} {Nature}\ }\textbf {\bibinfo {volume} {569}},\ \bibinfo
  {pages} {355} (\bibinfo {year} {2019})}\BibitemShut {NoStop}%
\bibitem [{\citenamefont {Bapat}\ and\ \citenamefont
  {Jordan}(2019)}]{bapat2019bang}%
  \BibitemOpen
  \bibfield  {author} {\bibinfo {author} {\bibfnamefont {A.}~\bibnamefont
  {Bapat}}\ and\ \bibinfo {author} {\bibfnamefont {S.}~\bibnamefont {Jordan}},\
  }\href@noop {} {\bibfield  {journal} {\bibinfo  {journal} {Quantum
  Information \& Computation}\ }\textbf {\bibinfo {volume} {19}},\ \bibinfo
  {pages} {424} (\bibinfo {year} {2019})}\BibitemShut {NoStop}%
\bibitem [{\citenamefont {Harmalkar}\ \emph {et~al.}(2020)\citenamefont
  {Harmalkar}, \citenamefont {Lamm},\ and\ \citenamefont
  {Lawrence}}]{harmalkar2020quantum}%
  \BibitemOpen
  \bibfield  {author} {\bibinfo {author} {\bibfnamefont {S.}~\bibnamefont
  {Harmalkar}}, \bibinfo {author} {\bibfnamefont {H.}~\bibnamefont {Lamm}}, \
  and\ \bibinfo {author} {\bibfnamefont {S.}~\bibnamefont {Lawrence}},\
  }\href@noop {} {\bibfield  {journal} {\bibinfo  {journal} {arXiv preprint
  arXiv:2001.11490}\ } (\bibinfo {year} {2020})}\BibitemShut {NoStop}%
\bibitem [{\citenamefont {Gustafson}\ and\ \citenamefont
  {Lamm}(2020)}]{gustafson2020toward}%
  \BibitemOpen
  \bibfield  {author} {\bibinfo {author} {\bibfnamefont {E.~J.}\ \bibnamefont
  {Gustafson}}\ and\ \bibinfo {author} {\bibfnamefont {H.}~\bibnamefont
  {Lamm}},\ }\href@noop {} {\bibfield  {journal} {\bibinfo  {journal} {arXiv
  preprint arXiv:2011.11677}\ } (\bibinfo {year} {2020})}\BibitemShut {NoStop}%
\bibitem [{\citenamefont {Choi}\ and\ \citenamefont
  {Lee}(2020)}]{choi2020rodeo}%
  \BibitemOpen
  \bibfield  {author} {\bibinfo {author} {\bibfnamefont {K.}~\bibnamefont
  {Choi}}\ and\ \bibinfo {author} {\bibfnamefont {D.}~\bibnamefont {Lee}},\
  }\href@noop {} {\bibfield  {journal} {\bibinfo  {journal} {arXiv preprint
  arXiv:2009.04092}\ } (\bibinfo {year} {2020})}\BibitemShut {NoStop}%
\bibitem [{\citenamefont {Brassard}\ \emph {et~al.}(2002)\citenamefont
  {Brassard}, \citenamefont {Hoyer}, \citenamefont {Mosca},\ and\ \citenamefont
  {Tapp}}]{brassard2002quantum}%
  \BibitemOpen
  \bibfield  {author} {\bibinfo {author} {\bibfnamefont {G.}~\bibnamefont
  {Brassard}}, \bibinfo {author} {\bibfnamefont {P.}~\bibnamefont {Hoyer}},
  \bibinfo {author} {\bibfnamefont {M.}~\bibnamefont {Mosca}}, \ and\ \bibinfo
  {author} {\bibfnamefont {A.}~\bibnamefont {Tapp}},\ }\href@noop {} {\bibfield
   {journal} {\bibinfo  {journal} {Contemporary Mathematics}\ }\textbf
  {\bibinfo {volume} {305}},\ \bibinfo {pages} {53} (\bibinfo {year}
  {2002})}\BibitemShut {NoStop}%
\bibitem [{\citenamefont {Berry}\ \emph {et~al.}(2014)\citenamefont {Berry},
  \citenamefont {Childs}, \citenamefont {Cleve}, \citenamefont {Kothari},\ and\
  \citenamefont {Somma}}]{Berry_2014}%
  \BibitemOpen
  \bibfield  {author} {\bibinfo {author} {\bibfnamefont {D.~W.}\ \bibnamefont
  {Berry}}, \bibinfo {author} {\bibfnamefont {A.~M.}\ \bibnamefont {Childs}},
  \bibinfo {author} {\bibfnamefont {R.}~\bibnamefont {Cleve}}, \bibinfo
  {author} {\bibfnamefont {R.}~\bibnamefont {Kothari}}, \ and\ \bibinfo
  {author} {\bibfnamefont {R.~D.}\ \bibnamefont {Somma}},\ }\href {\doibase
  10.1145/2591796.2591854} {\bibfield  {journal} {\bibinfo  {journal}
  {Proceedings of the 46th Annual ACM Symposium on Theory of Computing - STOC
  ’14}\ } (\bibinfo {year} {2014}),\ 10.1145/2591796.2591854}\BibitemShut
  {NoStop}%
\bibitem [{\citenamefont {Grover}\ and\ \citenamefont
  {Rudolph}(2002)}]{grover2002creating}%
  \BibitemOpen
  \bibfield  {author} {\bibinfo {author} {\bibfnamefont {L.}~\bibnamefont
  {Grover}}\ and\ \bibinfo {author} {\bibfnamefont {T.}~\bibnamefont
  {Rudolph}},\ }\href@noop {} {\bibfield  {journal} {\bibinfo  {journal} {arXiv
  preprint quant-ph/0208112}\ } (\bibinfo {year} {2002})}\BibitemShut {NoStop}%
\bibitem [{\citenamefont {Kaye}\ and\ \citenamefont
  {Mosca}(2004)}]{kaye2004quantum}%
  \BibitemOpen
  \bibfield  {author} {\bibinfo {author} {\bibfnamefont {P.}~\bibnamefont
  {Kaye}}\ and\ \bibinfo {author} {\bibfnamefont {M.}~\bibnamefont {Mosca}},\
  }\href@noop {} {\bibfield  {journal} {\bibinfo  {journal} {arXiv preprint
  quant-ph/0407102}\ } (\bibinfo {year} {2004})}\BibitemShut {NoStop}%
\bibitem [{\citenamefont {Tranter}\ \emph {et~al.}(2018)\citenamefont
  {Tranter}, \citenamefont {Love}, \citenamefont {Mintert},\ and\ \citenamefont
  {Coveney}}]{tranter2018comparison}%
  \BibitemOpen
  \bibfield  {author} {\bibinfo {author} {\bibfnamefont {A.}~\bibnamefont
  {Tranter}}, \bibinfo {author} {\bibfnamefont {P.~J.}\ \bibnamefont {Love}},
  \bibinfo {author} {\bibfnamefont {F.}~\bibnamefont {Mintert}}, \ and\
  \bibinfo {author} {\bibfnamefont {P.~V.}\ \bibnamefont {Coveney}},\
  }\href@noop {} {\bibfield  {journal} {\bibinfo  {journal} {Journal of
  chemical theory and computation}\ }\textbf {\bibinfo {volume} {14}},\
  \bibinfo {pages} {5617} (\bibinfo {year} {2018})}\BibitemShut {NoStop}%
\bibitem [{\citenamefont {Kitaev}\ and\ \citenamefont
  {Webb}(2008)}]{kitaev2008wavefunction}%
  \BibitemOpen
  \bibfield  {author} {\bibinfo {author} {\bibfnamefont {A.}~\bibnamefont
  {Kitaev}}\ and\ \bibinfo {author} {\bibfnamefont {W.~A.}\ \bibnamefont
  {Webb}},\ }\href@noop {} {\bibfield  {journal} {\bibinfo  {journal} {arXiv
  preprint arXiv:0801.0342}\ } (\bibinfo {year} {2008})}\BibitemShut {NoStop}%
\bibitem [{\citenamefont {Vedral}\ \emph {et~al.}(1996)\citenamefont {Vedral},
  \citenamefont {Barenco},\ and\ \citenamefont {Ekert}}]{vedral1996quantum}%
  \BibitemOpen
  \bibfield  {author} {\bibinfo {author} {\bibfnamefont {V.}~\bibnamefont
  {Vedral}}, \bibinfo {author} {\bibfnamefont {A.}~\bibnamefont {Barenco}}, \
  and\ \bibinfo {author} {\bibfnamefont {A.}~\bibnamefont {Ekert}},\
  }\href@noop {} {\bibfield  {journal} {\bibinfo  {journal} {Physical Review
  A}\ }\textbf {\bibinfo {volume} {54}},\ \bibinfo {pages} {147} (\bibinfo
  {year} {1996})}\BibitemShut {NoStop}%
\bibitem [{\citenamefont {Draper}(2000)}]{draper2000addition}%
  \BibitemOpen
  \bibfield  {author} {\bibinfo {author} {\bibfnamefont {T.~G.}\ \bibnamefont
  {Draper}},\ }\href@noop {} {\bibfield  {journal} {\bibinfo  {journal} {arXiv
  preprint quant-ph/0008033}\ } (\bibinfo {year} {2000})}\BibitemShut {NoStop}%
\bibitem [{\citenamefont {Cao}\ \emph {et~al.}(2013)\citenamefont {Cao},
  \citenamefont {Papageorgiou}, \citenamefont {Petras}, \citenamefont {Traub},\
  and\ \citenamefont {Kais}}]{Cao_2013}%
  \BibitemOpen
  \bibfield  {author} {\bibinfo {author} {\bibfnamefont {Y.}~\bibnamefont
  {Cao}}, \bibinfo {author} {\bibfnamefont {A.}~\bibnamefont {Papageorgiou}},
  \bibinfo {author} {\bibfnamefont {I.}~\bibnamefont {Petras}}, \bibinfo
  {author} {\bibfnamefont {J.}~\bibnamefont {Traub}}, \ and\ \bibinfo {author}
  {\bibfnamefont {S.}~\bibnamefont {Kais}},\ }\href {\doibase
  10.1088/1367-2630/15/1/013021} {\bibfield  {journal} {\bibinfo  {journal}
  {New Journal of Physics}\ }\textbf {\bibinfo {volume} {15}},\ \bibinfo
  {pages} {013021} (\bibinfo {year} {2013})}\BibitemShut {NoStop}%
\bibitem [{\citenamefont {Muñoz-Coreas}\ and\ \citenamefont
  {Thapliyal}(2018)}]{munozcoreas2018tcount}%
  \BibitemOpen
  \bibfield  {author} {\bibinfo {author} {\bibfnamefont {E.}~\bibnamefont
  {Muñoz-Coreas}}\ and\ \bibinfo {author} {\bibfnamefont {H.}~\bibnamefont
  {Thapliyal}},\ }\href@noop {} {\  (\bibinfo {year} {2018})},\ \Eprint
  {http://arxiv.org/abs/1712.08254} {arXiv:1712.08254 [quant-ph]} \BibitemShut
  {NoStop}%
\bibitem [{\citenamefont {Bhaskar}\ \emph {et~al.}(2015)\citenamefont
  {Bhaskar}, \citenamefont {Hadfield}, \citenamefont {Papageorgiou},\ and\
  \citenamefont {Petras}}]{bhaskar2015quantum}%
  \BibitemOpen
  \bibfield  {author} {\bibinfo {author} {\bibfnamefont {M.~K.}\ \bibnamefont
  {Bhaskar}}, \bibinfo {author} {\bibfnamefont {S.}~\bibnamefont {Hadfield}},
  \bibinfo {author} {\bibfnamefont {A.}~\bibnamefont {Papageorgiou}}, \ and\
  \bibinfo {author} {\bibfnamefont {I.}~\bibnamefont {Petras}},\ }\href@noop {}
  {\  (\bibinfo {year} {2015})},\ \Eprint {http://arxiv.org/abs/1511.08253}
  {arXiv:1511.08253 [quant-ph]} \BibitemShut {NoStop}%
\bibitem [{\citenamefont {H{\"a}ner}\ \emph {et~al.}(2018)\citenamefont
  {H{\"a}ner}, \citenamefont {Roetteler},\ and\ \citenamefont
  {Svore}}]{Hner2018OptimizingQC}%
  \BibitemOpen
  \bibfield  {author} {\bibinfo {author} {\bibfnamefont {T.}~\bibnamefont
  {H{\"a}ner}}, \bibinfo {author} {\bibfnamefont {M.}~\bibnamefont
  {Roetteler}}, \ and\ \bibinfo {author} {\bibfnamefont {K.}~\bibnamefont
  {Svore}},\ }\href@noop {} {\bibfield  {journal} {\bibinfo  {journal} {ArXiv}\
  }\textbf {\bibinfo {volume} {abs/1805.12445}} (\bibinfo {year}
  {2018})}\BibitemShut {NoStop}%
\bibitem [{\citenamefont {Zalka}(1998)}]{Zalka:1996st}%
  \BibitemOpen
  \bibfield  {author} {\bibinfo {author} {\bibfnamefont {C.}~\bibnamefont
  {Zalka}},\ }\href {\doibase 10.1098/rspa.1998.0162} {\bibfield  {journal}
  {\bibinfo  {journal} {Proc. Roy. Soc. Lond. A}\ }\textbf {\bibinfo {volume}
  {454}},\ \bibinfo {pages} {313} (\bibinfo {year} {1998})},\ \Eprint
  {http://arxiv.org/abs/quant-ph/9603026} {arXiv:quant-ph/9603026} \BibitemShut
  {NoStop}%
\bibitem [{\citenamefont {Nieto}\ and\ \citenamefont
  {Truax}(1997)}]{nieto1997holstein}%
  \BibitemOpen
  \bibfield  {author} {\bibinfo {author} {\bibfnamefont {M.~M.}\ \bibnamefont
  {Nieto}}\ and\ \bibinfo {author} {\bibfnamefont {D.~R.}\ \bibnamefont
  {Truax}},\ }\href@noop {} {\bibfield  {journal} {\bibinfo  {journal}
  {Fortschritte der Physik/Progress of Physics}\ }\textbf {\bibinfo {volume}
  {45}},\ \bibinfo {pages} {145} (\bibinfo {year} {1997})}\BibitemShut
  {NoStop}%
\bibitem [{\citenamefont {Marshall}\ \emph {et~al.}(2015)\citenamefont
  {Marshall}, \citenamefont {Pooser}, \citenamefont {Siopsis},\ and\
  \citenamefont {Weedbrook}}]{marshall2015quantum}%
  \BibitemOpen
  \bibfield  {author} {\bibinfo {author} {\bibfnamefont {K.}~\bibnamefont
  {Marshall}}, \bibinfo {author} {\bibfnamefont {R.}~\bibnamefont {Pooser}},
  \bibinfo {author} {\bibfnamefont {G.}~\bibnamefont {Siopsis}}, \ and\
  \bibinfo {author} {\bibfnamefont {C.}~\bibnamefont {Weedbrook}},\ }\href@noop
  {} {\bibfield  {journal} {\bibinfo  {journal} {Physical Review A}\ }\textbf
  {\bibinfo {volume} {92}},\ \bibinfo {pages} {063825} (\bibinfo {year}
  {2015})}\BibitemShut {NoStop}%
\bibitem [{\citenamefont {Shaw}\ \emph {et~al.}(2020)\citenamefont {Shaw},
  \citenamefont {Lougovski}, \citenamefont {Stryker},\ and\ \citenamefont
  {Wiebe}}]{shaw2020quantum}%
  \BibitemOpen
  \bibfield  {author} {\bibinfo {author} {\bibfnamefont {A.~F.}\ \bibnamefont
  {Shaw}}, \bibinfo {author} {\bibfnamefont {P.}~\bibnamefont {Lougovski}},
  \bibinfo {author} {\bibfnamefont {J.~R.}\ \bibnamefont {Stryker}}, \ and\
  \bibinfo {author} {\bibfnamefont {N.}~\bibnamefont {Wiebe}},\ }\href@noop {}
  {\bibfield  {journal} {\bibinfo  {journal} {arXiv preprint arXiv:2002.11146}\
  } (\bibinfo {year} {2020})}\BibitemShut {NoStop}%
\bibitem [{\citenamefont {Klco}\ and\ \citenamefont
  {Savage}(2019{\natexlab{b}})}]{Klco:2018zqz}%
  \BibitemOpen
  \bibfield  {author} {\bibinfo {author} {\bibfnamefont {N.}~\bibnamefont
  {Klco}}\ and\ \bibinfo {author} {\bibfnamefont {M.~J.}\ \bibnamefont
  {Savage}},\ }\href {\doibase 10.1103/PhysRevA.99.052335} {\bibfield
  {journal} {\bibinfo  {journal} {Phys. Rev.}\ }\textbf {\bibinfo {volume}
  {A99}},\ \bibinfo {pages} {052335} (\bibinfo {year} {2019}{\natexlab{b}})},\
  \Eprint {http://arxiv.org/abs/1808.10378} {arXiv:1808.10378 [quant-ph]}
  \BibitemShut {NoStop}%
\bibitem [{\citenamefont {Brun}(2019)}]{brun2019quantum}%
  \BibitemOpen
  \bibfield  {author} {\bibinfo {author} {\bibfnamefont {T.~A.}\ \bibnamefont
  {Brun}},\ }\href@noop {} {\  (\bibinfo {year} {2019})},\ \Eprint
  {http://arxiv.org/abs/1910.03672} {arXiv:1910.03672 [quant-ph]} \BibitemShut
  {NoStop}%
\bibitem [{\citenamefont {Stryker}(2019)}]{stryker2019oracles}%
  \BibitemOpen
  \bibfield  {author} {\bibinfo {author} {\bibfnamefont {J.~R.}\ \bibnamefont
  {Stryker}},\ }\href@noop {} {\bibfield  {journal} {\bibinfo  {journal}
  {Physical Review A}\ }\textbf {\bibinfo {volume} {99}},\ \bibinfo {pages}
  {042301} (\bibinfo {year} {2019})}\BibitemShut {NoStop}%
\bibitem [{\citenamefont {Tran}\ \emph
  {et~al.}(2020{\natexlab{a}})\citenamefont {Tran}, \citenamefont {Su},
  \citenamefont {Carney},\ and\ \citenamefont {Taylor}}]{tran2020faster}%
  \BibitemOpen
  \bibfield  {author} {\bibinfo {author} {\bibfnamefont {M.~C.}\ \bibnamefont
  {Tran}}, \bibinfo {author} {\bibfnamefont {Y.}~\bibnamefont {Su}}, \bibinfo
  {author} {\bibfnamefont {D.}~\bibnamefont {Carney}}, \ and\ \bibinfo {author}
  {\bibfnamefont {J.~M.}\ \bibnamefont {Taylor}},\ }\href@noop {} {\bibfield
  {journal} {\bibinfo  {journal} {arXiv preprint arXiv:2006.16248}\ } (\bibinfo
  {year} {2020}{\natexlab{a}})}\BibitemShut {NoStop}%
\bibitem [{\citenamefont {Lamm}\ \emph
  {et~al.}(2020{\natexlab{b}})\citenamefont {Lamm}, \citenamefont {Lawrence},\
  and\ \citenamefont {Yamauchi}}]{lamm2020suppressing}%
  \BibitemOpen
  \bibfield  {author} {\bibinfo {author} {\bibfnamefont {H.}~\bibnamefont
  {Lamm}}, \bibinfo {author} {\bibfnamefont {S.}~\bibnamefont {Lawrence}}, \
  and\ \bibinfo {author} {\bibfnamefont {Y.}~\bibnamefont {Yamauchi}},\
  }\href@noop {} {\bibfield  {journal} {\bibinfo  {journal} {arXiv preprint
  arXiv:2005.12688}\ } (\bibinfo {year} {2020}{\natexlab{b}})}\BibitemShut
  {NoStop}%
\bibitem [{\citenamefont {Halimeh}\ \emph {et~al.}(2020)\citenamefont
  {Halimeh}, \citenamefont {Lang}, \citenamefont {Mildenberger}, \citenamefont
  {Jiang},\ and\ \citenamefont {Hauke}}]{halimeh2020gauge}%
  \BibitemOpen
  \bibfield  {author} {\bibinfo {author} {\bibfnamefont {J.~C.}\ \bibnamefont
  {Halimeh}}, \bibinfo {author} {\bibfnamefont {H.}~\bibnamefont {Lang}},
  \bibinfo {author} {\bibfnamefont {J.}~\bibnamefont {Mildenberger}}, \bibinfo
  {author} {\bibfnamefont {Z.}~\bibnamefont {Jiang}}, \ and\ \bibinfo {author}
  {\bibfnamefont {P.}~\bibnamefont {Hauke}},\ }\href@noop {} {\bibfield
  {journal} {\bibinfo  {journal} {arXiv preprint arXiv:2007.00668}\ } (\bibinfo
  {year} {2020})}\BibitemShut {NoStop}%
\bibitem [{\citenamefont {Tran}\ \emph
  {et~al.}(2020{\natexlab{b}})\citenamefont {Tran}, \citenamefont {Chu},
  \citenamefont {Su}, \citenamefont {Childs},\ and\ \citenamefont
  {Gorshkov}}]{tran2020destructive}%
  \BibitemOpen
  \bibfield  {author} {\bibinfo {author} {\bibfnamefont {M.~C.}\ \bibnamefont
  {Tran}}, \bibinfo {author} {\bibfnamefont {S.-K.}\ \bibnamefont {Chu}},
  \bibinfo {author} {\bibfnamefont {Y.}~\bibnamefont {Su}}, \bibinfo {author}
  {\bibfnamefont {A.}~\bibnamefont {Childs}}, \ and\ \bibinfo {author}
  {\bibfnamefont {A.~V.}\ \bibnamefont {Gorshkov}},\ }\href@noop {} {\
  (\bibinfo {year} {2020}{\natexlab{b}})}\BibitemShut {NoStop}%
\bibitem [{\citenamefont {Cuccaro}\ \emph {et~al.}(2004)\citenamefont
  {Cuccaro}, \citenamefont {Draper}, \citenamefont {Kutin},\ and\ \citenamefont
  {Moulton}}]{cuccaro2004new}%
  \BibitemOpen
  \bibfield  {author} {\bibinfo {author} {\bibfnamefont {S.~A.}\ \bibnamefont
  {Cuccaro}}, \bibinfo {author} {\bibfnamefont {T.~G.}\ \bibnamefont {Draper}},
  \bibinfo {author} {\bibfnamefont {S.~A.}\ \bibnamefont {Kutin}}, \ and\
  \bibinfo {author} {\bibfnamefont {D.~P.}\ \bibnamefont {Moulton}},\
  }\href@noop {} {\bibfield  {journal} {\bibinfo  {journal} {arXiv preprint
  quant-ph/0410184}\ } (\bibinfo {year} {2004})}\BibitemShut {NoStop}%
\bibitem [{\citenamefont {Oliveira}\ and\ \citenamefont
  {Ramos}(2007)}]{oliveira2007quantum}%
  \BibitemOpen
  \bibfield  {author} {\bibinfo {author} {\bibfnamefont {D.~S.}\ \bibnamefont
  {Oliveira}}\ and\ \bibinfo {author} {\bibfnamefont {R.~V.}\ \bibnamefont
  {Ramos}},\ }\href@noop {} {\bibfield  {journal} {\bibinfo  {journal} {Quantum
  Comput. Comput}\ }\textbf {\bibinfo {volume} {7}},\ \bibinfo {pages} {17}
  (\bibinfo {year} {2007})}\BibitemShut {NoStop}%
\bibitem [{\citenamefont {Xia}\ \emph {et~al.}(2018)\citenamefont {Xia},
  \citenamefont {Li}, \citenamefont {Zhang}, \citenamefont {Liang},\ and\
  \citenamefont {Xin}}]{xia2018efficient}%
  \BibitemOpen
  \bibfield  {author} {\bibinfo {author} {\bibfnamefont {H.}~\bibnamefont
  {Xia}}, \bibinfo {author} {\bibfnamefont {H.}~\bibnamefont {Li}}, \bibinfo
  {author} {\bibfnamefont {H.}~\bibnamefont {Zhang}}, \bibinfo {author}
  {\bibfnamefont {Y.}~\bibnamefont {Liang}}, \ and\ \bibinfo {author}
  {\bibfnamefont {J.}~\bibnamefont {Xin}},\ }\href@noop {} {\bibfield
  {journal} {\bibinfo  {journal} {International Journal of Theoretical
  Physics}\ }\textbf {\bibinfo {volume} {57}},\ \bibinfo {pages} {3727}
  (\bibinfo {year} {2018})}\BibitemShut {NoStop}%
\bibitem [{\citenamefont {Guerreschi}(2019)}]{Guerreschi_2019}%
  \BibitemOpen
  \bibfield  {author} {\bibinfo {author} {\bibfnamefont {G.~G.}\ \bibnamefont
  {Guerreschi}},\ }\href {\doibase 10.1103/physreva.99.022306} {\bibfield
  {journal} {\bibinfo  {journal} {Physical Review A}\ }\textbf {\bibinfo
  {volume} {99}} (\bibinfo {year} {2019}),\
  10.1103/physreva.99.022306}\BibitemShut {NoStop}%
\bibitem [{\citenamefont {Paetznick}\ and\ \citenamefont
  {Svore}(2014)}]{paetznick2014repeatuntilsuccess}%
  \BibitemOpen
  \bibfield  {author} {\bibinfo {author} {\bibfnamefont {A.}~\bibnamefont
  {Paetznick}}\ and\ \bibinfo {author} {\bibfnamefont {K.~M.}\ \bibnamefont
  {Svore}},\ }\href@noop {} {\enquote {\bibinfo {title} {Repeat-until-success:
  Non-deterministic decomposition of single-qubit unitaries},}\ } (\bibinfo
  {year} {2014}),\ \Eprint {http://arxiv.org/abs/1311.1074} {arXiv:1311.1074
  [quant-ph]} \BibitemShut {NoStop}%
\bibitem [{\citenamefont {Mueller}(1981)}]{mueller1981multiplicity}%
  \BibitemOpen
  \bibfield  {author} {\bibinfo {author} {\bibfnamefont {A.~H.}\ \bibnamefont
  {Mueller}},\ }\href@noop {} {\bibfield  {journal} {\bibinfo  {journal}
  {Physics Letters B}\ }\textbf {\bibinfo {volume} {104}},\ \bibinfo {pages}
  {161} (\bibinfo {year} {1981})}\BibitemShut {NoStop}%
\bibitem [{\citenamefont {Dokshitzer}(1991)}]{dokshitzer1991basics}%
  \BibitemOpen
  \bibfield  {author} {\bibinfo {author} {\bibfnamefont {Y.}~\bibnamefont
  {Dokshitzer}},\ }\href@noop {} {\emph {\bibinfo {title} {Basics of
  perturbative QCD}}}\ (\bibinfo  {publisher} {Atlantica S{\'e}guier
  Fronti{\`e}res},\ \bibinfo {year} {1991})\BibitemShut {NoStop}%
\bibitem [{\citenamefont {Webber}(2000)}]{webber1999fragmentation}%
  \BibitemOpen
  \bibfield  {author} {\bibinfo {author} {\bibfnamefont {B.}~\bibnamefont
  {Webber}},\ }\href {\doibase 10.1142/S0217751X00005334} {\bibfield  {journal}
  {\bibinfo  {journal} {eConf}\ }\textbf {\bibinfo {volume} {C990809}},\
  \bibinfo {pages} {577} (\bibinfo {year} {2000})},\ \Eprint
  {http://arxiv.org/abs/hep-ph/9912292} {arXiv:hep-ph/9912292} \BibitemShut
  {NoStop}%
\bibitem [{\citenamefont {Andersson}(2005)}]{andersson2005lund}%
  \BibitemOpen
  \bibfield  {author} {\bibinfo {author} {\bibfnamefont {B.}~\bibnamefont
  {Andersson}},\ }\href@noop {} {\emph {\bibinfo {title} {The lund model}}},\
  Vol.~\bibinfo {volume} {7}\ (\bibinfo  {publisher} {Cambridge University
  Press},\ \bibinfo {year} {2005})\BibitemShut {NoStop}%
\bibitem [{\citenamefont {Cleve}\ \emph {et~al.}(1998)\citenamefont {Cleve},
  \citenamefont {Ekert}, \citenamefont {Macchiavello},\ and\ \citenamefont
  {Mosca}}]{cleve1998quantum}%
  \BibitemOpen
  \bibfield  {author} {\bibinfo {author} {\bibfnamefont {R.}~\bibnamefont
  {Cleve}}, \bibinfo {author} {\bibfnamefont {A.}~\bibnamefont {Ekert}},
  \bibinfo {author} {\bibfnamefont {C.}~\bibnamefont {Macchiavello}}, \ and\
  \bibinfo {author} {\bibfnamefont {M.}~\bibnamefont {Mosca}},\ }\href@noop {}
  {\bibfield  {journal} {\bibinfo  {journal} {Proceedings of the Royal Society
  of London. Series A: Mathematical, Physical and Engineering Sciences}\
  }\textbf {\bibinfo {volume} {454}},\ \bibinfo {pages} {339} (\bibinfo {year}
  {1998})}\BibitemShut {NoStop}%
\bibitem [{\citenamefont {Abrams}\ and\ \citenamefont
  {Lloyd}(1999)}]{abrams1999quantum}%
  \BibitemOpen
  \bibfield  {author} {\bibinfo {author} {\bibfnamefont {D.~S.}\ \bibnamefont
  {Abrams}}\ and\ \bibinfo {author} {\bibfnamefont {S.}~\bibnamefont {Lloyd}},\
  }\href@noop {} {\bibfield  {journal} {\bibinfo  {journal} {Physical Review
  Letters}\ }\textbf {\bibinfo {volume} {83}},\ \bibinfo {pages} {5162}
  (\bibinfo {year} {1999})}\BibitemShut {NoStop}%
\bibitem [{\citenamefont {Knill}\ \emph {et~al.}(2007)\citenamefont {Knill},
  \citenamefont {Ortiz},\ and\ \citenamefont {Somma}}]{knill2007optimal}%
  \BibitemOpen
  \bibfield  {author} {\bibinfo {author} {\bibfnamefont {E.}~\bibnamefont
  {Knill}}, \bibinfo {author} {\bibfnamefont {G.}~\bibnamefont {Ortiz}}, \ and\
  \bibinfo {author} {\bibfnamefont {R.~D.}\ \bibnamefont {Somma}},\ }\href@noop
  {} {\bibfield  {journal} {\bibinfo  {journal} {Physical Review A}\ }\textbf
  {\bibinfo {volume} {75}},\ \bibinfo {pages} {012328} (\bibinfo {year}
  {2007})}\BibitemShut {NoStop}%
\bibitem [{\citenamefont {Roggero}\ and\ \citenamefont
  {Baroni}(2020)}]{roggero2020short}%
  \BibitemOpen
  \bibfield  {author} {\bibinfo {author} {\bibfnamefont {A.}~\bibnamefont
  {Roggero}}\ and\ \bibinfo {author} {\bibfnamefont {A.}~\bibnamefont
  {Baroni}},\ }\href@noop {} {\bibfield  {journal} {\bibinfo  {journal}
  {Physical Review A}\ }\textbf {\bibinfo {volume} {101}},\ \bibinfo {pages}
  {022328} (\bibinfo {year} {2020})}\BibitemShut {NoStop}%
\bibitem [{\citenamefont {Mueller}\ and\ \citenamefont
  {Venugopalan}(2019{\natexlab{a}})}]{Mueller:2019gjj}%
  \BibitemOpen
  \bibfield  {author} {\bibinfo {author} {\bibfnamefont {N.}~\bibnamefont
  {Mueller}}\ and\ \bibinfo {author} {\bibfnamefont {R.}~\bibnamefont
  {Venugopalan}},\ }\href {\doibase 10.1103/PhysRevD.99.056003} {\bibfield
  {journal} {\bibinfo  {journal} {Phys. Rev.}\ }\textbf {\bibinfo {volume}
  {D99}},\ \bibinfo {pages} {056003} (\bibinfo {year} {2019}{\natexlab{a}})},\
  \Eprint {http://arxiv.org/abs/1901.10492} {arXiv:1901.10492 [hep-th]}
  \BibitemShut {NoStop}%
\bibitem [{\citenamefont {Brice{\~n}o}\ \emph {et~al.}(2020)\citenamefont
  {Brice{\~n}o}, \citenamefont {Guerrero}, \citenamefont {Hansen},\ and\
  \citenamefont {Sturzu}}]{briceno2020role}%
  \BibitemOpen
  \bibfield  {author} {\bibinfo {author} {\bibfnamefont {R.~A.}\ \bibnamefont
  {Brice{\~n}o}}, \bibinfo {author} {\bibfnamefont {J.~V.}\ \bibnamefont
  {Guerrero}}, \bibinfo {author} {\bibfnamefont {M.~T.}\ \bibnamefont
  {Hansen}}, \ and\ \bibinfo {author} {\bibfnamefont {A.}~\bibnamefont
  {Sturzu}},\ }\href@noop {} {\bibfield  {journal} {\bibinfo  {journal} {arXiv
  preprint arXiv:2007.01155}\ } (\bibinfo {year} {2020})}\BibitemShut {NoStop}%
\bibitem [{\citenamefont {Robin}\ \emph {et~al.}(2020)\citenamefont {Robin},
  \citenamefont {Savage},\ and\ \citenamefont {Pillet}}]{Robin:2020aeh}%
  \BibitemOpen
  \bibfield  {author} {\bibinfo {author} {\bibfnamefont {C.}~\bibnamefont
  {Robin}}, \bibinfo {author} {\bibfnamefont {M.~J.}\ \bibnamefont {Savage}}, \
  and\ \bibinfo {author} {\bibfnamefont {N.}~\bibnamefont {Pillet}},\
  }\href@noop {} {\  (\bibinfo {year} {2020})},\ \Eprint
  {http://arxiv.org/abs/2007.09157} {arXiv:2007.09157 [nucl-th]} \BibitemShut
  {NoStop}%
\bibitem [{\citenamefont {Kharzeev}\ and\ \citenamefont
  {Levin}(2017)}]{kharzeev2017deep}%
  \BibitemOpen
  \bibfield  {author} {\bibinfo {author} {\bibfnamefont {D.~E.}\ \bibnamefont
  {Kharzeev}}\ and\ \bibinfo {author} {\bibfnamefont {E.~M.}\ \bibnamefont
  {Levin}},\ }\href@noop {} {\bibfield  {journal} {\bibinfo  {journal}
  {Physical Review D}\ }\textbf {\bibinfo {volume} {95}},\ \bibinfo {pages}
  {114008} (\bibinfo {year} {2017})}\BibitemShut {NoStop}%
\bibitem [{\citenamefont {Hagiwara}\ \emph {et~al.}(2018)\citenamefont
  {Hagiwara}, \citenamefont {Hatta}, \citenamefont {Xiao},\ and\ \citenamefont
  {Yuan}}]{hagiwara2018classical}%
  \BibitemOpen
  \bibfield  {author} {\bibinfo {author} {\bibfnamefont {Y.}~\bibnamefont
  {Hagiwara}}, \bibinfo {author} {\bibfnamefont {Y.}~\bibnamefont {Hatta}},
  \bibinfo {author} {\bibfnamefont {B.-W.}\ \bibnamefont {Xiao}}, \ and\
  \bibinfo {author} {\bibfnamefont {F.}~\bibnamefont {Yuan}},\ }\href@noop {}
  {\bibfield  {journal} {\bibinfo  {journal} {Physical Review D}\ }\textbf
  {\bibinfo {volume} {97}},\ \bibinfo {pages} {094029} (\bibinfo {year}
  {2018})}\BibitemShut {NoStop}%
\bibitem [{\citenamefont {Kovner}\ \emph {et~al.}(2019)\citenamefont {Kovner},
  \citenamefont {Lublinsky},\ and\ \citenamefont
  {Serino}}]{kovner2019entanglement}%
  \BibitemOpen
  \bibfield  {author} {\bibinfo {author} {\bibfnamefont {A.}~\bibnamefont
  {Kovner}}, \bibinfo {author} {\bibfnamefont {M.}~\bibnamefont {Lublinsky}}, \
  and\ \bibinfo {author} {\bibfnamefont {M.}~\bibnamefont {Serino}},\
  }\href@noop {} {\bibfield  {journal} {\bibinfo  {journal} {Physics Letters
  B}\ }\textbf {\bibinfo {volume} {792}},\ \bibinfo {pages} {4} (\bibinfo
  {year} {2019})}\BibitemShut {NoStop}%
\bibitem [{\citenamefont {Tu}\ \emph {et~al.}(2020)\citenamefont {Tu},
  \citenamefont {Kharzeev},\ and\ \citenamefont {Ullrich}}]{tu2020einstein}%
  \BibitemOpen
  \bibfield  {author} {\bibinfo {author} {\bibfnamefont {Z.}~\bibnamefont
  {Tu}}, \bibinfo {author} {\bibfnamefont {D.~E.}\ \bibnamefont {Kharzeev}}, \
  and\ \bibinfo {author} {\bibfnamefont {T.}~\bibnamefont {Ullrich}},\
  }\href@noop {} {\bibfield  {journal} {\bibinfo  {journal} {Physical Review
  Letters}\ }\textbf {\bibinfo {volume} {124}},\ \bibinfo {pages} {062001}
  (\bibinfo {year} {2020})}\BibitemShut {NoStop}%
\bibitem [{\citenamefont {Berges}\ \emph
  {et~al.}(2018{\natexlab{a}})\citenamefont {Berges}, \citenamefont
  {Floerchinger},\ and\ \citenamefont {Venugopalan}}]{Berges:2017zws}%
  \BibitemOpen
  \bibfield  {author} {\bibinfo {author} {\bibfnamefont {J.}~\bibnamefont
  {Berges}}, \bibinfo {author} {\bibfnamefont {S.}~\bibnamefont
  {Floerchinger}}, \ and\ \bibinfo {author} {\bibfnamefont {R.}~\bibnamefont
  {Venugopalan}},\ }\href {\doibase 10.1016/j.physletb.2018.01.068} {\bibfield
  {journal} {\bibinfo  {journal} {Phys. Lett. B}\ }\textbf {\bibinfo {volume}
  {778}},\ \bibinfo {pages} {442} (\bibinfo {year} {2018}{\natexlab{a}})},\
  \Eprint {http://arxiv.org/abs/1707.05338} {arXiv:1707.05338 [hep-ph]}
  \BibitemShut {NoStop}%
\bibitem [{\citenamefont {Berges}\ \emph
  {et~al.}(2018{\natexlab{b}})\citenamefont {Berges}, \citenamefont
  {Floerchinger},\ and\ \citenamefont {Venugopalan}}]{Berges:2017hne}%
  \BibitemOpen
  \bibfield  {author} {\bibinfo {author} {\bibfnamefont {J.}~\bibnamefont
  {Berges}}, \bibinfo {author} {\bibfnamefont {S.}~\bibnamefont
  {Floerchinger}}, \ and\ \bibinfo {author} {\bibfnamefont {R.}~\bibnamefont
  {Venugopalan}},\ }\href {\doibase 10.1007/JHEP04(2018)145} {\bibfield
  {journal} {\bibinfo  {journal} {JHEP}\ }\textbf {\bibinfo {volume} {04}},\
  \bibinfo {pages} {145} (\bibinfo {year} {2018}{\natexlab{b}})},\ \Eprint
  {http://arxiv.org/abs/1712.09362} {arXiv:1712.09362 [hep-th]} \BibitemShut
  {NoStop}%
\bibitem [{\citenamefont {Bauer}\ \emph {et~al.}(2019)\citenamefont {Bauer},
  \citenamefont {De~Jong}, \citenamefont {Nachman},\ and\ \citenamefont
  {Provasoli}}]{Bauer:2019qxa}%
  \BibitemOpen
  \bibfield  {author} {\bibinfo {author} {\bibfnamefont {C.~W.}\ \bibnamefont
  {Bauer}}, \bibinfo {author} {\bibfnamefont {W.~A.}\ \bibnamefont {De~Jong}},
  \bibinfo {author} {\bibfnamefont {B.}~\bibnamefont {Nachman}}, \ and\
  \bibinfo {author} {\bibfnamefont {D.}~\bibnamefont {Provasoli}},\ }\href@noop
  {} {\  (\bibinfo {year} {2019})},\ \Eprint {http://arxiv.org/abs/1904.03196}
  {arXiv:1904.03196 [hep-ph]} \BibitemShut {NoStop}%
\bibitem [{\citenamefont {Beane}\ and\ \citenamefont
  {Ehlers}(2020)}]{beane2020chiral}%
  \BibitemOpen
  \bibfield  {author} {\bibinfo {author} {\bibfnamefont {S.~R.}\ \bibnamefont
  {Beane}}\ and\ \bibinfo {author} {\bibfnamefont {P.~J.}\ \bibnamefont
  {Ehlers}},\ }\href@noop {} {\bibfield  {journal} {\bibinfo  {journal} {Modern
  Physics Letters A}\ }\textbf {\bibinfo {volume} {35}},\ \bibinfo {pages}
  {2050048} (\bibinfo {year} {2020})}\BibitemShut {NoStop}%
\bibitem [{\citenamefont {Tarasov}\ and\ \citenamefont
  {Venugopalan}(2020)}]{Tarasov:2020cwl}%
  \BibitemOpen
  \bibfield  {author} {\bibinfo {author} {\bibfnamefont {A.}~\bibnamefont
  {Tarasov}}\ and\ \bibinfo {author} {\bibfnamefont {R.}~\bibnamefont
  {Venugopalan}},\ }\href@noop {} {\  (\bibinfo {year} {2020})},\ \Eprint
  {http://arxiv.org/abs/2008.08104} {arXiv:2008.08104 [hep-ph]} \BibitemShut
  {NoStop}%
\bibitem [{\citenamefont {Wegner}(1994)}]{wegner1994flow}%
  \BibitemOpen
  \bibfield  {author} {\bibinfo {author} {\bibfnamefont {F.}~\bibnamefont
  {Wegner}},\ }\href@noop {} {\bibfield  {journal} {\bibinfo  {journal}
  {Annalen der physik}\ }\textbf {\bibinfo {volume} {506}},\ \bibinfo {pages}
  {77} (\bibinfo {year} {1994})}\BibitemShut {NoStop}%
\bibitem [{\citenamefont {Perry}\ and\ \citenamefont
  {Wilson}(1993)}]{Perry:1993gp}%
  \BibitemOpen
  \bibfield  {author} {\bibinfo {author} {\bibfnamefont {R.~J.}\ \bibnamefont
  {Perry}}\ and\ \bibinfo {author} {\bibfnamefont {K.~G.}\ \bibnamefont
  {Wilson}},\ }\href {\doibase 10.1016/0550-3213(93)90363-T} {\bibfield
  {journal} {\bibinfo  {journal} {Nucl. Phys. B}\ }\textbf {\bibinfo {volume}
  {403}},\ \bibinfo {pages} {587} (\bibinfo {year} {1993})}\BibitemShut
  {NoStop}%
\bibitem [{\citenamefont {G{\l}azek}\ and\ \citenamefont
  {Wilson}(1993)}]{glazek1993renormalization}%
  \BibitemOpen
  \bibfield  {author} {\bibinfo {author} {\bibfnamefont {S.~D.}\ \bibnamefont
  {G{\l}azek}}\ and\ \bibinfo {author} {\bibfnamefont {K.~G.}\ \bibnamefont
  {Wilson}},\ }\href@noop {} {\bibfield  {journal} {\bibinfo  {journal}
  {Physical Review D}\ }\textbf {\bibinfo {volume} {48}},\ \bibinfo {pages}
  {5863} (\bibinfo {year} {1993})}\BibitemShut {NoStop}%
\bibitem [{\citenamefont {Perry}(1994)}]{perry1994renormalization}%
  \BibitemOpen
  \bibfield  {author} {\bibinfo {author} {\bibfnamefont {R.~J.}\ \bibnamefont
  {Perry}},\ }\href@noop {} {\bibfield  {journal} {\bibinfo  {journal} {Annals
  of Physics}\ }\textbf {\bibinfo {volume} {232}},\ \bibinfo {pages} {116}
  (\bibinfo {year} {1994})}\BibitemShut {NoStop}%
\bibitem [{\citenamefont {Beane}\ and\ \citenamefont
  {Farrell}(2020)}]{Beane:2020wjl}%
  \BibitemOpen
  \bibfield  {author} {\bibinfo {author} {\bibfnamefont {S.~R.}\ \bibnamefont
  {Beane}}\ and\ \bibinfo {author} {\bibfnamefont {R.~C.}\ \bibnamefont
  {Farrell}},\ }\href@noop {} {\  (\bibinfo {year} {2020})},\ \Eprint
  {http://arxiv.org/abs/2011.01278} {arXiv:2011.01278 [hep-th]} \BibitemShut
  {NoStop}%
\bibitem [{\citenamefont {Bravyi}\ \emph {et~al.}(2011)\citenamefont {Bravyi},
  \citenamefont {DiVincenzo},\ and\ \citenamefont
  {Loss}}]{bravyi2011schrieffer}%
  \BibitemOpen
  \bibfield  {author} {\bibinfo {author} {\bibfnamefont {S.}~\bibnamefont
  {Bravyi}}, \bibinfo {author} {\bibfnamefont {D.~P.}\ \bibnamefont
  {DiVincenzo}}, \ and\ \bibinfo {author} {\bibfnamefont {D.}~\bibnamefont
  {Loss}},\ }\href@noop {} {\bibfield  {journal} {\bibinfo  {journal} {Annals
  of physics}\ }\textbf {\bibinfo {volume} {326}},\ \bibinfo {pages} {2793}
  (\bibinfo {year} {2011})}\BibitemShut {NoStop}%
\bibitem [{\citenamefont {Otterbach}\ \emph {et~al.}(2017)\citenamefont
  {Otterbach}, \citenamefont {Manenti}, \citenamefont {Alidoust}, \citenamefont
  {Bestwick}, \citenamefont {Block}, \citenamefont {Bloom}, \citenamefont
  {Caldwell}, \citenamefont {Didier}, \citenamefont {Fried}, \citenamefont
  {Hong} \emph {et~al.}}]{otterbach2017unsupervised}%
  \BibitemOpen
  \bibfield  {author} {\bibinfo {author} {\bibfnamefont {J.}~\bibnamefont
  {Otterbach}}, \bibinfo {author} {\bibfnamefont {R.}~\bibnamefont {Manenti}},
  \bibinfo {author} {\bibfnamefont {N.}~\bibnamefont {Alidoust}}, \bibinfo
  {author} {\bibfnamefont {A.}~\bibnamefont {Bestwick}}, \bibinfo {author}
  {\bibfnamefont {M.}~\bibnamefont {Block}}, \bibinfo {author} {\bibfnamefont
  {B.}~\bibnamefont {Bloom}}, \bibinfo {author} {\bibfnamefont
  {S.}~\bibnamefont {Caldwell}}, \bibinfo {author} {\bibfnamefont
  {N.}~\bibnamefont {Didier}}, \bibinfo {author} {\bibfnamefont {E.~S.}\
  \bibnamefont {Fried}}, \bibinfo {author} {\bibfnamefont {S.}~\bibnamefont
  {Hong}},  \emph {et~al.},\ }\href@noop {} {\bibfield  {journal} {\bibinfo
  {journal} {arXiv preprint arXiv:1712.05771}\ } (\bibinfo {year}
  {2017})}\BibitemShut {NoStop}%
\bibitem [{\citenamefont {Peruzzo}\ \emph {et~al.}(2014)\citenamefont
  {Peruzzo}, \citenamefont {McClean}, \citenamefont {Shadbolt}, \citenamefont
  {Yung}, \citenamefont {Zhou}, \citenamefont {Love}, \citenamefont
  {Aspuru-Guzik},\ and\ \citenamefont {O’brien}}]{peruzzo2014variational}%
  \BibitemOpen
  \bibfield  {author} {\bibinfo {author} {\bibfnamefont {A.}~\bibnamefont
  {Peruzzo}}, \bibinfo {author} {\bibfnamefont {J.}~\bibnamefont {McClean}},
  \bibinfo {author} {\bibfnamefont {P.}~\bibnamefont {Shadbolt}}, \bibinfo
  {author} {\bibfnamefont {M.-H.}\ \bibnamefont {Yung}}, \bibinfo {author}
  {\bibfnamefont {X.-Q.}\ \bibnamefont {Zhou}}, \bibinfo {author}
  {\bibfnamefont {P.~J.}\ \bibnamefont {Love}}, \bibinfo {author}
  {\bibfnamefont {A.}~\bibnamefont {Aspuru-Guzik}}, \ and\ \bibinfo {author}
  {\bibfnamefont {J.~L.}\ \bibnamefont {O’brien}},\ }\href@noop {} {\bibfield
   {journal} {\bibinfo  {journal} {Nature communications}\ }\textbf {\bibinfo
  {volume} {5}},\ \bibinfo {pages} {4213} (\bibinfo {year} {2014})}\BibitemShut
  {NoStop}%
\bibitem [{\citenamefont {Kitaev}(1995)}]{kitaev1995quantum}%
  \BibitemOpen
  \bibfield  {author} {\bibinfo {author} {\bibfnamefont {A.~Y.}\ \bibnamefont
  {Kitaev}},\ }\href@noop {} {\bibfield  {journal} {\bibinfo  {journal} {arXiv
  preprint quant-ph/9511026}\ } (\bibinfo {year} {1995})}\BibitemShut {NoStop}%
\bibitem [{\citenamefont {Farhi}\ \emph {et~al.}(2000)\citenamefont {Farhi},
  \citenamefont {Goldstone}, \citenamefont {Gutmann},\ and\ \citenamefont
  {Sipser}}]{farhi2000quantum}%
  \BibitemOpen
  \bibfield  {author} {\bibinfo {author} {\bibfnamefont {E.}~\bibnamefont
  {Farhi}}, \bibinfo {author} {\bibfnamefont {J.}~\bibnamefont {Goldstone}},
  \bibinfo {author} {\bibfnamefont {S.}~\bibnamefont {Gutmann}}, \ and\
  \bibinfo {author} {\bibfnamefont {M.}~\bibnamefont {Sipser}},\ }\href@noop {}
  {\bibfield  {journal} {\bibinfo  {journal} {arXiv preprint quant-ph/0001106}\
  } (\bibinfo {year} {2000})}\BibitemShut {NoStop}%
\bibitem [{\citenamefont {Farhi}\ \emph {et~al.}(2014)\citenamefont {Farhi},
  \citenamefont {Goldstone},\ and\ \citenamefont {Gutmann}}]{farhi2014quantum}%
  \BibitemOpen
  \bibfield  {author} {\bibinfo {author} {\bibfnamefont {E.}~\bibnamefont
  {Farhi}}, \bibinfo {author} {\bibfnamefont {J.}~\bibnamefont {Goldstone}}, \
  and\ \bibinfo {author} {\bibfnamefont {S.}~\bibnamefont {Gutmann}},\
  }\href@noop {} {\bibfield  {journal} {\bibinfo  {journal} {arXiv preprint
  arXiv:1411.4028}\ } (\bibinfo {year} {2014})}\BibitemShut {NoStop}%
\bibitem [{\citenamefont {Motta}\ \emph {et~al.}(2020)\citenamefont {Motta},
  \citenamefont {Sun}, \citenamefont {Tan}, \citenamefont {O’Rourke},
  \citenamefont {Ye}, \citenamefont {Minnich}, \citenamefont {Brand{\~a}o},\
  and\ \citenamefont {Chan}}]{motta2020determining}%
  \BibitemOpen
  \bibfield  {author} {\bibinfo {author} {\bibfnamefont {M.}~\bibnamefont
  {Motta}}, \bibinfo {author} {\bibfnamefont {C.}~\bibnamefont {Sun}}, \bibinfo
  {author} {\bibfnamefont {A.~T.}\ \bibnamefont {Tan}}, \bibinfo {author}
  {\bibfnamefont {M.~J.}\ \bibnamefont {O’Rourke}}, \bibinfo {author}
  {\bibfnamefont {E.}~\bibnamefont {Ye}}, \bibinfo {author} {\bibfnamefont
  {A.~J.}\ \bibnamefont {Minnich}}, \bibinfo {author} {\bibfnamefont {F.~G.}\
  \bibnamefont {Brand{\~a}o}}, \ and\ \bibinfo {author} {\bibfnamefont
  {G.~K.-L.}\ \bibnamefont {Chan}},\ }\href@noop {} {\bibfield  {journal}
  {\bibinfo  {journal} {Nature Physics}\ }\textbf {\bibinfo {volume} {16}},\
  \bibinfo {pages} {205} (\bibinfo {year} {2020})}\BibitemShut {NoStop}%
\bibitem [{\citenamefont {Hansen}\ \emph {et~al.}(2017)\citenamefont {Hansen},
  \citenamefont {Meyer},\ and\ \citenamefont {Robaina}}]{Hansen:2017mnd}%
  \BibitemOpen
  \bibfield  {author} {\bibinfo {author} {\bibfnamefont {M.~T.}\ \bibnamefont
  {Hansen}}, \bibinfo {author} {\bibfnamefont {H.~B.}\ \bibnamefont {Meyer}}, \
  and\ \bibinfo {author} {\bibfnamefont {D.}~\bibnamefont {Robaina}},\ }\href
  {\doibase 10.1103/PhysRevD.96.094513} {\bibfield  {journal} {\bibinfo
  {journal} {Phys. Rev. D}\ }\textbf {\bibinfo {volume} {96}},\ \bibinfo
  {pages} {094513} (\bibinfo {year} {2017})},\ \Eprint
  {http://arxiv.org/abs/1704.08993} {arXiv:1704.08993 [hep-lat]} \BibitemShut
  {NoStop}%
\bibitem [{\citenamefont {Girvin}(2011)}]{girvin2011circuit}%
  \BibitemOpen
  \bibfield  {author} {\bibinfo {author} {\bibfnamefont {S.~M.}\ \bibnamefont
  {Girvin}},\ }\href@noop {} {\bibfield  {journal} {\bibinfo  {journal}
  {Quantum machines: measurement and control of engineered quantum systems}\
  }\textbf {\bibinfo {volume} {113}},\ \bibinfo {pages} {2} (\bibinfo {year}
  {2011})}\BibitemShut {NoStop}%
\bibitem [{\citenamefont {Childs}\ \emph {et~al.}(2021)\citenamefont {Childs},
  \citenamefont {Su}, \citenamefont {Tran}, \citenamefont {Wiebe},\ and\
  \citenamefont {Zhu}}]{Childs_2021}%
  \BibitemOpen
  \bibfield  {author} {\bibinfo {author} {\bibfnamefont {A.~M.}\ \bibnamefont
  {Childs}}, \bibinfo {author} {\bibfnamefont {Y.}~\bibnamefont {Su}}, \bibinfo
  {author} {\bibfnamefont {M.~C.}\ \bibnamefont {Tran}}, \bibinfo {author}
  {\bibfnamefont {N.}~\bibnamefont {Wiebe}}, \ and\ \bibinfo {author}
  {\bibfnamefont {S.}~\bibnamefont {Zhu}},\ }\href {\doibase
  10.1103/physrevx.11.011020} {\bibfield  {journal} {\bibinfo  {journal}
  {Physical Review X}\ }\textbf {\bibinfo {volume} {11}} (\bibinfo {year}
  {2021}),\ 10.1103/physrevx.11.011020}\BibitemShut {NoStop}%
\bibitem [{\citenamefont {Steane}(1996)}]{steane1996multiple}%
  \BibitemOpen
  \bibfield  {author} {\bibinfo {author} {\bibfnamefont {A.}~\bibnamefont
  {Steane}},\ }\href@noop {} {\bibfield  {journal} {\bibinfo  {journal}
  {Proceedings of the Royal Society of London. Series A: Mathematical, Physical
  and Engineering Sciences}\ }\textbf {\bibinfo {volume} {452}},\ \bibinfo
  {pages} {2551} (\bibinfo {year} {1996})}\BibitemShut {NoStop}%
\bibitem [{\citenamefont {Steane}(2007)}]{steane2007tutorial}%
  \BibitemOpen
  \bibfield  {author} {\bibinfo {author} {\bibfnamefont {A.~M.}\ \bibnamefont
  {Steane}},\ }in\ \href@noop {} {\emph {\bibinfo {booktitle}
  {PROCEEDINGS-INTERNATIONAL SCHOOL OF PHYSICS ENRICO FERMI}}},\ Vol.\ \bibinfo
  {volume} {162}\ (\bibinfo {organization} {IOS Press; Ohmsha; 1999},\ \bibinfo
  {year} {2007})\ p.~\bibinfo {pages} {1}\BibitemShut {NoStop}%
\bibitem [{\citenamefont {Calderbank}\ and\ \citenamefont
  {Shor}(1996)}]{calderbank1996good}%
  \BibitemOpen
  \bibfield  {author} {\bibinfo {author} {\bibfnamefont {A.~R.}\ \bibnamefont
  {Calderbank}}\ and\ \bibinfo {author} {\bibfnamefont {P.~W.}\ \bibnamefont
  {Shor}},\ }\href@noop {} {\bibfield  {journal} {\bibinfo  {journal} {Physical
  Review A}\ }\textbf {\bibinfo {volume} {54}},\ \bibinfo {pages} {1098}
  (\bibinfo {year} {1996})}\BibitemShut {NoStop}%
\bibitem [{\citenamefont {Yeter-Aydeniz}\ and\ \citenamefont
  {Siopsis}(2018{\natexlab{b}})}]{Yeter-Aydeniz:2017ubh}%
  \BibitemOpen
  \bibfield  {author} {\bibinfo {author} {\bibfnamefont {K.}~\bibnamefont
  {Yeter-Aydeniz}}\ and\ \bibinfo {author} {\bibfnamefont {G.}~\bibnamefont
  {Siopsis}},\ }\href {\doibase 10.1103/PhysRevD.97.036004} {\bibfield
  {journal} {\bibinfo  {journal} {Phys. Rev.}\ }\textbf {\bibinfo {volume}
  {D97}},\ \bibinfo {pages} {036004} (\bibinfo {year} {2018}{\natexlab{b}})},\
  \Eprint {http://arxiv.org/abs/1709.02355} {arXiv:1709.02355 [quant-ph]}
  \BibitemShut {NoStop}%
\bibitem [{\citenamefont {Macridin}\ \emph
  {et~al.}(2018{\natexlab{a}})\citenamefont {Macridin}, \citenamefont
  {Spentzouris}, \citenamefont {Amundson},\ and\ \citenamefont
  {Harnik}}]{Macridin:2018gdw}%
  \BibitemOpen
  \bibfield  {author} {\bibinfo {author} {\bibfnamefont {A.}~\bibnamefont
  {Macridin}}, \bibinfo {author} {\bibfnamefont {P.}~\bibnamefont
  {Spentzouris}}, \bibinfo {author} {\bibfnamefont {J.}~\bibnamefont
  {Amundson}}, \ and\ \bibinfo {author} {\bibfnamefont {R.}~\bibnamefont
  {Harnik}},\ }\href {\doibase 10.1103/PhysRevLett.121.110504} {\bibfield
  {journal} {\bibinfo  {journal} {Phys. Rev. Lett.}\ }\textbf {\bibinfo
  {volume} {121}},\ \bibinfo {pages} {110504} (\bibinfo {year}
  {2018}{\natexlab{a}})},\ \Eprint {http://arxiv.org/abs/1802.07347}
  {arXiv:1802.07347 [quant-ph]} \BibitemShut {NoStop}%
\bibitem [{\citenamefont {Macridin}\ \emph
  {et~al.}(2018{\natexlab{b}})\citenamefont {Macridin}, \citenamefont
  {Spentzouris}, \citenamefont {Amundson},\ and\ \citenamefont
  {Harnik}}]{Macridin:2018oli}%
  \BibitemOpen
  \bibfield  {author} {\bibinfo {author} {\bibfnamefont {A.}~\bibnamefont
  {Macridin}}, \bibinfo {author} {\bibfnamefont {P.}~\bibnamefont
  {Spentzouris}}, \bibinfo {author} {\bibfnamefont {J.}~\bibnamefont
  {Amundson}}, \ and\ \bibinfo {author} {\bibfnamefont {R.}~\bibnamefont
  {Harnik}},\ }\href {\doibase 10.1103/PhysRevA.98.042312} {\bibfield
  {journal} {\bibinfo  {journal} {Phys. Rev.}\ }\textbf {\bibinfo {volume}
  {A98}},\ \bibinfo {pages} {042312} (\bibinfo {year} {2018}{\natexlab{b}})},\
  \Eprint {http://arxiv.org/abs/1805.09928} {arXiv:1805.09928 [quant-ph]}
  \BibitemShut {NoStop}%
\bibitem [{\citenamefont {Mueller}\ and\ \citenamefont
  {Venugopalan}(2019{\natexlab{b}})}]{mueller2019constructing}%
  \BibitemOpen
  \bibfield  {author} {\bibinfo {author} {\bibfnamefont {N.}~\bibnamefont
  {Mueller}}\ and\ \bibinfo {author} {\bibfnamefont {R.}~\bibnamefont
  {Venugopalan}},\ }\href@noop {} {\bibfield  {journal} {\bibinfo  {journal}
  {Physical Review D}\ }\textbf {\bibinfo {volume} {99}},\ \bibinfo {pages}
  {056003} (\bibinfo {year} {2019}{\natexlab{b}})}\BibitemShut {NoStop}%
\bibitem [{\citenamefont {Brower}\ \emph {et~al.}(1999)\citenamefont {Brower},
  \citenamefont {Chandrasekharan},\ and\ \citenamefont
  {Wiese}}]{brower1999qcd}%
  \BibitemOpen
  \bibfield  {author} {\bibinfo {author} {\bibfnamefont {R.}~\bibnamefont
  {Brower}}, \bibinfo {author} {\bibfnamefont {S.}~\bibnamefont
  {Chandrasekharan}}, \ and\ \bibinfo {author} {\bibfnamefont {U.-J.}\
  \bibnamefont {Wiese}},\ }\href@noop {} {\bibfield  {journal} {\bibinfo
  {journal} {Physical Review D}\ }\textbf {\bibinfo {volume} {60}},\ \bibinfo
  {pages} {094502} (\bibinfo {year} {1999})}\BibitemShut {NoStop}%
\bibitem [{\citenamefont {Banerjee}\ \emph {et~al.}(2013)\citenamefont
  {Banerjee}, \citenamefont {B{\"o}gli}, \citenamefont {Dalmonte},
  \citenamefont {Rico}, \citenamefont {Stebler}, \citenamefont {Wiese},\ and\
  \citenamefont {Zoller}}]{banerjee2013atomic}%
  \BibitemOpen
  \bibfield  {author} {\bibinfo {author} {\bibfnamefont {D.}~\bibnamefont
  {Banerjee}}, \bibinfo {author} {\bibfnamefont {M.}~\bibnamefont {B{\"o}gli}},
  \bibinfo {author} {\bibfnamefont {M.}~\bibnamefont {Dalmonte}}, \bibinfo
  {author} {\bibfnamefont {E.}~\bibnamefont {Rico}}, \bibinfo {author}
  {\bibfnamefont {P.}~\bibnamefont {Stebler}}, \bibinfo {author} {\bibfnamefont
  {U.-J.}\ \bibnamefont {Wiese}}, \ and\ \bibinfo {author} {\bibfnamefont
  {P.}~\bibnamefont {Zoller}},\ }\href@noop {} {\bibfield  {journal} {\bibinfo
  {journal} {Physical review letters}\ }\textbf {\bibinfo {volume} {110}},\
  \bibinfo {pages} {125303} (\bibinfo {year} {2013})}\BibitemShut {NoStop}%
\bibitem [{\citenamefont {Zohar}\ \emph {et~al.}(2015)\citenamefont {Zohar},
  \citenamefont {Cirac},\ and\ \citenamefont {Reznik}}]{zohar2015quantum}%
  \BibitemOpen
  \bibfield  {author} {\bibinfo {author} {\bibfnamefont {E.}~\bibnamefont
  {Zohar}}, \bibinfo {author} {\bibfnamefont {J.~I.}\ \bibnamefont {Cirac}}, \
  and\ \bibinfo {author} {\bibfnamefont {B.}~\bibnamefont {Reznik}},\
  }\href@noop {} {\bibfield  {journal} {\bibinfo  {journal} {Reports on
  Progress in Physics}\ }\textbf {\bibinfo {volume} {79}},\ \bibinfo {pages}
  {014401} (\bibinfo {year} {2015})}\BibitemShut {NoStop}%
\bibitem [{\citenamefont {Klco}\ \emph {et~al.}(2020)\citenamefont {Klco},
  \citenamefont {Stryker},\ and\ \citenamefont {Savage}}]{Klco:2019evd}%
  \BibitemOpen
  \bibfield  {author} {\bibinfo {author} {\bibfnamefont {N.}~\bibnamefont
  {Klco}}, \bibinfo {author} {\bibfnamefont {J.~R.}\ \bibnamefont {Stryker}}, \
  and\ \bibinfo {author} {\bibfnamefont {M.~J.}\ \bibnamefont {Savage}},\
  }\href {\doibase 10.1103/PhysRevD.101.074512} {\bibfield  {journal} {\bibinfo
   {journal} {Phys. Rev. D}\ }\textbf {\bibinfo {volume} {101}},\ \bibinfo
  {pages} {074512} (\bibinfo {year} {2020})},\ \Eprint
  {http://arxiv.org/abs/1908.06935} {arXiv:1908.06935 [quant-ph]} \BibitemShut
  {NoStop}%
\bibitem [{\citenamefont {Kasper}\ \emph {et~al.}(2020)\citenamefont {Kasper},
  \citenamefont {Juzeliunas}, \citenamefont {Lewenstein}, \citenamefont
  {Jendrzejewski},\ and\ \citenamefont {Zohar}}]{kasper2020jaynes}%
  \BibitemOpen
  \bibfield  {author} {\bibinfo {author} {\bibfnamefont {V.}~\bibnamefont
  {Kasper}}, \bibinfo {author} {\bibfnamefont {G.}~\bibnamefont {Juzeliunas}},
  \bibinfo {author} {\bibfnamefont {M.}~\bibnamefont {Lewenstein}}, \bibinfo
  {author} {\bibfnamefont {F.}~\bibnamefont {Jendrzejewski}}, \ and\ \bibinfo
  {author} {\bibfnamefont {E.}~\bibnamefont {Zohar}},\ }\href@noop {}
  {\bibfield  {journal} {\bibinfo  {journal} {arXiv preprint arXiv:2006.01258}\
  } (\bibinfo {year} {2020})}\BibitemShut {NoStop}%
\bibitem [{\citenamefont {Davoudi}\ \emph {et~al.}(2020)\citenamefont
  {Davoudi}, \citenamefont {Raychowdhury},\ and\ \citenamefont
  {Shaw}}]{Davoudi:2020yln}%
  \BibitemOpen
  \bibfield  {author} {\bibinfo {author} {\bibfnamefont {Z.}~\bibnamefont
  {Davoudi}}, \bibinfo {author} {\bibfnamefont {I.}~\bibnamefont
  {Raychowdhury}}, \ and\ \bibinfo {author} {\bibfnamefont {A.}~\bibnamefont
  {Shaw}},\ }\href@noop {} {\  (\bibinfo {year} {2020})},\ \Eprint
  {http://arxiv.org/abs/2009.11802} {arXiv:2009.11802 [hep-lat]} \BibitemShut
  {NoStop}%
\bibitem [{\citenamefont {Dasgupta}\ and\ \citenamefont
  {Raychowdhury}(2020)}]{dasgupta2020cold}%
  \BibitemOpen
  \bibfield  {author} {\bibinfo {author} {\bibfnamefont {R.}~\bibnamefont
  {Dasgupta}}\ and\ \bibinfo {author} {\bibfnamefont {I.}~\bibnamefont
  {Raychowdhury}},\ }\href@noop {} {\bibfield  {journal} {\bibinfo  {journal}
  {arXiv preprint arXiv:2009.13969}\ } (\bibinfo {year} {2020})}\BibitemShut
  {NoStop}%
\bibitem [{\citenamefont {Accardi}\ \emph {et~al.}(2016)\citenamefont
  {Accardi}, \citenamefont {Albacete}, \citenamefont {Anselmino}, \citenamefont
  {Armesto}, \citenamefont {Aschenauer}, \citenamefont {Bacchetta},
  \citenamefont {Boer}, \citenamefont {Brooks}, \citenamefont {Burton},
  \citenamefont {Chang} \emph {et~al.}}]{accardi2016electron}%
  \BibitemOpen
  \bibfield  {author} {\bibinfo {author} {\bibfnamefont {A.}~\bibnamefont
  {Accardi}}, \bibinfo {author} {\bibfnamefont {J.}~\bibnamefont {Albacete}},
  \bibinfo {author} {\bibfnamefont {M.}~\bibnamefont {Anselmino}}, \bibinfo
  {author} {\bibfnamefont {N.}~\bibnamefont {Armesto}}, \bibinfo {author}
  {\bibfnamefont {E.}~\bibnamefont {Aschenauer}}, \bibinfo {author}
  {\bibfnamefont {A.}~\bibnamefont {Bacchetta}}, \bibinfo {author}
  {\bibfnamefont {D.}~\bibnamefont {Boer}}, \bibinfo {author} {\bibfnamefont
  {W.}~\bibnamefont {Brooks}}, \bibinfo {author} {\bibfnamefont
  {T.}~\bibnamefont {Burton}}, \bibinfo {author} {\bibfnamefont {N.-B.}\
  \bibnamefont {Chang}},  \emph {et~al.},\ }\href@noop {} {\bibfield  {journal}
  {\bibinfo  {journal} {The European Physical Journal A}\ }\textbf {\bibinfo
  {volume} {52}},\ \bibinfo {pages} {268} (\bibinfo {year} {2016})}\BibitemShut
  {NoStop}%
\bibitem [{\citenamefont {Zhang}\ \emph {et~al.}(2020)\citenamefont {Zhang},
  \citenamefont {Xing}, \citenamefont {Yan}, \citenamefont {Wang},\ and\
  \citenamefont {Zhu}}]{Zhang:2020uqo}%
  \BibitemOpen
  \bibfield  {author} {\bibinfo {author} {\bibfnamefont {D.-B.}\ \bibnamefont
  {Zhang}}, \bibinfo {author} {\bibfnamefont {H.}~\bibnamefont {Xing}},
  \bibinfo {author} {\bibfnamefont {H.}~\bibnamefont {Yan}}, \bibinfo {author}
  {\bibfnamefont {E.}~\bibnamefont {Wang}}, \ and\ \bibinfo {author}
  {\bibfnamefont {S.-L.}\ \bibnamefont {Zhu}},\ }\href@noop {} {\  (\bibinfo
  {year} {2020})},\ \Eprint {http://arxiv.org/abs/2011.01431} {arXiv:2011.01431
  [quant-ph]} \BibitemShut {NoStop}%
\bibitem [{\citenamefont {Krasnitz}\ and\ \citenamefont
  {Venugopalan}(1999)}]{krasnitz1999non}%
  \BibitemOpen
  \bibfield  {author} {\bibinfo {author} {\bibfnamefont {A.}~\bibnamefont
  {Krasnitz}}\ and\ \bibinfo {author} {\bibfnamefont {R.}~\bibnamefont
  {Venugopalan}},\ }\href@noop {} {\bibfield  {journal} {\bibinfo  {journal}
  {Nuclear Physics B}\ }\textbf {\bibinfo {volume} {557}},\ \bibinfo {pages}
  {237} (\bibinfo {year} {1999})}\BibitemShut {NoStop}%
\bibitem [{\citenamefont {Berges}\ \emph {et~al.}(2012)\citenamefont {Berges},
  \citenamefont {Schlichting},\ and\ \citenamefont
  {Sexty}}]{berges2012overpopulated}%
  \BibitemOpen
  \bibfield  {author} {\bibinfo {author} {\bibfnamefont {J.}~\bibnamefont
  {Berges}}, \bibinfo {author} {\bibfnamefont {S.}~\bibnamefont {Schlichting}},
  \ and\ \bibinfo {author} {\bibfnamefont {D.}~\bibnamefont {Sexty}},\
  }\href@noop {} {\bibfield  {journal} {\bibinfo  {journal} {Physical Review
  D}\ }\textbf {\bibinfo {volume} {86}},\ \bibinfo {pages} {074006} (\bibinfo
  {year} {2012})}\BibitemShut {NoStop}%
\bibitem [{\citenamefont {Mace}\ \emph {et~al.}(2020)\citenamefont {Mace},
  \citenamefont {Mueller}, \citenamefont {Schlichting},\ and\ \citenamefont
  {Sharma}}]{mace2020chiral}%
  \BibitemOpen
  \bibfield  {author} {\bibinfo {author} {\bibfnamefont {M.}~\bibnamefont
  {Mace}}, \bibinfo {author} {\bibfnamefont {N.}~\bibnamefont {Mueller}},
  \bibinfo {author} {\bibfnamefont {S.}~\bibnamefont {Schlichting}}, \ and\
  \bibinfo {author} {\bibfnamefont {S.}~\bibnamefont {Sharma}},\ }\href@noop {}
  {\bibfield  {journal} {\bibinfo  {journal} {Physical Review Letters}\
  }\textbf {\bibinfo {volume} {124}},\ \bibinfo {pages} {191604} (\bibinfo
  {year} {2020})}\BibitemShut {NoStop}%
\bibitem [{\citenamefont {Turchette}\ \emph {et~al.}(1995)\citenamefont
  {Turchette}, \citenamefont {Hood}, \citenamefont {Lange}, \citenamefont
  {Mabuchi},\ and\ \citenamefont {Kimble}}]{Phase_shift}%
  \BibitemOpen
  \bibfield  {author} {\bibinfo {author} {\bibfnamefont {Q.}~\bibnamefont
  {Turchette}}, \bibinfo {author} {\bibfnamefont {C.}~\bibnamefont {Hood}},
  \bibinfo {author} {\bibfnamefont {W.}~\bibnamefont {Lange}}, \bibinfo
  {author} {\bibfnamefont {H.}~\bibnamefont {Mabuchi}}, \ and\ \bibinfo
  {author} {\bibfnamefont {H.}~\bibnamefont {Kimble}},\ }\href {\doibase
  10.1103/PhysRevLett.75.4710} {\bibfield  {journal} {\bibinfo  {journal}
  {Phys. Rev. Lett.}\ }\textbf {\bibinfo {volume} {75}},\ \bibinfo {pages}
  {4710} (\bibinfo {year} {1995})},\ \Eprint
  {http://arxiv.org/abs/quant-ph/9511008} {arXiv:quant-ph/9511008} \BibitemShut
  {NoStop}%
\bibitem [{\citenamefont {Kaye}(2005)}]{Kaye2005OptimizedQI}%
  \BibitemOpen
  \bibfield  {author} {\bibinfo {author} {\bibfnamefont {P.}~\bibnamefont
  {Kaye}},\ }\href@noop {} {\bibfield  {journal} {\bibinfo  {journal} {Quantum
  Inf. Comput.}\ }\textbf {\bibinfo {volume} {5}},\ \bibinfo {pages} {474}
  (\bibinfo {year} {2005})},\ \Eprint {http://arxiv.org/abs/quant-ph/0407095}
  {arXiv:quant-ph/0407095 [quant-ph]} \BibitemShut {NoStop}%
\bibitem [{\citenamefont {Kaye}(2004)}]{Kaye2004ReversibleAC}%
  \BibitemOpen
  \bibfield  {author} {\bibinfo {author} {\bibfnamefont {P.}~\bibnamefont
  {Kaye}},\ }\href@noop {} {\  (\bibinfo {year} {2004})},\ \Eprint
  {http://arxiv.org/abs/quant-ph/0408173} {arXiv:quant-ph/0408173 [quant-ph]}
  \BibitemShut {NoStop}%
\bibitem [{\citenamefont {Zakharov}\ \emph {et~al.}(1992)\citenamefont
  {Zakharov}, \citenamefont {Lvov},\ and\ \citenamefont
  {Falkovich}}]{Zakharov}%
  \BibitemOpen
  \bibfield  {author} {\bibinfo {author} {\bibfnamefont {V.}~\bibnamefont
  {Zakharov}}, \bibinfo {author} {\bibfnamefont {V.}~\bibnamefont {Lvov}}, \
  and\ \bibinfo {author} {\bibfnamefont {G.}~\bibnamefont {Falkovich}},\
  }\href@noop {} {\emph {\bibinfo {title} {Kolmogorov Spectra of Turbulence I:
  Wave Turbulence}}}\ (\bibinfo  {publisher} {Springer Verlag},\ \bibinfo
  {year} {1992})\BibitemShut {NoStop}%
\bibitem [{\citenamefont {Deng}\ \emph {et~al.}(2018)\citenamefont {Deng},
  \citenamefont {Schlichting}, \citenamefont {Venugopalan},\ and\ \citenamefont
  {Wang}}]{deng2018off}%
  \BibitemOpen
  \bibfield  {author} {\bibinfo {author} {\bibfnamefont {J.}~\bibnamefont
  {Deng}}, \bibinfo {author} {\bibfnamefont {S.}~\bibnamefont {Schlichting}},
  \bibinfo {author} {\bibfnamefont {R.}~\bibnamefont {Venugopalan}}, \ and\
  \bibinfo {author} {\bibfnamefont {Q.}~\bibnamefont {Wang}},\ }\href@noop {}
  {\bibfield  {journal} {\bibinfo  {journal} {Physical Review A}\ }\textbf
  {\bibinfo {volume} {97}},\ \bibinfo {pages} {053606} (\bibinfo {year}
  {2018})}\BibitemShut {NoStop}%
\bibitem [{\citenamefont {Bogner}\ \emph {et~al.}(2010)\citenamefont {Bogner},
  \citenamefont {Furnstahl},\ and\ \citenamefont {Schwenk}}]{bogner2010low}%
  \BibitemOpen
  \bibfield  {author} {\bibinfo {author} {\bibfnamefont {S.}~\bibnamefont
  {Bogner}}, \bibinfo {author} {\bibfnamefont {R.}~\bibnamefont {Furnstahl}}, \
  and\ \bibinfo {author} {\bibfnamefont {A.}~\bibnamefont {Schwenk}},\
  }\href@noop {} {\bibfield  {journal} {\bibinfo  {journal} {Progress in
  Particle and Nuclear Physics}\ }\textbf {\bibinfo {volume} {65}},\ \bibinfo
  {pages} {94} (\bibinfo {year} {2010})}\BibitemShut {NoStop}%
\end{thebibliography}%
%
%
%
\appendix
%
%
\section{Single-particle digitization scheme}\label{app:detailsSingleParticle}
In this Appendix, we provide additional details on the single-particle digitization strategy introduced in Section \ref{sec:strategies_for_quantum_simulation},
based on mapping single particle states to a chain of spins \Eqs{eq:stateBinaryDecomp}{eq:emptystate}, where
\begin{align}\label{eq:BosondecompositionFull}
a_{\mathbf{q}}^\dagger \equiv \frac{1}{\sqrt{M}} \sum_{i=0}^{M-1} a_{\mathbf{q}}^{(i)\dagger} \, ,
\end{align}
and similarly for $a_{\mathbf{q}}$. Here $a_{\mathbf{q}}^{(i)\dagger}$, $a_{\mathbf{q}}^{(i)}$ are ``hard-core boson'' creation (annihilation) operators
which can be written as a product of spin raising (lowering) operators $S^\pm =1/2( \sigma^x \pm i\sigma^y)$. A simple example is a digitization with $N=4$ qubits per particle register in $d=1$ dimensions, where there are eight ``occupied'' states with $\textbf{q}\in [-7/2,7/2]$,
\begin{align}
| \pm 1/2 \rangle \equiv | \da\da; \ua/\da ; \ua\rangle  \,,
\quad
| \pm 3/2 \rangle \equiv | \da\ua; \ua/\da ; \ua\rangle \,, 
\quad
| \pm 5/2 \rangle \equiv | \ua\da; \ua/\da ; \ua\rangle  
\,,\quad
| \pm 7/2 \rangle \equiv | \ua\ua; \ua/\da ; \ua\rangle  \,,
\end{align}
and the empty state $| \Omega \rangle = | \da\da; \da; \da\rangle$. Fock operators are then
\begin{align}\label{eq:a_S_map}
&a_{- 1/2}^{(i)\dagger}  \equiv  S_0^+ \,, 
\qquad
a_{- 3/2}^{(i)\dagger}  \equiv  S_2^+ S_0^+ \,,
\qquad
a_{- 5/2}^{(i)\dagger}  \equiv  S_3^+ S_0^+ \,,
\qquad
a_{- 7/2}^{(i)\dagger}  \equiv  S_3^+ S_2^+ S_0^+  \, ,
\end{align}
where $a^{(i)\dagger}_{+|\textbf{q}|} = S^+_1 a^{(i)\dagger}_{-|\textbf{q}|}$. We label the k=0,\dots,3 qubits from right to left so that $k=0$ labels the occupation number qubit, $k=1$ the sign qubit and $k=2,3$ are the binary decomposition of $|\textbf{q}|$. We use the identical map for states in the position representation.

One can check that $a_{\textbf{q}}^{(i)\dagger} | \Omega^{(i)} \rangle = | \textbf{q}^{(i)} \rangle $ and $(a_{\textbf{q}}^{(i)\dagger})^2=(a_{\textbf{q}}^{(i)})^2=0$. Using \Eq{eq:BosondecompositionFull}, one can also show that $[a_{\textbf{q}}^\dagger,a_{\textbf{q}'}^\dagger]=[a_{\textbf{q}},a_{\textbf{q}'}]=0$ and 
\begin{align}\label{eq:algebra}
[a_{\textbf{q}},a_{\textbf{q}}^\dagger]&=\frac{1}{M}\sum_{i=0}^{M-1} \Big[    \{a^{(i)}_{\textbf{q}},a_{\textbf{q}}^{(i)\dagger}\}-2 a^{(i)\dagger}_{\textbf{q}}a^{(i)}_{\textbf{q}}\Big]
=1+O\left(\frac{\mathfrak{n}_{\textbf{q}}}{M}\right) \, ,
\end{align}
where $1$ is a unit matrix in the space spanned by $| \textbf{q} \rangle$ and $| \Omega \rangle$, as well as $[a_{\textbf{q}},a_{\textbf{q}'}^\dagger]=O({\mathfrak{n}_{\textbf{q}}}/{M})$ where $\mathfrak{n}_{\textbf{q}}$ is the occupation number of the mode $\textbf{q}$.
\section{Details of State Preparation}\label{app:StochasticStatePrep}
 We will present here details of the initial state preparation algorithm in  Section \ref{sec:InitialStatePrep}. We begin by discussing the preparation of a wavepacket superposition via the algorithm of~\cite{grover2002creating,kaye2004quantum} and
contrast it with a simpler, albeit less general, variant. For simplicity, we work in $d=1$, and use the standard binary representation, not the ``inverted'' one used in the main text. Assuming a symmetric distribution in momentum, the first Hadamard operation on the sign qubit creates and equal superposition of negative and positive momenta. Below, we illustrate
the algorithm acting on the qubits representing the absolute value of
momentum $p$.

Starting from a fiducial state with all the qubits in $|0\rangle$, our algorithm applies the following (per qubit) operation\footnote{We now work in the standard binary representation.}
\begin{equation}
|0\rangle \to \cos \theta_k|0\rangle  +  \sin \theta_k|0\rangle
\end{equation}
for all $k=0,\dots,n_Q-1$ qubits, so that to each $|1\rangle$ gets multiplied with a sine and to each $|0\rangle$ with a cosine. Then the final state (for $n_Q$ qubit) reads 
\begin{align}
|0,0,\cdots,0\rangle\to \sum_{p=0}^{2^{n_Q}-1}\left\{ \prod_{k=0}^{n_Q-1}    \left(\cos^{(1-p_k)}(\theta_k)\sin^{(p_k)}(\theta_k)\right)\right\} |p\rangle=\sum_{p=0}^{2^{n_Q}-1} \psi_p|p\rangle\equiv |\Psi\rangle\, ,
\end{align}
where $|p\rangle$ here stands for the $n_Q$ qubits storing the absolute value of a single particle and $p_k\in\{0,1\}$.

Adjusting the map $k\rightarrow \theta_k$ classically, one can reproduce
a wide range of distributions. For example, choosing $\theta_k = \pi/4$ up to some $k'$ and $\theta_k = 0$ thereafter would produce a step function.
While this distribution is localized (in momentum space), its Fourier conjugate is $\sin(x)/x$ (in position space), which falls off only polynomial, and is thus undesired.

One can however produce sufficiently smooth distributions that fall off exponentially in position and momentum space. Simple examples of this are shown in
the figure below, where for illustration we have chosen the following maps
\begin{align}\label{eq:a}
\theta_k^{\rm linear}&=\frac{\pi}{4}-\epsilon+\frac{2\epsilon-\frac{\pi}{4}}{n_Q-1} k \, ,    \\
\theta_k^{\rm quadratic}&=\frac{\pi}{4}-\epsilon+ \left(2\epsilon-\frac{\pi}{4}-c_0(n_Q - 1)^2\right)\frac{k}{n_Q-1} +c_0 k^2  \, , \\
\theta_k^{\rm cubic}&=\frac{\pi}{4}-\epsilon+ \left(2\epsilon-\frac{\pi}{4}-c_1(n_Q - 1)^2-c_2(n_Q - 1)^3\right)\frac{k}{n_Q-1} +c_1 k^2+c_2 k^3  \, .\label{eq:c}
\end{align}
The $c_i$ parameters are adjusted such that the resulting distribution is smoothed (in the sense of having less and smaller peaks); we took $c_0=-0.01325$, $c_1=-0.0195$, $c_2=.0005905$, the numerical regulator $\epsilon=0.015$ and $n_Q=10$. These maps are fixed at the initial point $p=0$ where $\theta_0=\pi/4-\epsilon$ and $\theta_{n_Q-1}=\epsilon$ is the smallest possible value.  The resulting distributions decay exponentially, $\sim \exp(-p/\sigma)$ as was desired. We note that the use of these simple low order polynomials leads to a roughness of the curves, which can be smoothed by use of
higher order polynomials as is shown in the figure.

 \begin{figure}[h]
\begin{center}
\includegraphics[width=0.42\textwidth]{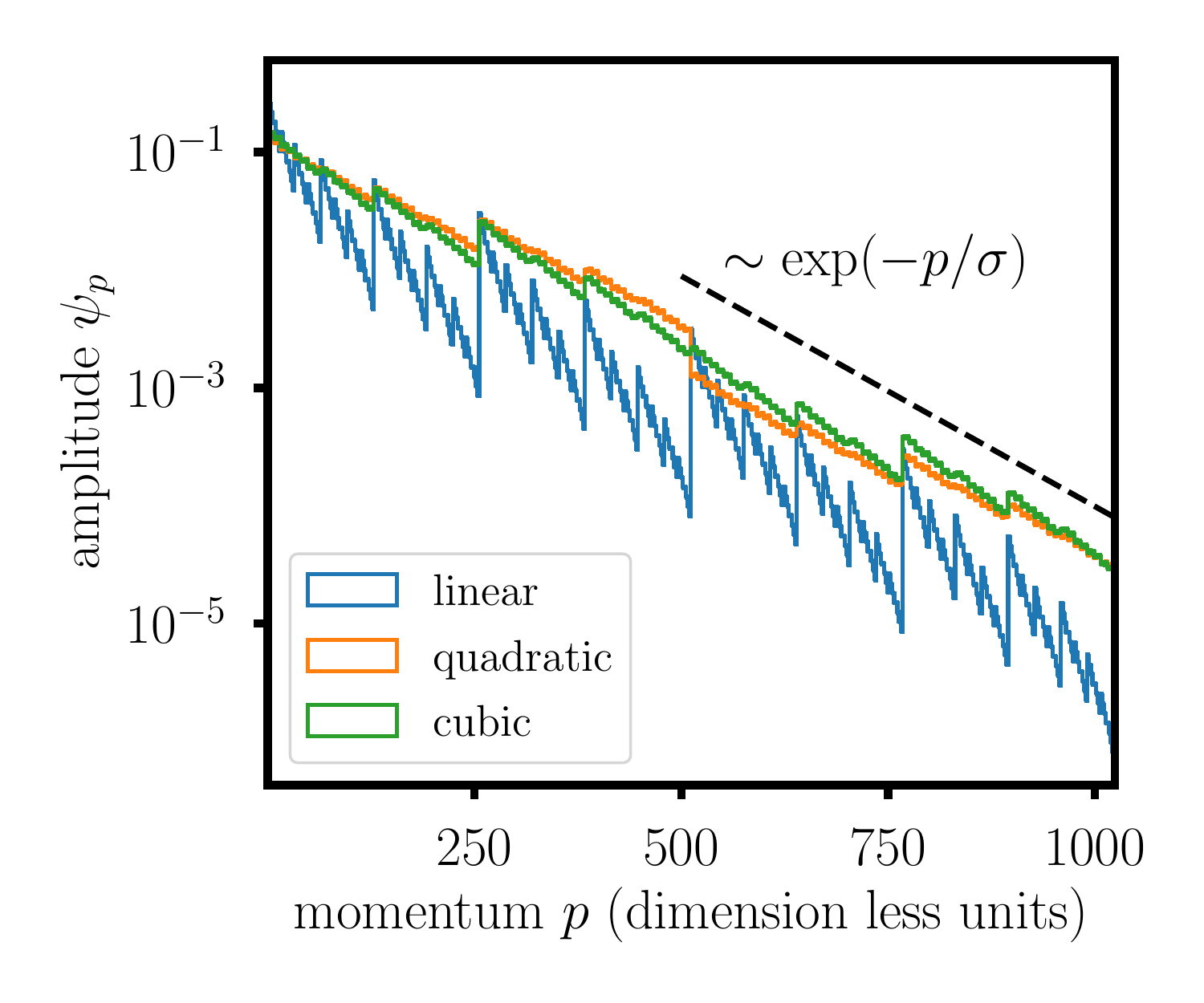}
\end{center}
\caption{Induced distributions for $\psi_p$ using the above polynomial maps. For comparison, we additionally plot an off-set exponentially decaying distribution, with $\sigma=100$.} 
\label{fig:reply}
\end{figure}
s
While the analytic maps \Eqs{eq:a}{eq:c}, which we worked out with pen and paper, provide some insight, in practice one would use a (classical) numerical optimization procedure
to determine the optimal map, without assuming a functional form, for a given target distribution.
In this case, the difference of our approach to that of~\cite{grover2002creating,kaye2004quantum} is that
the angles $\theta_k$ are determined simultaneously (`globally') 
while in~\cite{grover2002creating,kaye2004quantum} they are determined sequentially (the latter also requiring controlled operations). Our motivation for this ``classical out-sourcing'' was the relative simplicity of our approach over that of~\cite{grover2002creating,kaye2004quantum} in light of available near-term resources. However, being satisfied with our approach producing distributions relevant for our problem,  we do not know if it is also capable of producing more general distributions realizable with ~\cite{grover2002creating,kaye2004quantum}

We continue here with details of the Bose-symmetrization procedure 
discussed in section \ref{sec:InitialStatePrep}. The idea behind the algorithm is to start from an unsymmetrized state, work out all permutations of particle registers
that together give the symmetrized state (a simple combinatorial problem) and then use an ancilla register in a Bell superposition. Every 
state in this superposition is interpreted as the binary representation of a number labeling the respective Bose-permutations of the initial unsymmetrized state.
Each combination may then be used as the control qubits to execute a unique swap operation. 

A simple but non-trivial example is the case of $\mathfrak{n}=2$
initial wave packets in $M=3$ registers, where the Bose symmetrized state, obtained from the initial unsymmetrized state $\ket{\Omega, \Psi_1,\Psi_0}$, reads
\begin{align}\label{eq:desiredBose}
\frac{1}{\sqrt{6}}\Big[&\ket{\Omega,\Psi_1,\Psi_0}+\ket{\Omega,\Psi_0,\Psi_1}
+\ket{\Psi_0,\Psi_1,\Omega}+\ket{\Psi_1,\Psi_0,\Omega}
+\ket{\Psi_0,\Omega,\Psi_1}
+\ket{\Psi_1,\Omega,\Psi_0}
\Big]\, .
\end{align}
Following the recipe given in the main text, the number of possible Bose permutation for this $M$ and $\mathfrak{n}$ is not a power of two.
Using $s=3$ ancilla qubits in a Bell superposition in fact gives $2^3=8$ different permutations. Because of this 
the following state is generated:
\begin{align}\label{eq:bose_swap}
\ket{\Omega,\Psi_1,\Psi_0}\to \frac{1}{\sqrt{8}}\Big[&\ket{\Omega,\Psi_1,\Psi_0}\ket{0,0,0}+\ket{\Omega,\Psi_0,\Psi_1}\ket{0,0,1}
+\ket{\Psi_0,\Psi_1,\Omega}\ket{1,0,0}+\ket{\Psi_1,\Psi_0,\Omega}\ket{1,0,1}
\nn
+&\ket{\Psi_0,\Omega,\Psi_1}(\ket{0,1,1}+\ket{1,1,0})
+\ket{\Psi_1,\Omega,\Psi_0}(\ket{0,1,0}+\ket{1,1,1})
\Big]\, ,
\end{align}
where states $\ket{\Psi_0,\Omega,\Psi_1}$ and $\ket{\Psi_1,\Omega,\Psi_0}$ are now twice as likely as any other state. These unwanted permutations can be eliminated by introducing a single ancilla $\ket{0}$, and flipping it to 
$\ket{1}$ if either $| 1,1,0 \rangle$ or $|1,1,1\rangle$ is detected by a simple controlled $\sigma^x$ gate. If the ancilla 
is then measured in the $\ket{0}$ state, \Eq{eq:bose_swap} collapses onto \Eq{eq:desiredBose} with probability given by ratio of the number of desired terms in \Eq{eq:bose_swap} to the total number of states, in this specific example $p_{\rm success}=6/8$.

Note that although the number of basic gate operations depends on the number of measurements one needs to perform in order to eliminate all undesired states -- in the previous example at least two --  $p_{\rm success}$ only depends on $\{\mathfrak{n},M\}$. In the example above, if each of the two undesired states had been eliminated separately, the probability of preparing the correct symmetrized state would be $p_{\rm success}=(7/8)\times (6/7)=6/8$, as promised. In general, it is easily recognized that 
\begin{align}
p_{\rm success}=\frac{\mathcal{N}}{2^s} >\frac{1}{2}  \,  ,
\end{align}
with $\mathcal{N}=M!/(M-\mathfrak{n})!$ the number of Bose-permutations one needs to generate and $s$ an integer such that $2^s$ is the closest power of two to $\mathcal{N}$ from above,  $s=\lceil{\log_2(M!/(M-\mathfrak{n})!) }\rceil=O(\log (M^\mathfrak{n}))$.

\begin{figure}[h]
\begin{center}
\includegraphics[width=0.5\textwidth]{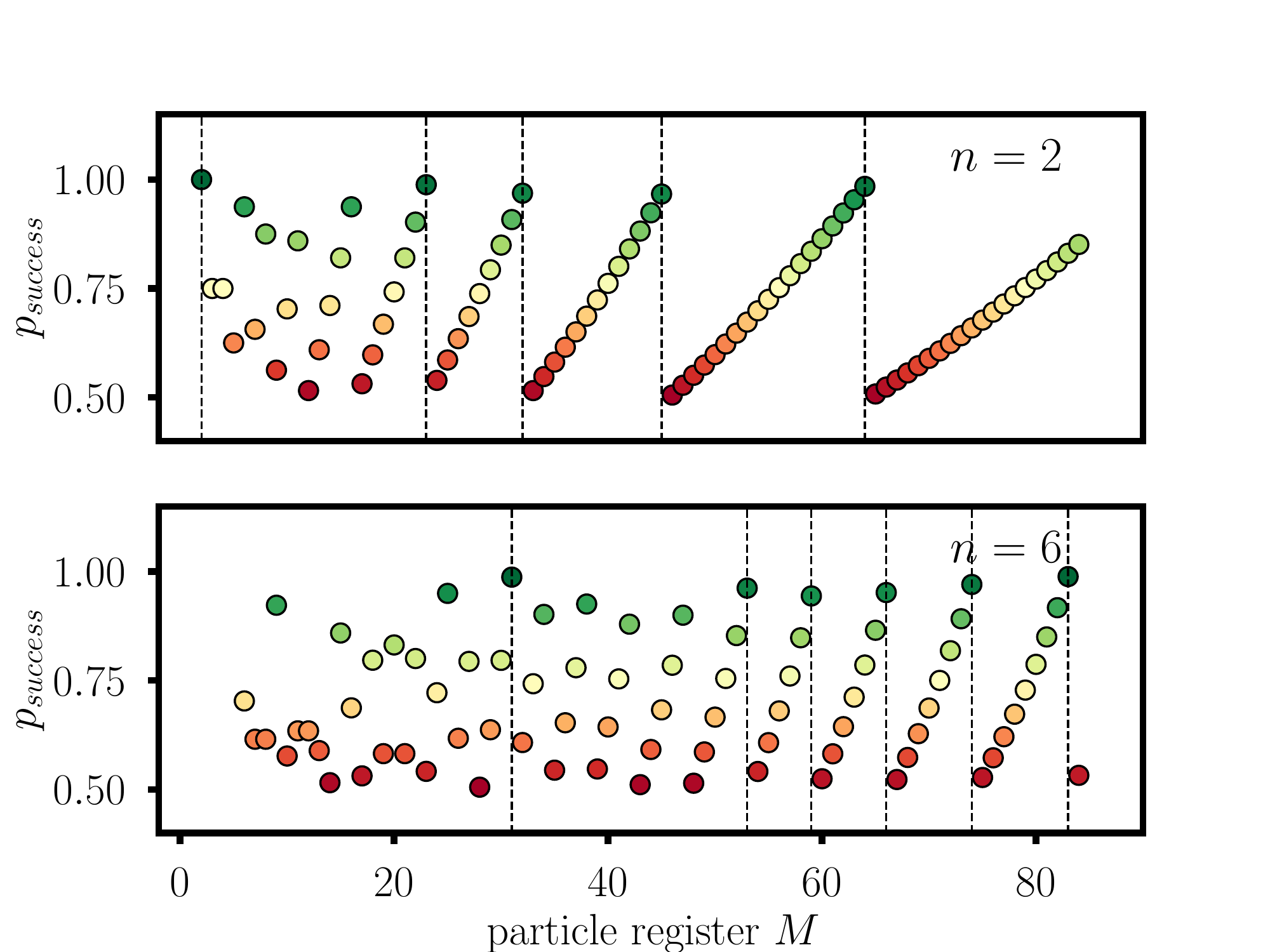}
\end{center}
\caption{Probability of preparing the correct Bose symmetric state $p_{\rm success}$ as a function of the number for single particle registers $M$, for $n=2$ (top) and $n=6$ (bottom) initial single particle states. The graduation in color between red and green is to guide the eye from low ($\approx 0.5$) to high ($\approx 1.0$) success probability, vertical dashed lines indicate values of $M$ that maximize $p_{\rm success}$.}
\label{fig:Prob_B}
\end{figure}

In \Fig{fig:Prob_B} we give some values for $n=2$ and $n=6$ as a function of $M$, with $6$ being the number of `particles' one would need to represent the quantum numbers of the proton in an extension of this work. Shown is the total probability of sucess for given choices of $M$, with the graduation in color from red to green guiding the eye from low ($\approx 0.5$) to high ($\approx 1.0$) probability. 
One would like to choose $M$ as large as possible, to minimize truncation effects, but this may not always be possible due to limited resources.  
However, as indicated by vertical dashed lines, one can always choose $M$ optimally in a reasonable range, so that  $p_{\rm success}$ is maximized.
%
%
%
%

The next step is to un-compute the $s$ ancilla qubits, as described in the main text. For $\mathfrak{n}=2$ this can be done using the occupation number qubit, as well as sign qubit,
because the initial wavepackets have opposite momentum in order to be able to interact.  The major difference for $\mathfrak{n}>2$ is that it is not sufficient to only use  sign and occupation number qubits alone to un-compute the ancilla qubits. In this case, one must also use $r$ of the qubits making up the momentum $\textbf{q}$ (or position $\xb$ after the respective transformation). Because the wavepackets are assumed widely separated, a small
number of qubits should suffice to un-compute the ancillas. The cost of un-computing the ancillas would increase from $ O(M^\mathfrak{n})$, to $\sim O(M^\mathfrak{n}r)\ll O(M^\mathfrak{n}\log(\mathcal{V}))$, where
$r\ll \log_2(\mathcal{V})$ is the number of qubits representing the momentum/position of each wavepacket which differ uniquely from each other. One then un-computes the $s'\le s$ ancillas
that are in the $|1\rangle$ state. Since one can choose $s'$ to be very small (compared to $s$) its contribution to the overall scaling estimate is subleading. 
Overall, the algorithm uses $s$ Hadamard gates to prepare the ancilla register, and $\mathcal{O}(2^s\log{\mathcal{V}})\sim \mathcal{O}(M^{\mathfrak{n}}\log{\mathcal{V}})$ controlled swap operations, and the un-computation of the ancilla register requires $O(M^{\mathfrak{n}})$ operations; the overall gate complexity is $\mathcal{O}(M^{\mathfrak{n}}\log{\mathcal{V}})$.

\section{Details of the Kinetic term}\label{app:kineticterm}
\begin{figure}[h]
\begin{center}
\includegraphics[width=0.75\textwidth]{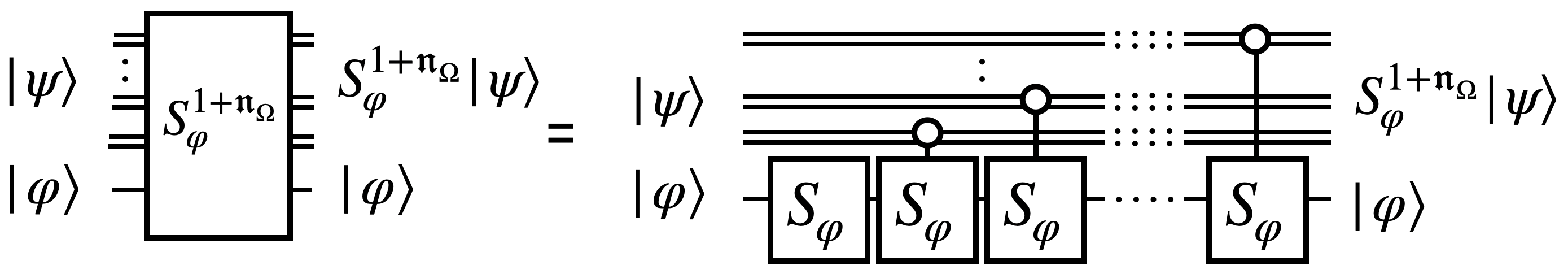}
\end{center}
\caption{Circuit implementing the final step in the time evolution dictated by $H_0$. The first \fbox{$S_{\scriptscriptstyle\varphi}$} gate contributes with $S_\varphi$ to the phase, while the last $M$ gates only contribute if controlled by a particle register in the vacuum state, thus generating the term proportional to $\mathfrak{n}_\Omega$ in the phase.}
\label{fig:kineticcircuit3}
\end{figure}

In this Appendix, we discuss the implementation of the gates \fbox{$\omega$} and \fbox{${\scriptstyle S}_{\scriptscriptstyle\varphi}^{\scriptscriptstyle 1+\mathfrak{n}_{\scriptscriptstyle \Omega}}$}, necessary for the algorithm introduced in Section \ref{sec:TimeEvolution}. 
The gate \fbox{$\omega$} takes as an input two registers, one of which is a particle register $\ket{\textbf{q}}$ and the other an ancilla register of $l$ qubits in the state $\ket{0^{\otimes l}}$. Under the action of this gate, the state $\ket{\textbf{q}}\otimes \ket{0^{\otimes l}}$ transforms to $\ket{\textbf{q}}\otimes \ket{\omega_\textbf{q}}$. Assuming that an efficient classical algorithm exists to compute $\omega_{\textbf{q}}$ for any $\textbf{q}$, and ensuring that for $| \Omega \rangle$, $\omega_\Omega =0$ (using the occupation number qubit as control), we treat \fbox{$\omega$} as a quantum oracle.
The gate implementing \Eq{eq:FreeTimeEvoDetails} is given in
\Fig{fig:kineticcircuit3} and consists on the sequential application of single controlled gates \fbox{${\scriptstyle S}_{\scriptscriptstyle\varphi}$} which takes the state $\ket{\psi}\otimes \ket{\varphi}$ to $\exp\left(-i\frac{\delta}{M} \varphi\right)\ket{\psi}\otimes \ket{\varphi}$.
%
%
%
%
This set of operations uses conditional single qubit phase shift gates \fbox{$C_{\scriptscriptstyle\phi}$} \cite{NielsenChuang,Phase_shift}, with matrix representation 
\begin{equation}\label{eq:qft_app}
C_\phi\equiv\begin{pmatrix}
1 & 0 \\
0 & e^{i\phi} 
\end{pmatrix}   \, ,
\end{equation}
where $\phi=-\frac{\delta}{M}2^d$ ($0\leq d \leq l-1$) chosen accordingly to the binary decomposition of $\varphi$. The full multi-qubit gate is constructed as a product of single qubit gates.
%
%
%
%
\section{Details of the Squeezing transformation}\label{app:squeezing}
In this Section, we will show that the operator $S$ realizes \Eq{eq:squeezingTrafo}. First, note that
\begin{align}\label{eq:app_sq_1}
 S  a_\textbf{q} S^\dagger = \prod_{\pb,\pb^\prime} e^ {-z_\pb(a_\pb^\dagger a_{-\pb}^\dagger - a_{-\pb} A_{\pb})}  a_\textbf{q}
 e^{z_\pb^\prime(a_{\pb^\prime}^{\dagger} a_{-\pb^\prime}^{\dagger} - a_{-\pb^\prime} a_{\pb^\prime})} \,.
\end{align}
Taking into account that $a_\pb$ and $a_\pb^\dagger$ obey the canonical commutation relations, \Eq{eq:app_sq_1} takes the form
\begin{align}
e^X a_\textbf{q} e^{-X} = \sum_{k=0}^\infty \frac{1}{k!} \underbrace{[X,[X,\dots [X,a_\textbf{q}]]}_{\text{k times}}\dots] \, ,
\end{align}
where $X\equiv -z_\textbf{q}(a_\textbf{q}^\dagger a_{-\textbf{q}}^\dagger - a_{-\textbf{q}} a_{\textbf{q}})$. Using the simple identities
\begin{align}
[X,a_\textbf{q}] = z_\textbf{q} a_{-\textbf{q}}^\dagger\,,\qquad [X,a^\dagger_{-\textbf{q}}] = z_\textbf{q} a_{\textbf{q}}\,,
\end{align}
it follows directly that for $z_\textbf{q}<0$
\begin{align}
e^X a_\textbf{q} e^{-X}  &=  \sum_{k=0}^\infty \frac{ (z_\textbf{q})^{2k}}{(2k)!} a_\textbf{q} +  \sum_{k=0}^\infty \frac{(z_\textbf{q})^{2k+1}}{(2k+1)!} a_{-\textbf{q}}^\dagger
=\cosh(z_\textbf{q}) a_\textbf{q} + \frac{z_\textbf{q}}{|z_\textbf{q}|} \sinh(z_\textbf{q}) a_{-\textbf{q}}^\dagger \, .
\end{align}

\begin{figure}[h]
\begin{center}
\includegraphics[width=0.7\textwidth]{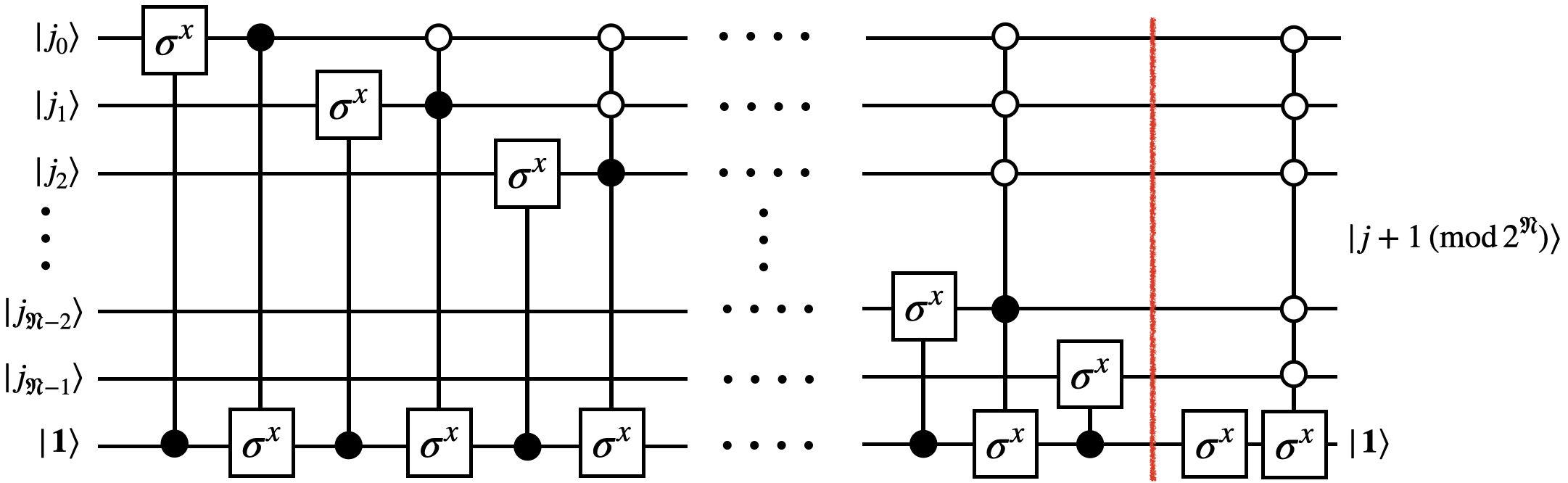}
\end{center}
\caption{Circuit implementing the bit increment operator $I_\mathfrak{N}$, introduced by Kaye~\cite{Kaye2005OptimizedQI,Kaye2004ReversibleAC}. The number of elementary quantum gate operations required scales as $O(\mathfrak{N}^2)$ for $\mathfrak{N}\geq3$, leading to the polylogarithm scaling mentioned in the main text. }
\label{fig:IN}
\end{figure}

In the implementation of the squeezing operation introduced in the main text, we made use of the bit increment  operator $I_\mathfrak{N}$, that performs the transformation $\ket{j}\to \ket{j+1 \, ({\rm mod \, 2}^\mathfrak{N})}$, where $\ket{j}=\ket{j_0,j_1,\cdots,j_{\mathfrak{N}-2},j_{\mathfrak{N}-1}}$ and $j_i\in \{0,1\}$ for any $i$. A decomposition of $I_\mathfrak{N}$ in terms of usual quantum gates is given in \Fig{fig:IN}, an alternative formulation is given in Eq.(47) of~\cite{shaw2020quantum}.

The implementation of $I_\mathfrak{N}$ in terms of \Fig{fig:IN} uses the fact that unitary increments in the binary basis consist in consecutively flipping all qubits, i.e. $\ket{0}\to \ket{1}$ and $\ket{1}\to\ket{0}$, while keeping track of the first time the state $\ket{0}$ is given as an input qubit. To do this, a flag ancilla qubit is prepared in the $\ket{1}$ state and it is only flipped back to $\ket{0}$ just after one performs the transformation $\ket{0}\to\ket{1}$ (on an input qubit); all possible remaining qubit flips are skipped. This operation is performed by the circuit detailed to the left of the vertical red (color online) line in \Fig{fig:IN}. In the end, one un-computes the ancilla back to the state $\ket{1}$ via a single $\sigma^x$ gate. The special (boundary) case $\ket{1,1,\cdots,1}\otimes\ket{1}\to\ket{0,0,\cdots,0}\otimes\ket{1}$ has the ancilla un-computed by the last gate in the diagram shown.
%
%
%

%
%
\section{Details of the Interaction term}\label{app:interactionterm}
In this Appendix, we will discuss how to explicitly construct the operator $U^{\rm diag}_{I,\mathbf{n}}$. We illustrate the algorithm for the simplest example $\mathbf{n}=-1/2$ and $M=4$.
 The generalization for all $\mathfrak{n}$ and $M$ is discussed below.
 
 For this simple example, $\phi^{(i) \, \rm diag}_{-1/2}$ is simply the $\sigma^z$ operator acting only on the occupancy qubit of register $i$;  see \Eq{eq:a_S_map}. For $M=4$, $U^{\rm diag}_{I,-1/2}$ acts only on the respective occupancy qubits of the four particle registers.
Using the fact that $(\sigma^z)^2=1$, we can write $U^{\rm diag}_{I,-1/2}$ as
 \begin{align}\label{eq:eq_ap_phi4_1}
U^{\rm diag}_{I,-1/2}&\equiv 
\exp\left\{-i \Delta \sum_{s=0}^2 c_{s,-1/2}\mathcal{O}_{s,-1/2}\right\} \,.
\end{align}
The three distinct operators appearing in \Eq{eq:eq_ap_phi4_1} are $\mathcal{O}_{0,-1/2}=1^{\otimes 4} $, $\mathcal{O}_{1,-1/2}=(\sigma^z)^{\otimes 4} $ and $\mathcal{O}_{2,-1/2}=\mathcal{P}_{\small \Sigma}(1\otimes 1\otimes\sigma^z\otimes
\sigma^z)$, with coefficients $c_{0,-1/2}=4!(4+12)$, $c_{1,-1/2}=4!$ and $c_{2,-1/2}=4!(2+1)$. Here $\mathcal{P}_{\small \Sigma}(\hat{X})$ stands for the sum over all permutations of the operator $\hat{X}$ in the tensor product. Each operator is simply a product of standard Pauli $z$-rotations~\cite{NielsenChuang}. 
The generalization of \Eq{eq:eq_ap_phi4_1} to arbitrary $\mathbf{n}$ (and $M$) requires replacing $\sigma^z$ by its  higher dimensional analogue, given in Section \ref{sec:interactionpart}. For $M>4$ one has to repeat the algorithm for all $M(M-1)(M-2)(M-3)/4!\sim O(M^4)$ possible four-tuples formed out of $M$ registers.

%
%
\section{Details of the Renormalization procedure}\label{app:renormalization}
In this Appendix, we present some details of the renormalization procedure.
Concretely, for weak coupling \Eq{eq:RGgroupHamiltonian}
can be expanded as
\begin{align}
H^{\rm eff} &= H + [i\eta,H] + \frac{1}{2!}[i\eta[i\eta,H]]+\dots
= H_0 + H_I +[i\eta,H_0] + [i\eta,H_I] 
+ \frac{1}{2}[i\eta,[i\eta,H_0]] + O(\lambda^3)\,,
\end{align}
where $H=H_0+H_I$ and $H_I\sim O(\lambda)$, $\eta\sim O(\lambda)$. We label  eigenstates $H_0| \alpha, i\rangle = E_{\alpha,i} | \alpha, i\rangle$, where $\alpha = l,h$ denote low and high energy sectors (the computational basis states of Section \ref{sec:strategies_for_quantum_simulation}).
To block-diagonalize $H$ such that $\langle \alpha,i | H^{\rm eff} | \beta,j \rangle=0$ if $\alpha\neq \beta$, we require that the diagonal elements of $i\eta$ vanish, $\langle \alpha, i| i\eta | \alpha,j\rangle=0$, and we set 
$\langle \alpha, i | i\eta | \beta,j\rangle = {\langle \alpha,i | H_I | \beta,j\rangle}/({E_{\alpha,i}-E_{\beta,j}})$ for $\alpha\neq \beta$. With this, 
 the off-diagonal elements of $H^{\rm eff}$  cancel to $O(\lambda^2)$.
In this case, $H^{\rm eff} = H_0 + H_I +\frac{1}{2}[i\eta,H_I]+O(\lambda^3)$,
with the low energy matrix elements given by
\begin{align}\label{eq:perturbativeHamilt}
\langle l,i | H_{\rm eff} | l,j\rangle 
= \langle l,i| H | l,j \rangle + \frac{1}{2}\sum_{k} \langle l,i | H_I | h,k \rangle \langle h,k | H_I | l,j \rangle 
\, \Big[ \frac{1}{E_{l,i} - E_{h,k}} + \frac{1}{E_{l,j} - E_{h,k}}    \Big]\,.
\end{align}
The same transformation applies to any operator $\mathcal{O}_{\rm eff} = T\mathcal{O}T^\dagger$, which can be expressed as $\langle l,i | \mathcal{O}_{\rm eff} | l ,j\rangle  = \langle l,i | \mathcal{O} | l ,j\rangle  + \langle l,i |  \Delta \mathcal{O}| l,j\rangle$. For the matrix elements for an observable diagonal in the eigenbasis of $H_0$ (such as particle number), this reads as\footnote{This formalism is analogous to a Poisson bracket formalism invented in the context 
of weak wave turbulence in fluids~\cite{Zakharov}. Interestingly, it has been exploited recently to study the self-similar infrared behavior of a scalar $\phi^4$ theory far-off-equilibrium~\cite{deng2018off}.}
\begin{align}\label{eq:almostfinal}
\langle l,i |  \Delta \mathcal{O}| l,j\rangle& = \sum_k \Big\{ \frac{\langle l,i | H_I | h,k \rangle}{E_{l,i}-E_{h,k}}\frac{\langle h,k| H_I|l,j\rangle}{E_{h,l}-E_{l,j}} \,
\frac{1}{2}[\mathcal{O}_j^l + \mathcal{O}_i^l ]
-\frac{\langle l,i| H_I | h,k \rangle }{E_{h,k} - E_{l,j}}\mathcal{O}_k^h \frac{\langle h,k| H_I|l,j\rangle}{E_{l,i}-E_{h,k}}\Big\}\,,
\end{align}
where we abbreviated $\langle l, i | \mathcal{O} | l,j\rangle \equiv \mathcal{O}^{l}_i\delta_{ij}$.
The procedure outlined can in principle be continued to arbitrary order $O(\lambda^n)$.

The generalization of \Eq{eq:almostfinal} to an operator that is not diagonal in
the $H_0$ eigenbasis is
\begin{align}
\langle l,i |  \Delta \mathcal{O}| l,j\rangle &= \sum_k \Big\{  \frac{\langle l,i | H_I | h,k \rangle}{E_{l,i}-E_{h,k}} \langle h,k | \mathcal{O} | l,j \rangle 
 \langle l,i | \mathcal{O} | h,k \rangle \frac{\langle h,k | H_I | l,j \rangle}{E_{l,j}-E_{h,k}}
+\frac{1}{2} \sum_{k,m} \Big\{ \frac{\langle l,i| H_I| h,k \rangle}{E_{l,i}-E_{h,k}} \frac{\langle h,k | H_I| l,m \rangle}{E_{h,k}-E_{l,m}} \langle l,m | \mathcal{O}| l,j\rangle
\nn
&\quad- \frac{\langle l,i | H_I | h,k \rangle}{E_{l,i}-E_{h,k}} \langle h,k | \mathcal{O} | h,m\rangle \frac{\langle h,m| H_I | l,j\rangle}{E_{h,m}-E_{l,j}}
-\frac{\langle l,i|H_I|h,m\rangle}{E_{l,i}-E_{h,m}} \langle h,m| \mathcal{O} | h,k \rangle \frac{\langle h,k| H_I | l,j\rangle}{E_{h,k}-E_{l,j}}
\nn 
& \quad+\langle l,i| \mathcal{O} | l,m\rangle \frac{\langle l,m| H_I| h,k \rangle }{E_{l,m}-E_{h,k}}\frac{\langle h,k| H_I| l,j \rangle }{E_{h,k}-E_{l,j}}\Big\}\,.
\end{align}
To generalize the renormalization procedure beyond weak coupling, one may use Wegner's formulation of an infinitesimal operator renormalization group~\cite{wegner1994flow} whereby states inside an energy shell of width $\delta$ around the cutoff $\Lambda$ are integrated: $ H(\Lambda - n \delta) = T(n) H(\Lambda) T^\dagger (n)$
with $T(n)=\exp(i\eta(n))$, $H(\Lambda- N\delta) =H^{\rm eff}$ after a number of RG steps $N$, and $\eta(n) = [H_d(n),H(n)]$. Here $H_d(n)$ is the diagonal part of the Hamiltonian obtained after $n\le N$ steps. The Hamiltonian $H(\Lambda\rightarrow \infty)$ is usually not known,
and in practice one starts from an ansatz for $H_{ll}^{\rm eff}$ at finite $\Lambda$, such as \Eq{eq:latticeHamiltonian}, and takes the continuum limit as described in Section \ref{sec:Renormalization}. Classical numerical procedures have been derived from Wegner's operator RG~\cite{bogner2010low} and it would be interesting to explore their use in quantum computation.

\end{document}